\documentclass[12pt]{article}
\usepackage{amsmath,amssymb,amsfonts,color,graphicx,cite,feynarts,soul}
\usepackage[usenames,dvipsnames,svgnames,table]{xcolor}
\input paperdef

\graphicspath{{figs/}}

\evensidemargin \oddsidemargin
\allowdisplaybreaks

\begin{document}
\thispagestyle{empty}

\def\thefootnote{\fnsymbol{footnote}}

\vspace{0.5cm}

\begin{center}

{\large\sc {\bf 
Direct Chargino-Neutralino Production at the LHC:
}}

\vspace{0.4cm}

{\large\sc {\bf Interpreting the Exclusion Limits in the Complex MSSM
}}

\vspace{1cm}

{\sc
A.~Bharucha$^{1}$%
\footnote{email: Aoife.Bharucha@desy.de}%
, S.~Heinemeyer$^{2}$%
\footnote{email: Sven.Heinemeyer@cern.ch}%
, F.~von der Pahlen$^{2}$%
\footnote{email: pahlen@ifca.unican.es}%
\footnote{MultiDark Fellow}%
}

\vspace*{.7cm}

{\sl
$^1$II.~Institut f\"ur Theoretische Physik, Universit\"at Hamburg, Lupurer Chaussee 149, D--22761 Hamburg, Germany

\vspace{0.1cm}

$^2$Instituto de F\'isica de Cantabria (CSIC-UC), E-39005 Santander, Spain

}

\end{center}

\vspace*{0.1cm}

\begin{abstract}
\noindent
We re-assess the exclusion limits on the parameters describing the
supersymmetric (SUSY) electroweak sector of the MSSM obtained from
the search for direct chargino-neutralino production at the LHC.
We start from the published limits obtained for simplified models, where 
for the case of heavy sleptons the relevant branching ratio,
$\br(\neu{2} \to \neu{1} Z)$, is set to one. We
show how the decay mode $\neu{2} \to \neu{1} h$, 
which cannot be neglected in any realistic model once kinematically
allowed, substantially reduces the excluded
parameter region. We analyze the dependence of the excluded regions on
the phase of the gaugino soft SUSY-breaking mass parameter, $M_1$, on
the mass of the light scalar tau, $\mstaul$, on $\tb$ as well as on the
squark and slepton mass scales. Large reductions in the ranges 
of parameters excluded can be observed in all scenarios.
The branching ratios of charginos and neutralinos are evaluated
using a full NLO calculation for the complex MSSM. 
The size of the effects of the NLO calculation on the exclusion bounds
is investigated.
We furthermore assess the potential reach of the experimental analyses
after collecting $100\,\ifb$ at the LHC running at $13 \tev$.
\end{abstract}
\vspace*{.5cm}

\def\thefootnote{\arabic{footnote}}
\setcounter{page}{0}
\setcounter{footnote}{0}
\newpage

\newcommand{\DecCCh[1]}{\cha{2} \to \cha{1} h_{#1}}
\newcommand{\DecCCZ}{\cha{2} \to \cha{1} Z}
\newcommand{\DecCNH}[2]{\cha{#1} \to \neu{#2} H^\mp}
\newcommand{\DecCNW}[2]{\cha{#1} \to \neu{#2} W^\mp}
\newcommand{\DecCnSl}[3]{\cham{#1} \to \nu_{#2}\, \tilde{#2}_{#3}^{-}}
\newcommand{\DecClSn}[2]{\cham{#1} \to {#2}^{-}\, \tilde{\nu_{#2}}}
\newcommand{\DecCxy}[1]{\cha{#1} \to {\rm xy}}

\newcommand{\DecNCW}[2]{\neu{#1} \to \champ{#2} W^\pm}
\newcommand{\DecNCH}[2]{\neu{#1} \to \champ{#2} H^\pm}
\newcommand{\DecNlpmSl}[3]{\neu{#1} \to {#2}^\mp\, \tilde{#2}_{#3}^\pm}
\newcommand{\DecNnSnnbarSn}[2]{\neu{#1} \to \bar{\nu}_{#2}\, \tilde{\nu}_{#2}  /  {\nu}_{#2}\, \tilde{\nu}_{#2}^\dagger  }

\newcommand{\decCCh}{\DecCCh{k}}
\newcommand{\decCCZ}{\DecCCZ}
\newcommand{\decCNH}{\DecCNH{i}{j}}
\newcommand{\decCNW}{\DecCNW{i}{j}}
\newcommand{\decCnSl}{\DecCnSl{i}{l}{k}}
\newcommand{\decClSn}{\DecClSn{i}{l}}
\newcommand{\decCxy}{\DecCxy{i}}

\newcommand{\DecNxy}[1]{\neu{#1} \to {\rm xy}}
\newcommand{\DecNCmH}[2]{\neu{#1} \to \cham{#2} H^+}
\newcommand{\DecNCmW}[2]{\neu{#1} \to \cham{#2} W^+}
\newcommand{\DecNCpH}[2]{\neu{#1} \to \chap{#2} H^-}
\newcommand{\DecNCpW}[2]{\neu{#1} \to \chap{#2} W^-}
\newcommand{\DecNNh}[3]{\neu{#1} \to \neu{#2} h_{#3}}
\newcommand{\DecNNZ}[2]{\neu{#1} \to \neu{#2} Z}
\newcommand{\DecNnSn}[2]{\neu{#1} \to {\nu}_{#2}\, \tilde{\nu}_{#2}^\dagger}
\newcommand{\DecNlSl}[3]{\neu{#1} \to {#2}^-\, \tilde{#2}_{#3}^+}
\newcommand{\DecNbarnSn}[2]{\neu{#1} \to \bar{\nu}_{#2}\, \tilde{\nu}_{#2}}
\newcommand{\DecNlpSl}[3]{\neu{#1} \to {#2}^+\, \tilde{#2}_{#3}^-}

\newcommand{\decNCH}{\DecNCH{i}{j}}
\newcommand{\decNCW}{\DecNCW{i}{j}}
\newcommand{\decNlpmSl}{\DecNlpmSl{i}{\ell}{k}}
\newcommand{\decNnSnnbarSn}{\DecNnSnnbarSn{i}{\ell}}

\newcommand{\decNCmH}{\DecNCmH{i}{j}}
\newcommand{\decNCmW}{\DecNCmW{i}{j}}
\newcommand{\decNNh}{\DecNNh{i}{j}{k}}
\newcommand{\decNNZ}{\DecNNZ{i}{j}}
\newcommand{\decNnSn}{\DecNnSn{i}{\ell}}
\newcommand{\decNlSl}{\DecNlSl{i}{\ell}{k}}
\newcommand{\decNbarnSn}{\DecNbarnSn{i}{\ell}}
\newcommand{\decNlpSl}{\DecNlpSl{i}{\ell}{k}}

\newcommand{\decNxy}{\DecNxy{i}}

\newcommand{\DecCmCh}[1]{\cham{2} \to \cham{1} h_{#1}}
\newcommand{\DecCmCZ}{\cham{2} \to \cham{1} Z}
\newcommand{\DecCmNH}[2]{\cham{#1} \to \neu{#2} H^-}
\newcommand{\DecCmNW}[2]{\cham{#1} \to \neu{#2} W^-}
\newcommand{\DecCmnSl}[3]{\cham{#1} \to \bar{\nu}_{#2}\, \tilde{#2}_{#3}^-}
\newcommand{\DecCmlSn}[2]{\cham{#1} \to {#2}^-\, \tilde{\nu}_{#2}^\dagger}
\newcommand{\DecCmxy}[1]{\cham{#1} \to {\rm xy}}

\newcommand{\decCmCh}{\DecCmCh{k}}
\newcommand{\decCmCZ}{\DecCmCZ}
\newcommand{\decCmNH}{\DecCmNH{i}{j}}
\newcommand{\decCmNW}{\DecCmNW{i}{j}}
\newcommand{\decCmnSl}{\DecCmnSl{i}{l}{k}}
\newcommand{\decCmlSn}{\DecCmlSn{i}{l}}
\newcommand{\decCmxy}{\DecCmxy{i}}


\section{Introduction}

The LHC is actively searching for physics beyond the Standard
  Model (BSM). Many of those searches
rely on the predictions of specific
models. A well motivated model is the Minimal Supersymmetric 
Standard Model (MSSM)~\cite{mssm}, which provides a
framework in which such predictions can be made. 
Provided
$R$~parity is conserved~\cite{rpv}, the final particle of any SUSY
decay chain is the lightest supersymmetric particle (LSP), e.g.\ the
lightest neutralino. It was shown that this particle is a
natural candidate for Cold Dark Matter
(CDM)~\cite{cdm}. The MSSM also contains a rich Higgs
phenomenology, 
particularly relevant in light of the 
exciting recent discovery by ATLAS and CMS of a scalar resonance at
$\sim 125 \gev$~\cite{discovery}, as the requirement of an
additional Higgs doublet results in a total of five physical Higgs
bosons, the light and heavy $\cp$-even Higgs bosons, $h$ and $H$,
the $\cp$-odd $A$ and the charged Higgs bosons, $H^\pm$.

The search for supersymmetry (SUSY) at the LHC has not (yet) led to
a positive result. In particular,
bounds on the first and second generation squarks and the gluinos
from ATLAS and CMS are very roughly at the TeV scale, depending on
details of the 
assumed parameters, see e.g.~\cite{non-deg-squarks}. 
On the other hand, bounds on the
electroweak SUSY sector, where $\cha{1,2}$ and $\neu{1,2,3,4}$ denote
the charginos and neutralinos (i.e.\ the charged (neutral) SUSY
partners of the SM gauge and Higgs bosons) are substantially
weaker. Here it should be noted that 
models based on Grand Unified Theories (GUTs) naturally predict a
lighter electroweak 
spectrum (see \citere{guts} and references therein). Furthermore, the
anomalous magnetic moment of the muon shows a more than $\sim 4\si,$
deviation from the SM prediction, see \citere{fredl-gm2} and references
therein. Agreement of this measurement with the MSSM requires
charginos and neutralinos in the range of several hundreds of~GeV.%
~This provides a strong motivation for the search of these electroweak
particles, which
could be in the kinematic reach of the LHC.
One promising channel is
the {\em direct} production of a chargino and neutralino,
$pp \to \cha{1}\neu{2} \; (+X)$. Although the cross
sections are generically lower than for the direct production of
colored particles, the searches at ATLAS and CMS have lead to
several limits in the range of the order of several hundreds of~GeV,
see e.g.~\citeres{ATLAS:2012ku,CMS:2012ewa}, which are
{\em independent} of the mass scale of colored SUSY particles.
The highest sensitivity comes from multi-lepton final states, including
tau leptons, which offer the possibility to distinguish a signal from
the large hadronic background. As an additional advantage
the theoretical calculation is cleaner than that of the production of
electroweak SUSY particles via cascade decays of colored particles. 

Early on, studies had proposed the clean trilepton signal at the LHC, coming
either from intermediate sleptons (particularly of
the first and second generation) or gauge boson decays
to which the experiments are more sensitive (see e.g.~\citere{Baer:1994nr}). %
More recently
there have been several studies investigating the LHC
sensitivity to this decay mode, e.g.\ in the region where decays to
trileptons via $W$ and $Z$ bosons dominate
~\cite{Baer:2012wg}. 
Direct production of electroweak SUSY particles has also been investigated 
in more challenging scenarios 
where the lightest chargino and two neutralinos are higgsino-like, 
and thus nearly degenerate, such that their decay signals
are lost in the SM background~\cite{Baer:2011ec},
but a same sign diboson signal from gaugino production could be detectable for
wino masses up to $550\gev$ with $100~\ifb$ at
LHC14~\cite{Baer:2013yha}.
Other recent studies have focused on
improving the reach of searches using e.g.\ a kinematic observable, 
the visible transverse energy for $WZ+E_T^{\rm miss}$ final
states~\cite{Cabrera:2012gf}. Furthermore, there has been an
increasing interest in the $Wh+E_T^{\rm miss}$ final state~\cite{Baer:2012ts}, 
including a $h \to b \bar b$ decay,
for which the use of jet substructure was found to improve
results~\cite{Byakti:2012qk,Ghosh:2012mc}. This improvement is
particularly helpful in gauge-mediated scenarios, in which the GUT
relation is broken ($M_2 \nsim 2 M_1$), and boosted Higgs bosons could be
observed with $15~\ifb$ at LHC14
for values of $M_2\sim 200 \gev$ -- $300 \gev$~\cite{Byakti:2012qk}. 

As discussed before, ATLAS~\cite{ATLAS:2012ku,ATLAS:2012-154,ATLAS:2013-028,ATLAS:2013-035,ATLAS:2013-049} 
and CMS~\cite{CMS:2012ewa,CMS:2012ewb} are actively searching for the direct
production of charginos and neutralinos, in particular for the process
$pp \to \cha{1}\neu{2}$ with the subsequent decays 
$\cha{1} \to \neu{1} W^\pm$ and $\neu{2} \to \neu{1} Z$, 
resulting a three lepton signature.
These searches are performed mostly in so-called ``simplified models'',
where the branching ratios of the relevant SUSY particles are set
to one, assuming that all other potential decay modes are kinematically
forbidden.
The results are (often) presented in the $\mneu{2}$--$\mneu{1}$
parameter plane.

To compare with these experimental results, precise predictions for 
the $WZ+E_T^{\rm miss}$ final state are required, involving both 
calculations for the gaugino production cross section and
the branching ratios of the subsequent chargino and neutralino decays.
The production cross section in the $\cp$-conserving real MSSM
(rMSSM) was calculated at NLO and incorporated into the code 
{\tt Prospino~2.1}~\cite{prospinoNN}, as well as at NLL accuracy and
investigated in the context of the LHC8%
\footnote{With LHC$x$ we denote the LHC running at 
$\sqrt{s} = x \tev$.}%
~in \citere{Fuks:2012qx}. 
Chargino and neutralino cross sections for the LHC8 in the complex MSSM
have not been analyzed so far. 
Chargino and neutralino decays have been
calculated at the one-loop level in the
rMSSM \cite{BaroII,GRACESUSY,liebler,Drees:2006um} and in the
complex MSSM~\cite{ChaDecCPVYang,ChaDecCPVEberl,aoife-decays,LHCxC,LHCxN,Bharucha:2012ya},
where our evaluations 
are based on the first full one-loop (NLO) calculation (of all
non-hadronic decays) presented in \citeres{LHCxC,LHCxN}.
A phenomenological analysis in the complex MSSM, where these
state-of-the-art results are combined to make predictions for the LHC
is still lacking. 
Turning to the neutralino decays, in \citere{LHCxN} the NLO results for all possible 
neutralino decays were considered as a function of $\phiMe$,
under the assumption that colored particles are kinematically
excluded. It was found that a change of the phase of 
$\MOne = |\MOne| e^{i\phiMe}$ can
significantly alter the dominant decay mode when the decay modes to
neutralinos and Higgs bosons are allowed. The NLO corrections have been
found to be sizeable, particularly for channels involving Higgs
bosons. \bigskip

In this paper we define and analyze a set of scenarios for the
production and decay of charginos and neutralinos at the LHC8, 
where we take $\mneu{2}$ and
$\mneu{1}$ as free parameters.
The starting point is the scenario used by ATLAS to present their results
for $21~\ifb$ \cite{ATLAS:2013-035}, which so far constitutes the
most sensitive test of {\em direct} electroweak SUSY production. We
show the effect on the chargino and neutralino searches
of the inclusion of the decay
$\neu{2} \to \neu{1} h$ (with $\Mh \approx 125 \gev$). 
Direct and indirect effects from decays to the recently discovered Higgs
boson~\cite{discovery} 
must not be neglected in any realistic analysis. 
Subsequently, we deviate from
the ATLAS scenario in several ways, motivated by current limits on
the MSSM parameter space. In particular, we vary 
the phase of the gaugino soft SUSY-breaking mass parameter, $\phiMe$,
which has a strong impact on the branching ratios of the $\neu{2}$ and
thus on the limits of the exclusion regions in the
$\mneu{2}$--$\mneu{1}$ plane. We furthermore analyze the scenario with a
light scalar tau in the so-called $\Stau$-coannihilation region, where
the $\neu{1}$ provides a good CDM candidate.
We vary other parameters, such as $\tb$ (the ratio of the two vacuum
expectation values of the two Higgs doublets, $\tb = v_2/v_1$), 
the higgsino mass parameter $\mu$,
and the masses that set the scale for the scalar leptons or the scalar quarks.
By analyzing these variations we aim to provide a more realistic
interpretation of current ATLAS and CMS limits on the electroweak SUSY
particles. 
Finally we investigate which limits can be expected from the first 
$100~\ifb$ at the LHC13 (i.e.\ with $\sqrt{s} = 13 \tev$), which could
be obtained in the years 2015-2017.

The paper is organized as follows:
We begin with a short review of the relevant parameters and
couplings as well as the calculations employed in our analysis in
\refse{sec:calc}. Then in \refse{sec:bench} we review in more detail the
existing experimental analyses and define the various
scenarios in which the analysis will be performed. \refse{sec:results}
contains the numerical results, i.e.\ the re-interpretation of the
existing mass limits in the benchmark scenarios, as well as our
extrapolation to the LHC13. We conclude in \refse{sec:conc}.


\section{Details of the calculation}
\label{sec:calc}

In this section, after having introduced the necessary notation, we will
illustrate the dependence of the couplings on the fundamental parameters
via a simple expansion, and then go on to describe the details of 
the calculations employed in our (NLO) analysis in
\refse{sec:results}.


\subsection{Notation}

In the chargino case, two $2 \times 2$ matrices $\matr{U}$ and
$\matr{V}$ are necessary for the diagonalization of the chargino mass
matrix~$\matr{X}$, 
\begin{align}
\matr{M}_{\cham{}} = \matr{V}^* \, \matr{X}^\top \, \matr{U}^{\dagger} =
  \begin{pmatrix} m_{\tilde{\chi}^\pm_1} & 0 \\ 
                  0 & m_{\tilde{\chi}^\pm_2} \end{pmatrix}  \quad
\text{with} \quad
  \matr{X} =
  \begin{pmatrix}
    \MTwo & \sqrt{2} \sinb \MW \\
    \sqrt{2} \cosb \MW & \mu
\label{eq:X}
  \end{pmatrix}~,
\end{align}
where $\matr{M}_{\cham{}}$ is the diagonal mass matrix with the chargino
masses $\mcha{1}, \mcha{2}$ as entries, which are determined as the
(real and positive) singular values of $\matr{X}$ and
$\MW$ is the mass of the $W$~boson. 
The singular value decomposition of $\matr{X}$ also yields results for 
$\matr{U}$ and~$\matr{V}$. 

In the neutralino case, as the neutralino mass matrix $\matr{Y}$ is
symmetric, one $4 \times 4$~matrix is sufficient for the diagonalization
\begin{align}
\matr{M}_{\neu{}} = \matr{N}^* \, \matr{Y} \, \matr{N}^{\dagger} =
\text{\bf diag}(m_{\neu{1}}, m_{\neu{2}}, m_{\neu{3}}, m_{\neu{4}})
\end{align}
with
\begin{align}
\matr{Y} &=
  \begin{pmatrix}
    \MOne                  & 0                & -\MZ \, \sw \cosb
    & \MZ \, \sw \sinb \\ 
    0                      & \MTwo            & \quad \MZ \, \cw \cosb
    & -\MZ \, \cw \sinb \\ 
    -\MZ \, \sw \cosb      & \MZ \, \cw \cosb & 0
    & -\mu             \\ 
    \quad \MZ \, \sw \sinb & -\MZ \, \cw \sinb & -\mu              & 0
  \end{pmatrix}~.
\label{eq:Y}
\end{align}
$\MZ$ is the mass of the $Z$~boson,
$\cw = \MW/\MZ$ and $\sw = \sqrt{1 - \cw^2}$. 
The unitary 4$\times$4 matrix $\matr{N}$ and the physical neutralino
(tree-level) masses $\mneu{k}$ ($k = 1,2,3,4$) result from a numerical Takagi 
factorization \cite{Takagi} of $\matr{Y}$. 

When working in the complex MSSM
it should be noted that the results for physical observables are
affected only by certain combinations of the complex phases of the 
parameters. It is possible, for instance, to rotate the phase
$\phiMz$ away, which we adopt here. In this case 
the phase $\phimu$ is tightly constrained~\cite{plehnix}. 
Consequently, we take $\mu$ to be a real parameter.
Further note that in the case of the complex MSSM, the three neutral Higgs
bosons $h$, $H$ and $A$ mix at the loop
level~\cite{mhiggsCPXgen,Demir,mhiggsCPXRG1,mhiggsCPXFD1},
resulting in the (mass ordered) $\He$, $\Hz$ and $\Hd$,
which are not states of definite~$\cp$-parity.
In the following we denote the light Higgs with $\He$, independent
whether the parameters are chosen complex or real. The Higgs sector
predictions have been derived with 
{\tt FeynHiggs~2.9.4}~\cite{feynhiggs,mhiggslong,mhiggsAEC,mhcMSSMlong}.


\subsection{\boldmath{$\phiMe$} dependence of neutralino amplitudes}
\label{sec:amplitudes}

In this section we investigate the $\phiMe$ dependence of amplitudes for
$\neu2$ decays in the limit
$\mu\gg |\MOne|, \MTwo$; $\MHp\gg\MZ$; $\tb\gg 1$, 
which will be relevant for most of the analyzed benchmark scenarios.
The full $\neu{i}\neu{j}Z/\He$ 
couplings take the form (with $e$ denoting the electric charge, 
$\al_{\rm em} = e^2/(4\,\pi)$, and $\al$ is the angle that diagonalizes the
$\cp$-even Higgs sector at tree-level)
\begin{align}
C^L_{\neu{i}\neu{j}Z} 
&= -\frac{e}{2\cw\sw}[N_{{i}3}N_{{j}3}{^{*}} -  N_{{i}4}N_{{j}4}{^{*}}]
\label{eq:cnnz}
\\
C^L_{\neu{i}\neu{j}h_1} &= -\frac{e}{2\cw\sw}
\KKL
(\sina N_{{i}3}{^{*}}+\cosa N_{{i}4}{^{*}})
(\sw N_{{j}1}{^{*}} - \cw N_{{j}2}{^{*}}) 
+ ({i}\leftrightarrow {j})
\KKR
~,
\label{eq:cnnh1}
\end{align}
showing the left-handed (LH) parts, with the right-handed (RH)
parts following from hermiticity of the Lagrangian, see e.g.\ 
\citere{feynarts-mf},
\begin{align}
\label{eq:cLcR}
C^R_{\neu{i}\neu{j}Z}= 
-C^{L*}_{\neu{i}\neu{j}Z}, 
\quad 
C^R_{\neu{i}\neu{j}h_1} =
C^{L*}_{\neu{i}\neu{j}h_1} 
~.
\end{align}

In the limit of interest to us, the two lightest neutralinos are almost
purely bino and wino-like states, 
 $\neu{1}\sim\tilde{B}$, $\neu{2}\sim\tilde{W}$. 
Here we neglect the mixing between the bino and wino components,
which has a subleading effect in our approximation, such that 
$N_{12} \simeq N_{21}\simeq 0$, while $|N_{11}| \simeq |N_{22}| \simeq 1$. 
Note that in the Higgs decoupling limit~\cite{decoupling} 
one has $(\be - \al) \to \pi/2$. 
In this limit 
we obtain for \refeqs{eq:cnnz} and (\ref{eq:cnnh1})
\begin{align}
C^L_{\neu{1}\neu{2}Z} 
&\approx 
\frac{e}{2} \frac{\MZ^2 }{\mu^2}\exp\left(\frac{i\phiMe}{2}\right) 
~,
\label{eq:nnz.approx}
\\
C^L_{\neu{1}\neu{2}h_1}
&\approx
\frac{e}{2}\frac{\MZ}{\mu} 
\KL \frac{\MOne + \MTwo}{\mu} + \frac{4}{\tb} \KR
\exp\left(\frac{-i\phiMe}{2}\right) 
~, 
\label{eq:nnh.approx}
\end{align}
where the neglected terms are of higher order in $\MZ/\mu,\ M_{1/2}/\mu$
and $ 1/\tb$. 
\refeqs{eq:nnz.approx} and (\ref{eq:nnh.approx}) also show that 
the absolute value of the Higgs coupling is largest (smallest) for positive (negative) $\MOne$.
The partial decay widths, however,
also depend on the relative intrinsic $\cp$ factor 
$\eta_{12}$ 
of the neutralinos and that of the Higgs boson.
(the $Z$-boson is $\cp$-even). 
This effect leads to a larger (much larger near the threshold) dependence on
the $\cp$-phases than the one resulting from the change in the absolute value
of the couplings, provided $\mneu1 \neq 0$, as we illustrate below.

Using the relation (\ref{eq:cLcR}) between the LH and RH couplings we express
the  tree-level partial decay widths as
\pagebreak
\begin{align}
\label{NNZtree}
\Gtree(\DecNNZ{2}{1}) =& 
\frac{\beta^*({\neu{1},\neu{2},Z})}{16\pi \mneu{2}}
 |C^L_{\neu{1}\neu{2}Z}|^2  \\ &
\times 
\KL \mneu{2}^2+\mneu{1}^2-2\MZ^2+\frac{(\mneu{2}^2-\mneu{1}^2)^2}{\MZ^2} 
+ 6
\cos(\varphi_{\neu{1}\neu{2}Z})
\mneu{2}\mneu{1}
\KR
~, 
\non \\[.5em]
\label{NNHtree}
\Gtree(\DecNNh{2}{1}{1}) =& 
\frac{\beta^*({\neu{1},\neu{2},h_1})}{16\pi \mneu{2}}
 |C^L_{\neu{1}\neu{2}h_1}|^2
\non \\ &
\times
\KL 
\mneu{2}^2+\mneu{1}^2-m_{h_1}^2
+ 2
\cos(\varphi_{\neu{1}\neu{2}h_1})
\mneu{2}\mneu{1}
\KR
~,  
\end{align}
with
\begin{align}
\label{eq:cosphi}
\cos(\varphi_{\neu{1}\neu{2}Z})
&= \frac{2\re \KKKL C^{L*}_{\neu{1}\neu{2}Z} C^{R}_{\neu{1}\neu{2}Z} \KKKR}
    { |C^L_{\neu{1}\neu{2}Z}|^2 + |C^R_{\neu{1}\neu{2}Z}|^2
}~,\qquad 
\cos(\varphi_{\neu{1}\neu{2}h_1})
= \frac{2\re \KKKL C^{L*}_{\neu{1}\neu{2}h_1} C^{R}_{\neu{1}\neu{2}h_1} \KKKR}
    { |C^L_{\neu{1}\neu{2}h_1}|^2 + |C^R_{\neu{1}\neu{2}h_1}|^2
}~,
\end{align}
and
$\beta^*({a,b,c}) = \la^{1/2}(m_a^2,m_b^2,m_c^2)/m_a^2 $, 
where $\lambda(x,y,z)=(x-y-z)^2-4yz$.
The coefficients defined by \refeq{eq:cosphi} are related to the relative
$\cp$ phase factor of the particles involved.
In this section we will further assume, for the sake of simplicity, that $h_1$ is
$\cp$-even. However, the generalization to the general case is straightforward.
When $\MOne>0$ the relative $\cp$ parity of the neutralinos and the Higgs boson or $Z$ boson is positive,
allowing this process in s-wave.
When $\MOne<0$ the situation is the opposite, 
with negative relative $\cp$ parity. In this case only odd-values of the
total angular momentum allowed, leading to p-wave suppressed processes near
the corresponding decay thresholds.

In the limit of interest the partial decay widths, 
\refeqs{NNZtree} and (\ref{NNHtree}), are given by
\begin{align}
\label{NNZtree.approx}
\Gtree(\DecNNZ{2}{1}) \approx & 
\frac{{e^2}\beta^*({\neu{1},\neu{2},Z})}{64\pi \mneu{2}}
\frac{\MZ^4}{\mu^4} 
\non \\ &
\times
\KL \mneu{2}^2+\mneu{1}^2-2\MZ^2+\frac{(\mneu{2}^2-\mneu{1}^2)^2}{\MZ^2} 
+ 6
\cos(\phiMe)
\mneu{2}\mneu{1}
\KR
~, 
 \\ 
\label{NNHtree.approx}
\Gtree(\DecNNh{2}{1}{1}) \approx & 
\frac{{e^2}\beta^*({\neu{1},\neu{2},h_1})}{64\pi \mneu{2}}
\frac{\MZ^2}{\mu^2} 
\left|
\frac{\MOne + \MTwo}{\mu} + \frac{4}{\tb} 
\right|^2
\non \\ &
\times
\KL 
\mneu{2}^2+\mneu{1}^2-m_{h_1}^2
+ 2
\cos(\phiMe)
\mneu{2}\mneu{1}
\KR
~.
\end{align}
These expressions show explicitly the large $\phiMe$ dependence of the tree-level partial decay widths 
to the Higgs boson, both from the couplings as well as from the relative $\cp$ of initial and final states.


\subsection{Calculation}
\label{sec:calc-det} 

Here we briefly review the calculations used
for the direct production cross section of $\neu{2}\cha{1}$, 
and for the branching ratios for the subsequent decay of the neutralino
into a $Z$ boson and of the chargino into a $W$ boson and the LSP. 
The main production channels for $\cha{1}\neu{2}$ at the LHC, 
as well as their two-body decays to gauge and Higgs bosons and the neutralino decay to a tau-stau pair 
are shown in 
\reffis{fig:prod} and \ref{fig:dec}, respectively.

\begin{figure}[t!]
\begin{center}
\includegraphics[width=0.36\textwidth]{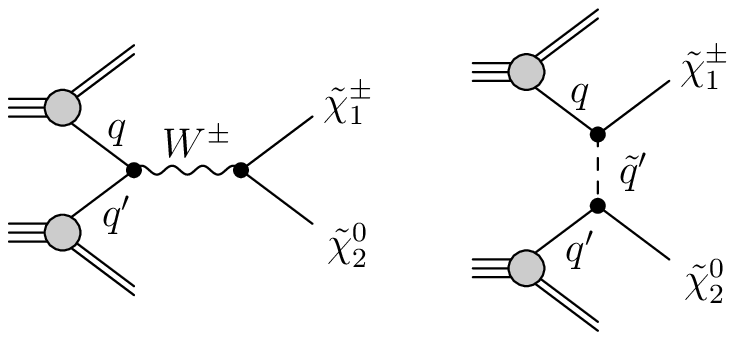}
\caption{
Main production channels for $\cha{1}\neu{2}$ at the LHC. 
Here $q$ and $q^\prime$ ($\tilde q^\prime$) denote quarks (squarks) of the first generation.}
\label{fig:prod}
\end{center}
\vspace{-1em}
\end{figure}

\begin{figure}[t!]
\begin{center}
\includegraphics[width=0.77\textwidth]{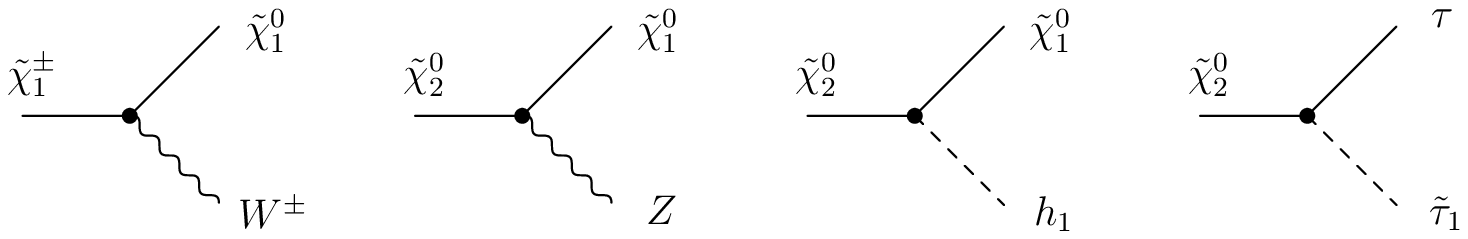}
\caption{
Two-body decay channels of $\cha{1}$ and $\neu{2}$ to gauge and Higgs bosons, 
as well as the neutralino decay to a tau-stau pair.
It should be noted that these channels are kinematically allowed only in parts of the MSSM parameter space.}
\label{fig:dec}
\end{center}
\vspace{-1em}
\end{figure}

The production of neutralinos and charginos at
the LHC is calculated using the program {\tt Prospino~2.1}~\cite{prospinoNN}. 
The effect of complex
parameters on these cross sections can only enter via chargino or
neutralino mixing effects. We have evaluated these cross sections at
the parton level to estimate its effect, which turns out to be
negligible in our analysis%
\footnote{
The same holds for the production $pp \to \chap1\cham1$.}%
.~Consequently, the {\tt Prospino} results can be taken over also for
the complex MSSM results. Small differences for the calculation of
$pp \to \chap1 \neu2$ and of $pp \to \cham1 \neu2$ are neglected.
The NLL corrections to the gaugino
production cross section calculated in \citere{Fuks:2012qx,resummino}
are not included, and we estimate their effects to be at the per-cent level.

The production is dominated by wino pair production, 
where the largest contribution is from the $s$-channel
gauge boson diagrams.
If one assumes that $\MOne<\MTwo$, as is the case when
the GUT relation for the gaugino mass parameters holds, 
then the neutralino with the largest wino component is either 
the second lightest neutralino (for $M_2<\mu$) or the heaviest one.
Therefore, $\neu{2}\champ{1}$ and $\chap{1}\cham{1}$ will have the
largest production cross sections. 
Note that although the $t$ and $u$-channel contribution
to pair production are suppressed due to squark propagators 
if one assumes the first generation squarks to be heavy,
the destructive interference of the $t$-channel with
left-handed squark exchange and the $s$-channel gauge boson channel 
can be significant, as will be discussed in \refse{sec:results}.

\medskip
In \citeres{LHCxC,LHCxN} 
we have calculated the full one-loop (NLO) corrections to the
branching ratios for all non-hadronic chargino and neutralino decays 
for arbitrary parameters in the complex MSSM. 
The calculation is based on 
{\tt FeynArts}/{\tt Formcalc}~\cite{feynarts,formcalc}, and the corresponding
model file conventions~\cite{feynarts-mf} are used throughout.
 In particular, the results
were analyzed and found to be reliable as a function of $\phiMe$. 
We will employ this NLO calculation for our investigations.
The benchmark scenarios defined in the
following section are such that the decays 
$\cha1 \to \neu1 W^\pm$ as well as $\neu2 \to \neu1 Z$, 
$\neu2 \to \neu1 \He$, $\neu2 \to \Staue^\pm \tau^\mp$ are the only
  relevant ones.
As analyzed in the previous subsection
the decays of a wino-like $\neu{2}$ to $\neu{1} h_i$ 
are most sensitive to $\phiMe$ due to the relative $\cp$ between the
bino-like $\neu{1}$ 
and the wino-like $\neu2$, which is controlled by $\phiMe$.
This, however, can be modified when
loop corrections are taken into account as
discussed in \refse{sec:nlo}. 
Furthermore, the NLO corrections are largest for decays to Higgs
bosons~\cite{LHCxN}
and thus have to be taken into account in a precision analysis.
The production cross sections and decay branching ratios have been
evaluated numerically using the OpenStack infrastructure as described
in \citere{Campos:2012vb}.


\section{Benchmark scenarios and experimental motivation}
\label{sec:bench}


\subsection{Overview of current experimental results}

In \citeres{CMS:2012ewa,ATLAS:2012ku,CMS:2012ewb,ATLAS:2012-154,ATLAS:2013-028,ATLAS:2013-035,ATLAS:2013-049},
ATLAS and CMS have studied the sensitivity to electroweak gaugino pair
production, particularly to the production of the second lightest
neutralino and lightest chargino via multi-lepton signatures. 
Here the chargino and neutralino
decay either via sleptons or via gauge bosons, depending on the slepton
masses, which are parameterized via $x$ (where
$m_{\tilde{l}}=\mneu{1}+x(\mneu{2}-\mneu{1})$). Exclusion limits are
then obtained within specific models, primarily simplified models which 
set all relevant branching ratios to one, assuming that all other
channels are kinematically forbidden.
The ATLAS results at $7\tev$ are presented in
\citere{ATLAS:2012ku} for $4.7~\ifb$, and the updated results
including $8\tev$ data are given in \citeres{ATLAS:2012-154} for $13~\ifb$
and \cite{ATLAS:2013-028,ATLAS:2013-035,ATLAS:2013-049} for up to
$21~\ifb$.
An update of CMS including the $8\tev$ data was published in
\citere{CMS:2012ewb} for $9.2~\ifb$, where opposite sign
(OS) dileptons inconsistent with a $Z$ boson are also studied. The
$7\tev$ results were published in \citere{CMS:2012ewa} for $4.98~\ifb$.

In addition to the $3$ lepton events
(electrons, muons and hadronically reconstructed taus) analyzed by ATLAS,
CMS also considers the case when one of the leptons is unidentified,
selecting events with
 same the sign (SS) lepton pairs $e\tau$, $\mu\tau$ and
$\tau \tau$.
Further, OS lepton pairs and 2 jets 
for on-shell $WZ$ and $ZZ$ events where one $Z$ decays to {$e^+e^-$} 
or {$\mu^+\mu^-$} and the other gauge boson decays hadronically are
considered. Simplified models are used to obtain exclusion limits, tuned
to search for decays via sleptons or gauge boson, mainly for $x=0.5$
but also for $x=0.05$ and $0.95$. Here models with different couplings
to $\tau$ leptons are considered, i.e.\ the sleptons may be left-handed
or right-handed or a mixture, such that the final state leptons are
predominantly light, flavor independent, or mostly taus.

ATLAS, on the other hand, presents its results for this channel by
combining 3-lepton (electrons or muons) searches in various signal
regions, the primary criterion being whether the invariant mass
same-flavor-opposite-sign (SFOS) lepton pair lies around the $Z$ boson
mass or not, thus defining $Z$-enriched and $Z$-depleted regions
respectively. By making requirements on the reconstructed mass of the
SFOS lepton pair ($m_{\rm SFOS}$) and on the transverse momentum ($p_T$)
of the third lepton, the depleted region is further subdivided into
regions targeting either small mass splittings between the neutralinos
(here the $Z$ is off-shell and this region is discussed later), mass
splittings close to the $Z$-boson mass, or decays via sleptons by
requiring high transverse momentum of the third-leading lepton. 
In the simplified models, a number of assumptions are made,
first and foremost that the neutralino and chargino are wino-like and
the lightest neutralino bino-like. As for the sleptons, either $x=0.5$
in which case the branching ratio to all sleptons is assumed to be
$1/6$, or $x$ is very large (where the precise value is not quoted)
such that the decay to sleptons may be ignored, and the branching ratio
to gauge bosons is assumed to be $1$. 


\subsection{Definition of benchmark scenarios}

Since so far only ATLAS reported an analysis using the full 2012 data set 
with numerical values for the excluded cross sections%
~\cite{ATLAS:2013-035},
we will use their results for our baseline analysis.
In order to interpret the ATLAS exclusions in terms of the complex MSSM,
we calculate the cross section in benchmark scenarios similar to those
used by ATLAS, including NLO corrections as described 
in \refse{sec:calc-det}.
We re-analyze the ATLAS
$95\%$~CL exclusion bounds in the simplified analyses in the
$\mneu{2}$--$\mneu{1}$ plane, taking $\MOne$ and $\MTwo$ as free
parameters with central values:
\begin{align}
 M_1=100 \gev \; \mbox{and} \; M_2=250 \gev.
\label{ATLpara1}
\end{align} 
The other parameters are chosen as in the ATLAS analysis
presented in \citere{ATLAS:2013-035}%
\footnote{
Not all parameters are clearly defined in \citere{ATLAS:2013-035}. We
select and choose our parameters to be as close to the original analysis
as possible.}%
, 
\begin{align}
\mu = 1 \tev, \; \tb = 6, \; \Msqez = \Msqd = \Msl = 2 \tev, \; 
\At = 2.8 \tev~.
\label{ATLpara}
\end{align}
$\Msqez$ denotes the diagonal soft SUSY-breaking parameter in the scalar
quark mass matrices of the first and second generation, similarly
$\Msqd$ for the third generation and $\Msl$ for all three generations of
scalar leptons. If all three mass scales are identical we also use the
abbreviation $\msusy := \Msqez = \Msqd = \Msl$. We will clearly indicate
where we deviate from the ``unification'' for scalar leptons.
$\At$ is the
trilinear coupling between stop quarks and Higgs bosons, which is
chosen to give the desired value of $\MHe$.
The other trilinear couplings, set to zero in
\citere{ATLAS:2012-154,ATLAS:2013-035}, 
we set to $\At$ for squarks and to zero for sleptons.
Setting also the $A_{q\neq t}$ to zero would have a minor impact on
our analysis.
The effect the large sfermion mass scale is a small destructive
interference of the $s$-channel amplitude with the $t$-channel squark
exchange. 
The large higgsino mass parameter $\mu$ results in a gaugino-like pair of
produced neutralino and chargino. The lightest Higgs boson mass (as
calculated with 
{\tt FeynHiggs~2.9.4}~\cite{feynhiggs,mhiggslong,mhiggsAEC,mhcMSSMlong})
is evaluated to be $\sim 125.5 \gev$, defining the value of $\At$ in
\refeq{ATLpara}. 
In order to scan the $\mneu{2}$--$\mneu{1}$ plane we 
use the ranges
\begin{align}
|\MOne| = 0 \ldots 200 \gev~, \quad \MTwo = 100 \ldots 400 \gev
\quad \mbox{with~} |\MOne| \le \MTwo~.
\end{align}

\medskip
The main aim of this paper, as discussed above, is the interpretation of
the ATLAS exclusion limits in several ``physics motivated'' benchmark
scenarios. Taking the parameters in \refeq{ATLpara} as our baseline
scenario, we deviate from it in the following directions.

\begin{enumerate}

\item We take $\phiMe$, the phase of $\MOne$, to be a free
 parameter. Note that for the considered central benchmark 
 scenario, as $\tb$ is low and $\msusy$ is high, 
 the full range is allowed by current electric dipole moment (EDM) 
constraints~\cite{Baker:2006ts,Regan:2002ta,Griffith:2009zz},
as verified explicitly via both
 \texttt{CPsuperH~2.3}~\cite{Lee:2012wa,Lee:2007gn,Lee:2003nta} and  
 \texttt{FeynHiggs~2.9.4}~\cite{feynhiggs,mhiggslong,mhiggsAEC,mhcMSSMlong}.

\item The variation of $\tb$ can have a strong impact on the
 couplings between the neutralinos and the Higgs boson,
see \refeq{eq:nnh.approx}. 
We therefore analyze the effect of variation of $\tb$ in the range
$\tb = 6 \ldots 20$.

\item 
 Although in general the sleptons are assigned a common mass
$\msusy$, in order to consider the possibility that neutralino decays
to sleptons could compete with the decays to $Z$ and Higgs bosons, 
we consider $\MstauR=|M_1|$, where $\MstauR$
denotes the ``right-handed'' soft SUSY-breaking parameter in the
scalar tau mass matrix, see Eq.~(87) in \citere{LHCxN}.
This scenario is motivated by the measured relic density of dark matter. 
For this choice of parameters one finds $\mneu{1} \lsim \mstaul$,
i.e.\ the stau co-annihilation region.
We have confirmed (using $\texttt{micrOMEGAs3.1}$~\cite{micromegas})
that in our scenario the relic density is in agreement%
\footnote{
Small changes in the parameters which can have a drastic impact on the
predicted CDM density, but only a small impact on the chargino/neutralino
phenomenology are not relevant.
}
~with the latest measurements presented 
by Planck~\cite{Ade:2013zuv}
earlier this year, $\Omega^{\rm DM} h^2 = 0.1199 \pm 0.0027$.

\item 
As shown in \refeqs{eq:nnz.approx}, (\ref{eq:nnh.approx})
the lighter neutralino-Higgs couplings depend
strongly on $\mu$, and consequently also the Born amplitudes. 
We investigate two scenarios: (i) $\MTwo < \mu$ with 
$\mu = 2000 \gev$. This scenario shows similar characteristics as
the ATLAS baseline scenario, but decouples the $\mu$
parameter further, as will be discussed briefly in \refse{sec:nlo}. 
(ii) $\MTwo > \mu$ with $\mu = 100 \ldots 400 \gev$. In this
scenario the lighter neutralinos and chargino possess a substantial higgsino
component.

\item Although $\msusy$ has a negligible impact on the decays of the
electroweak SUSY particles to gauge bosons,
it plays an important role in the production, as the $t$-channel squark
exchange and the $s$-channel gauge boson exchange amplitudes interfere
destructively. We consider the range from $1.2\tev$ to $3\tev$.
\end{enumerate}

The various scenarios are summarized in \refta{tab:para}

\newcommand{\Satlas}{$S_{\rm ATLAS}$}
\newcommand{\Scompl}{$S_{\rm ATLAS}^{\phiMe}$}
\newcommand{\nSgauge}{$S_{\tilde{H}\tilde{B}}$}
\newcommand{\Stb}{$S_{\rm ATLAS}^{\tan\be}$}
\newcommand{\Stbt}{$S_{\rm ATLAS}^{\tan\be=20}$}
\newcommand{\Ssusy}{$S_{\rm ATLAS}^{\rm SUSY}$}
\newcommand{\Sstau}{$S^{\rm DM}$}
\newcommand{\Satlasmu}{$S_{\rm ATLAS}^{\mu}$}

\newcommand{\oScompl}{$S_{\rm complex}$}
\newcommand{\Sgauge}{$S_{{\rm low-}\mu}$}
\newcommand{\oStb}{$S_{\tan\be}$}
\newcommand{\oSsusy}{$S_{\rm SUSY}$}
\newcommand{\oSstau}{$S_{\rm DM}$}

\begin{table}[ht!]
\renewcommand{\arraystretch}{1.5}
\BC
\begin{tabular}{|c||c|c|c|c|c|c|
|c|
}
\hline
Scenario & $\phiMe$ & $\mu$& $\tb$ & $\msusy$
&$M_{\tilde{\tau}_R}$ 
\\ \hline\hline
\Satlas & 
$ 0$          & $ 1000            $ & $ 6 $ 
& $ 2000 $ & $\msusy$               %
\\ \hline\hline
\Scompl  &  
$ 0 \ldots \pi$ & $ 1000            $ & $ 6 $ 
& $ 2000 $ & $ \msusy$              %
\\ \hline
\Stb  &  
$ 0              $ & $ 1000             $ & $ 6 \ldots 20$ 
& $ 2000 $ & $\msusy$               %
\\ \hline
\Satlasmu  & 
$ 0              $ & $ 2000             $ & $ 6 $ 
& $ 2000 $ & $\msusy$             %
\\ \hline
\Ssusy  & 
$ 0              $ & $ 1000             $ & $ 6 $ 
& $ 1200, 3000 $ & $\msusy$         %
\\ \hline\hline
\Sstau  &  
$ 0 \ldots \pi$ & $ 1000             $ & $ 6, 20$  
& $ 2000 $ & $|M_1|$                %
\\ \hline
\Sgauge  & 
$ 0              $ & $100$ \ldots $ 400 $ & $ 6 $ 
& $ 2000 $ & $ \msusy$                %
\\ \hline
\end{tabular}
\caption{
Parameters for benchmark scenarios (masses in $\gev$). 
We furthermore have for all scenarios: $|\MOne| = 0 \ldots 200 \gev$,
$\MTwo = 100 \ldots 400 \gev$ (with $|\MOne| \le \MTwo$), 
$M_3 = 1500 \gev$ (gluino mass parameter), 
except for the intermediate higgsinos scenario \Sgauge, where we set
$\MTwo = 500\gev$. 
The first (baseline)
scenario corresponds to the ATLAS analysis in \citere{ATLAS:2013-035}.
Our ``central benchmark scenario'' refers to the case $\MOne=100 \gev$ and
$\MTwo=250 \gev$. The value of $\At$ is adjusted
to ensure $\MHe \approx 125.5 \gev$.
}
\label{tab:para}
\EC
\renewcommand{\arraystretch}{1.0}
\end{table}

Besides interpreting the current results in the scenarios
summarized in \refta{tab:para} we also evaluate possible future limits
(assuming the absence of a signal). We analyze the following two
(future) scenarios:

\begin{itemize}

\item[$(i)$] A combination of ATLAS and CMS data, which for simplicity we
take as resulting in a doubling of the luminosity, i.e.\ assuming
$42~\ifb$ analyzed 
by ATLAS. The change in the experimental limit on the production cross
section (times branching ratio) is evaluated by assuming a purely
statistical effect, thus dividing the current limit by $\sqrt{2}$.

\item[$(ii)$] The first run at $\sqrt{s} = 13 \tev$ that could take place
in 2015-2017. We assume that ATLAS collects $100~\ifb$. The new limit is
evaluated from the existing limits by a simple rescaling of signal
and background cross sections. More details are given in
\refse{sec:lhc13}. 

\end{itemize}


\section{Interpretation of ATLAS exclusion limits}
\label{sec:results}

We re-analyze the ATLAS results of \citere{ATLAS:2013-035} in the scenarios
defined in \refta{tab:para}. In all scenarios the decay 
$\neu{2} \to \neu{1} \He$ is taken into account.
Production cross sections and branching ratios are evaluated as
described in \refse{sec:calc-det}.
Note that in the simplified model analyses, 
$\mneu{2}=\mcha{1}$, which is not the case in the MSSM. 
However, for lighter gauginos, corresponding to $|\mu|>|M_1|,M_2$,
this relation holds to a good approximation.
In our analysis, we 
chose to take the convention that our $\mneu{2}$ corresponds to the
ATLAS $\mneu{2}$, and our $\mcha{1}$ is calculated accordingly. Note
that in almost 
all of the parameter space explored the difference $\mneu{2}-\mcha{1}$ 
will be less than $1\gev$, where larger values are indicated explicitly.


\subsection{Effect of \boldmath{$\neu2$} decays to Higgs bosons and
  sign of \boldmath{$\MOne$}}
\label{sec:Scomp}

We begin by re-interpreting the ATLAS simplified model exclusion bounds, 
taking full NLO branching ratios into account. Including
$\neu{2} \to \neu{1} \He$ has a considerable 
impact on the $\br(\neu{2} \to \neu{1} Z)$, thus weakens the existing
conclusion limits. Keeping $M_1$ real, we analyze the excluded parameters 
allowing two possibilities for $\phiMe$, namely 0 and $\pi$.
(The case of a complex $\MOne$ will be addressed in \refse{sec:complex}.)

\begin{figure}[t!]
\begin{center}
\begin{tabular}{c}
\includegraphics[width=0.49\textwidth,height=7.5cm]{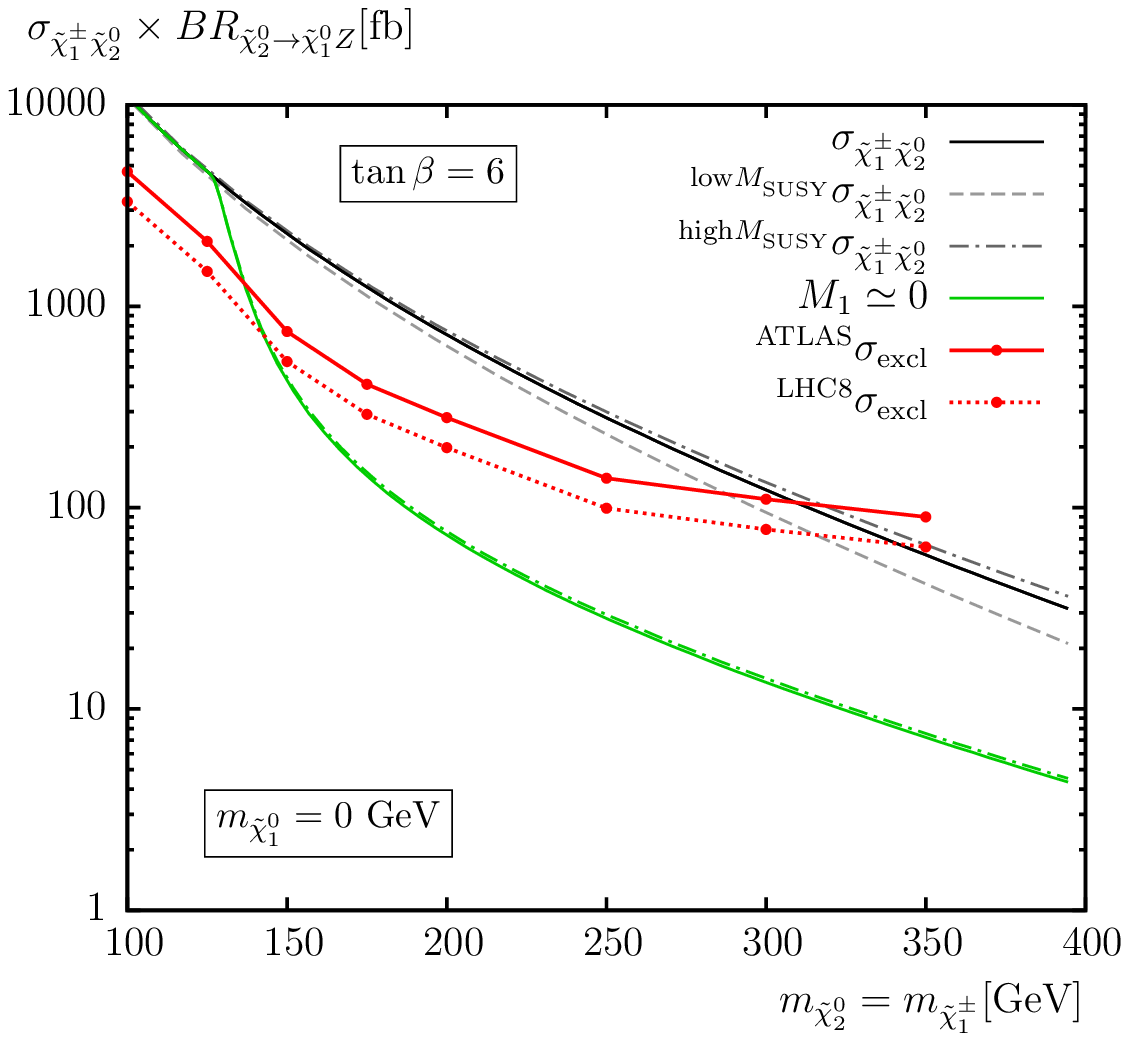}
\hspace{-4mm}
\includegraphics[width=0.49\textwidth,height=7.5cm]{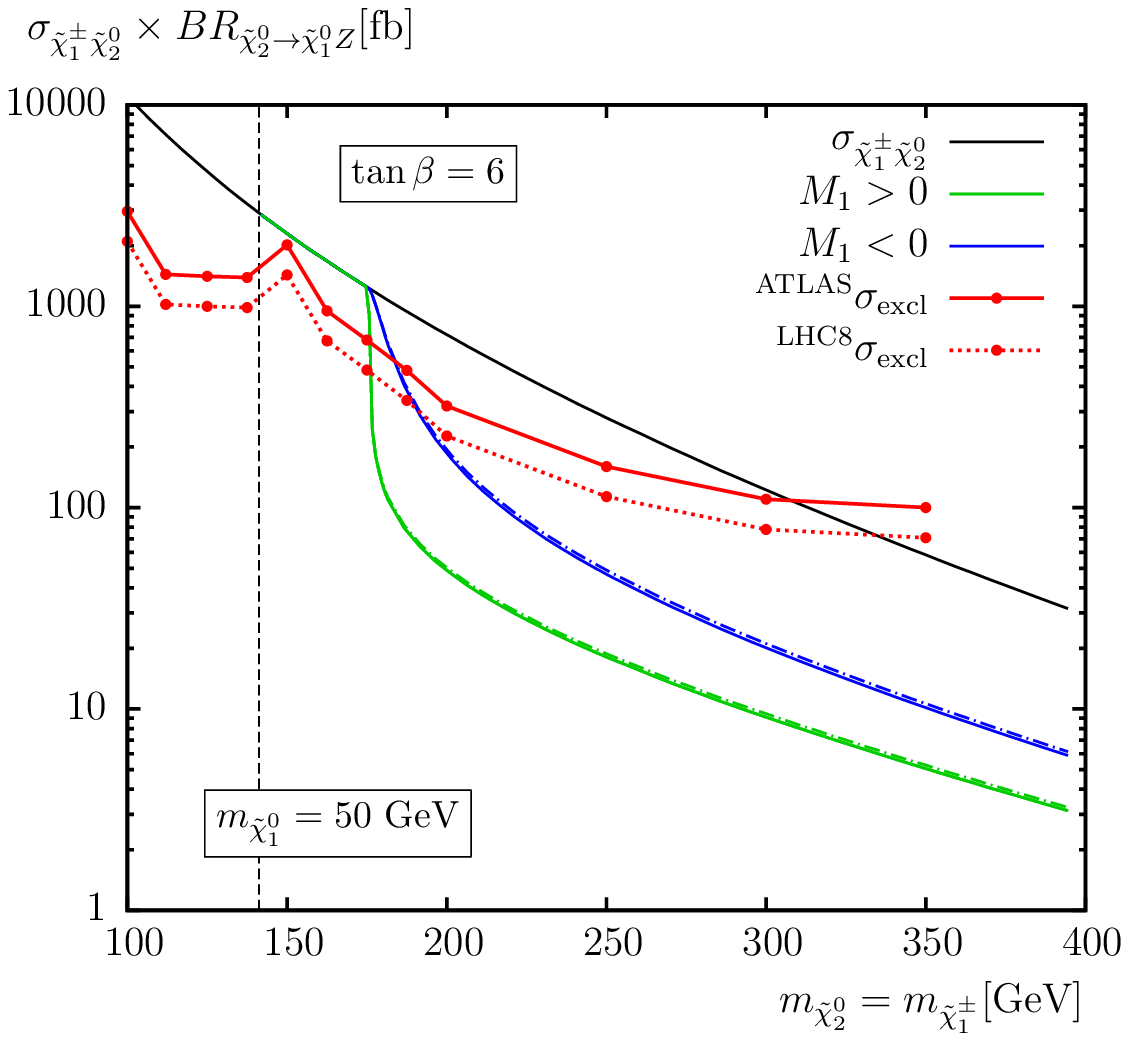}\\[2em]
\end{tabular}
\vspace{2em}
\begin{tabular}{c}
\includegraphics[width=0.49\textwidth,height=7.5cm]{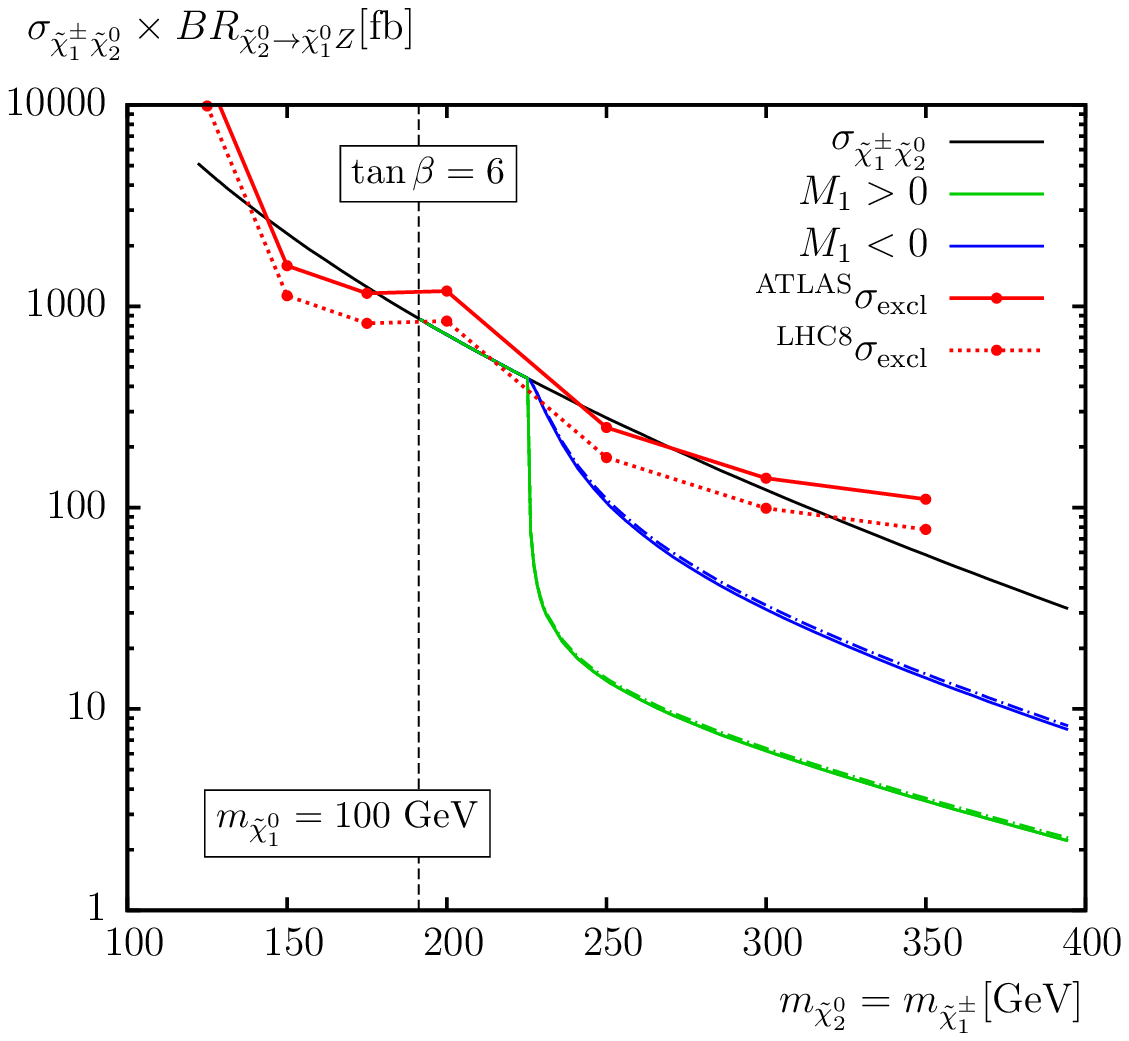}
\hspace{-4mm}
\includegraphics[width=0.49\textwidth,height=7.5cm]{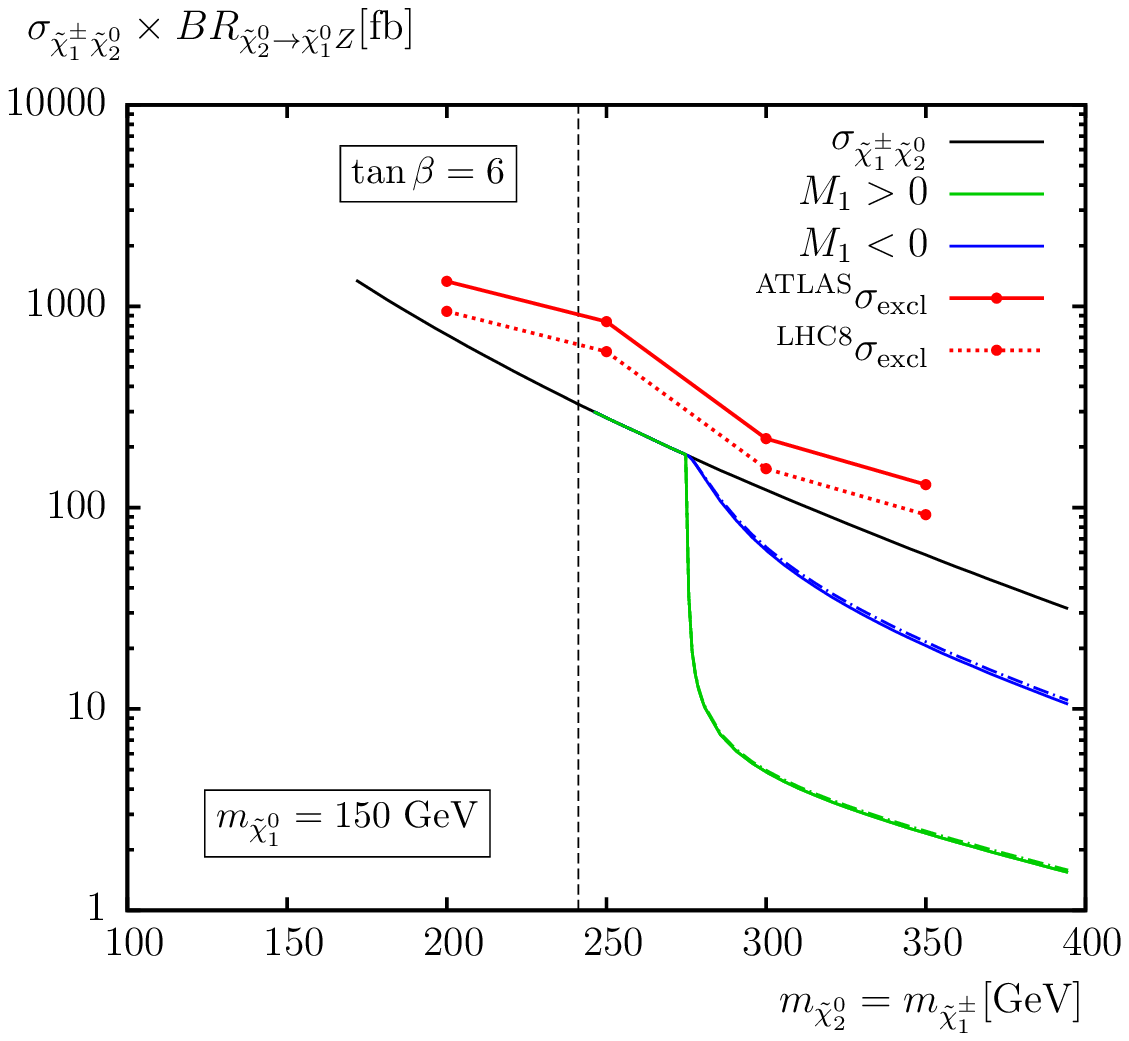} 
\end{tabular}
\caption{
$\cha{1},\neu{2}$ production cross section at the LHC8 (black)
times the $\br(\neu{2}\to Z\neu{1})$ 
at tree (solid) and one-loop level (dashed-dotted)
for 
$\phiMe = 0$ (green) and $\phiMe = \pi$ (blue). 
$M_1$ is chosen such that $\mneu{1} = 0, 50, 100, 150\gev$, in the top
left and right, bottom left and right plot, respectively. 
The $95\%$ CL exclusion cross sections from ATLAS
taken from Fig.~8 and 9 of \cite{ATLAS:2013-035} (red, solid) are projected for the 
combined full data-set of ATLAS and CMS (red, dashed).
Also shown in the upper left figure is the production cross section with
 $\msusy=1.2\tev$ (dashed, light gray)
and  $\msusy=3\tev$ (dot-dashed, dark gray).
The vertical dashed lines show the position of the $\neu{2}\to Z\neu{1}$
kinematical threshold.
}
\label{fig:xsbr.tb6.atlasexcl}
\end{center}
\end{figure}

In \reffi{fig:xsbr.tb6.atlasexcl} we show as a black line the production
cross section for $\cha{1}\neu{2}$ pair-production at the LHC8 as a
function of $\mneu{2}$ and for $\mneu{1} = 0, 50, 100, 150 \gev$ in the
upper right, upper left, lower right, lower left plot, respectively.%
\footnote{
It should be noted that a massless neutralino is not excluded by experimental
searches~\cite{masslessx}.}%
~The
parameters are chosen according to 
\Satlas, except for $\MOne $ and $\MTwo$ which are varied.
In order to avoid $\mneu{1}\approx\mneu{2}$
the lines in the lower plots
start at $\mneu2 = 125 \gev$ (left) and $\mneu2 = 175 \gev$ (right).
The production cross section times the 
$\br(\neu{2}\to Z\neu{1})$ is also shown for $\phiMe = 0$ (green) and 
$\phiMe = \pi$
(blue), except for $\mneu{1}\approx \MOne = 0.2\gev$ (upper left
plot), where the sign is irrelevant. 
The chargino decays with $100\%$ as $\cha{1} \to \neu{1} W^\pm$.
The dashed green and blue lines correspond to the tree-level evaluation,
whereas the solid lines are obtained by our full one-loop evaluation. The
difference in \reffi{fig:xsbr.tb6.atlasexcl}, however, is barely visible.
Below the kinematical threshold for $\neu2 \to \neu1 \He$ the 
green/blue lines are on top of the black line, i.e.\ 
$\br(\neu2 \to \neu1 Z) = 1$. The green/blue lines stop at the dashed
vertical line, indicating the kinematical threshold for $\neu2 \to \neu1 Z$.
The $95\%$~CL exclusion cross sections from ATLAS~\cite{ATLAS:2013-035} 
are given as red dots,
connected by solid red lines. These lines do not correspond to
true experimental analyses and are only indicative.
Also shown as dotted red line is the projection for a combination of
ATLAS and CMS data (see the end of the previous section and the
discussion below). 

The crossing point between the black and the red line corresponds to the
highest $\mneu{2}$ value that is excluded by ATLAS, see
\citere{ATLAS:2013-035}.  
However, taking into account the decay $\neu{2} \to \neu{1} \He$ as well
as a variation of the sign of $\MOne$, resulting in the green and
blue lines, moves 
the highest excluded $\mneu{2}$ to substantially smaller values. In the
case of $\mneu{1} = 0$, as shown in the upper left plot in
\reffi{fig:xsbr.tb6.atlasexcl}, the exclusion bound for $\mneu{2}$ moves from
$\sim 310 \gev$ down to $\sim 140 \gev$. 
For positive $\MOne$ the decay to a Higgs boson is enhanced, 
resulting in a reduced $\br(\neu{2} \to \neu{1} Z)$,
while a smaller reduction is obtained for negative $M_1$.
This can be understood from the dependence of the decay amplitudes
for $\neu2 \to \neu1 Z/\He$ as discussed in \refse{sec:amplitudes}.
From those expressions it is clear that the enhancement of the
decay to Higgs bosons increases with $M_1$.  
The corresponding strong variation of the partial decay widths leads to
a strong variation in $\br(\neu2 \to \neu1 Z)$.%
\footnote{In \citere{Baer:2012ts} it was pointed out that, in the large
$\mu$ regime, the gaugino pair production process with subsequent decay
of a neutralino to a Higgs boson and the LSP would have the highest
reach sensitivity for very large luminosities at LHC14. 
Notice that, for $|\MOne|\simeq \MTwo$, the phase of $\MOne$
could have a significant effect on this reach. 
}
~Consequently, the differences of the excluded $\mneu{2}$ values for
opposite signs of $\MOne$
becomes larger with increasing $\mneu{1}$, as is visible 
comparing the green and the blue curves in the four plots in
\reffi{fig:xsbr.tb6.atlasexcl}. 
The $\order{1/\tb}$ term in \refeq{NNHtree.approx}
further increases this ratio of couplings, suppressing the
branching ratio to $\neu{1} Z$, especially
for small $\tb$. We discuss this effect in \refse{sec:tb}, where we
compare the limits for $\tb=6$ (\reffi{fig:xsbr.tb6.atlasexcl})  and
$\tb=20$ (\reffi{fig:xsbr.tb20.atlasexcl}).  

\medskip
A similar conclusion holds for
the projected combination of ATLAS and CMS results, shown as red-dotted lines. 
The reduced statistical error leads to an increase of the
 excluded values of $\mneu{2}$  by \order{10 \gev} for 
$\mneu{1} = 0, 50 \gev$. For $\mneu{1} = 100 \gev$,
where the current analysis just barely sets a limit, 
the exclusion is larger, while for $\mneu{1} = 150 \gev$ neither the
current nor the combined analysis yield any exclusion limit.

As explained 
in \refse{sec:amplitudes}%
~a change in the phase of $M_1$ (here from $0$ to $\pi$) 
does not only change the couplings
but also has a dynamical effect on the decay processes, 
which depends on the relative $\cp$ of the two neutralinos and the Higgs
or gauge boson. 
The corresponding p-wave suppressed (s-wave) amplitude for opposite
(equal) relative $\cp$ of the neutralinos and the Higgs or $Z$ boson 
is most pronounced at the corresponding neutralino decay thresholds, while it
 becomes negligible for boosted Higgs bosons or gauge bosons%
\footnote{Analyses for neutralino decays to boosted Higgs bosons have
been presented in~\cite{Baer:2012ts,Byakti:2012qk,Ghosh:2012mc,Howe:2012xe}}.
The p-wave suppression effect in the $\MOne<0$ scenario, compared to the
s-wave  $\MOne>0$  decays is reflected in the softer rise of the green
curve at the threshold for the decay to the Higgs boson. 
Since both the $Z$ and the lightest Higgs boson are $\cp$-even%
\footnote{In case of a complex $\MOne$ the lightest Higgs will
receive a very small $\cp$-odd admixture.}%
, the effect cancels out in the branching ratio for larger mass differences.
Notice, however, that this will not be the case if there are additional
decay channels open, as will be discussed in \refse{sec:dm} for the
DM-motivated scenario \Sstau.

\bigskip
We also briefly investigate the effects of a variation of the overall
sfermion mass scale, $\msusy$, i.e.\ the \Ssusy\ scenario.  
In the upper left plot of \reffi{fig:xsbr.tb6.atlasexcl}, besides the
production cross section using the default value 
$\msusy (= \Msqez = \Msqd = \Msl) = 2 \tev$ shown
as a black line, we also show as dotted (dot-dashed) gray line the
$\cha{1}\neu{2}$ production cross section for $\Msqez = 1.2 (3) \tev$. 
Choosing the squarks of the first two families at (roughly) 
their experimental lower mass limit increases the destructive
interference of the $t$-channel 
squark exchange with the $s$-channel gauge boson exchange production channels. 
The effect is negligible for the smaller neutralino/chargino
masses where the $s$-channel 
dominates, while it results in a suppression of over $30\%$ for the
largest masses shown.  
Accordingly, we observe a smaller (larger) production cross section for
a smaller (larger) value of $\Msqez$. While the effects are small in
comparison to taking into account the decay to Higgs bosons (green
line), one can observe that the low $\Msqez$ value has an effect as
sizable as the combination of ATLAS and CMS data.
We will not investigate the effects of a variation of $\Msqez$ further.


\subsection{\boldmath{$\tb$} dependence}
\label{sec:tb}

As the change in $\tb$ has a negligible effect on the production
cross section, by definition the ATLAS limits do not depend on $\tb$. 
However, as seen in \refeqs{eq:nnh.approx},
the couplings of the neutralinos to
the Higgs bosons are strongly affected by $\tb$, resulting in a 
larger branching ratio of the second neutralino to a
$Z$ boson and the LSP. The experimentally excluded region changes
accordingly, with chargino and second lightest neutralino masses 
excluded up to higher masses than for $\tb = 6$.
This is illustrated in \reffi{fig:xsbr.tb20.atlasexcl}, 
where we show the production cross section
times the branching ratios 
for the same parameters as in \reffi{fig:xsbr.tb6.atlasexcl} except
for $\tb$,
which is increased from 6 to 20, i.e. for scenario \Stb.
The mass exclusion limits lie at $190 \gev$ and $200\gev$ for a massless LSP,
for the tree-level and NLO results, respectively. 

\begin{figure}[t!]
\begin{center}
\begin{tabular}{c}
\includegraphics[width=0.49\textwidth,height=7.5cm]{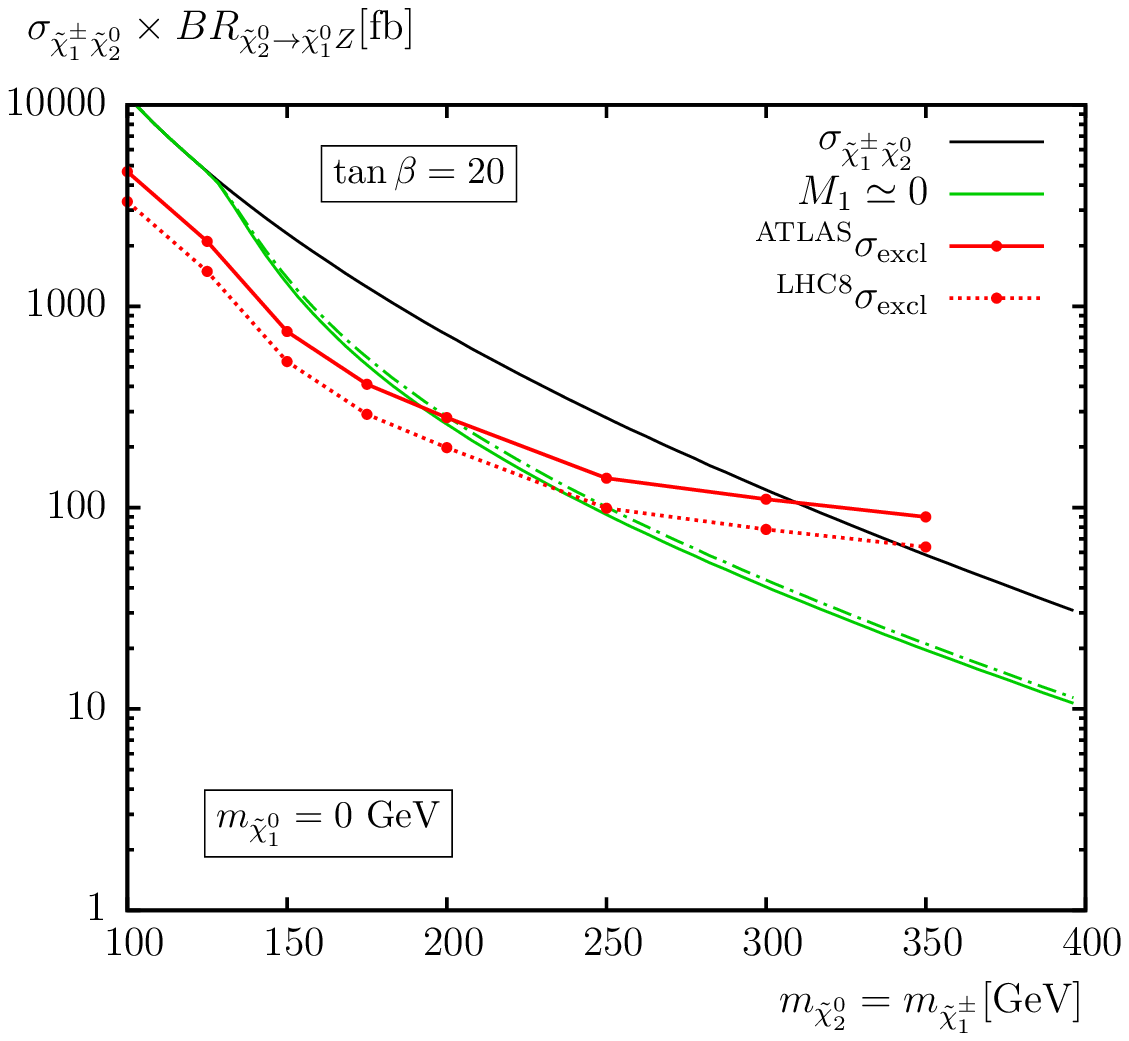}
\hspace{-4mm}
\includegraphics[width=0.49\textwidth,height=7.5cm]{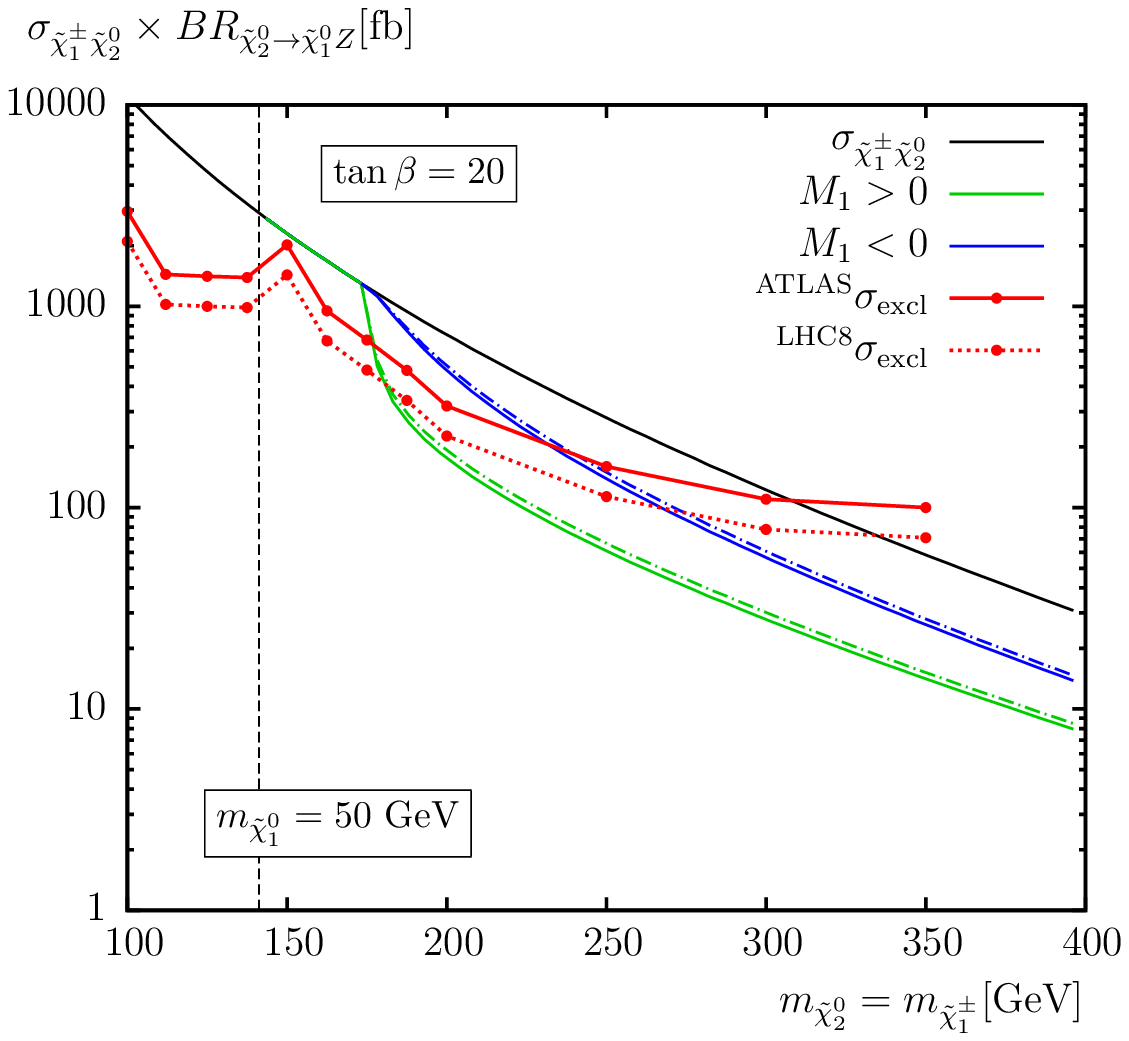} 
\end{tabular}
\begin{tabular}{c}
\includegraphics[width=0.49\textwidth,height=7.5cm]{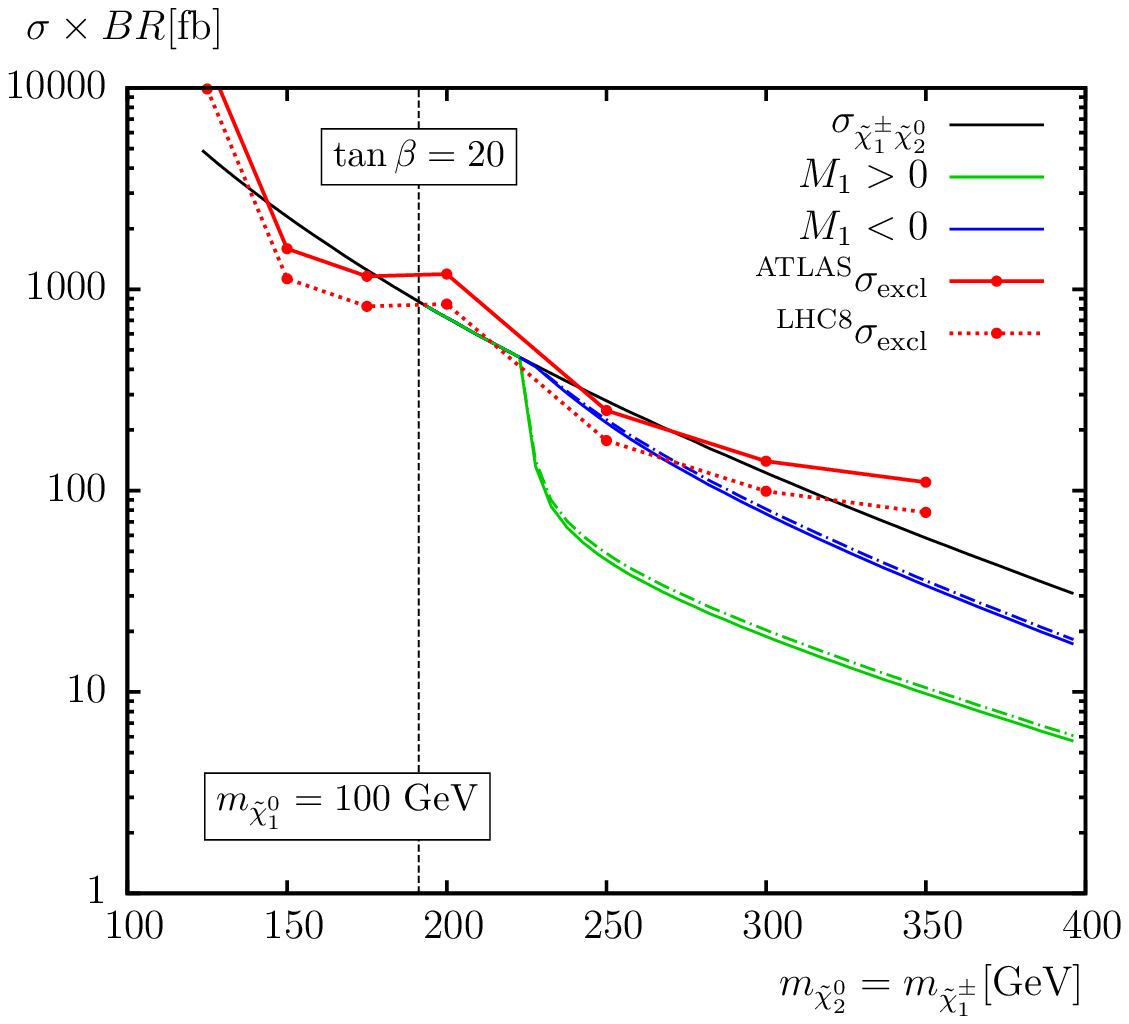}
\hspace{-4mm}
\includegraphics[width=0.49\textwidth,height=7.5cm]{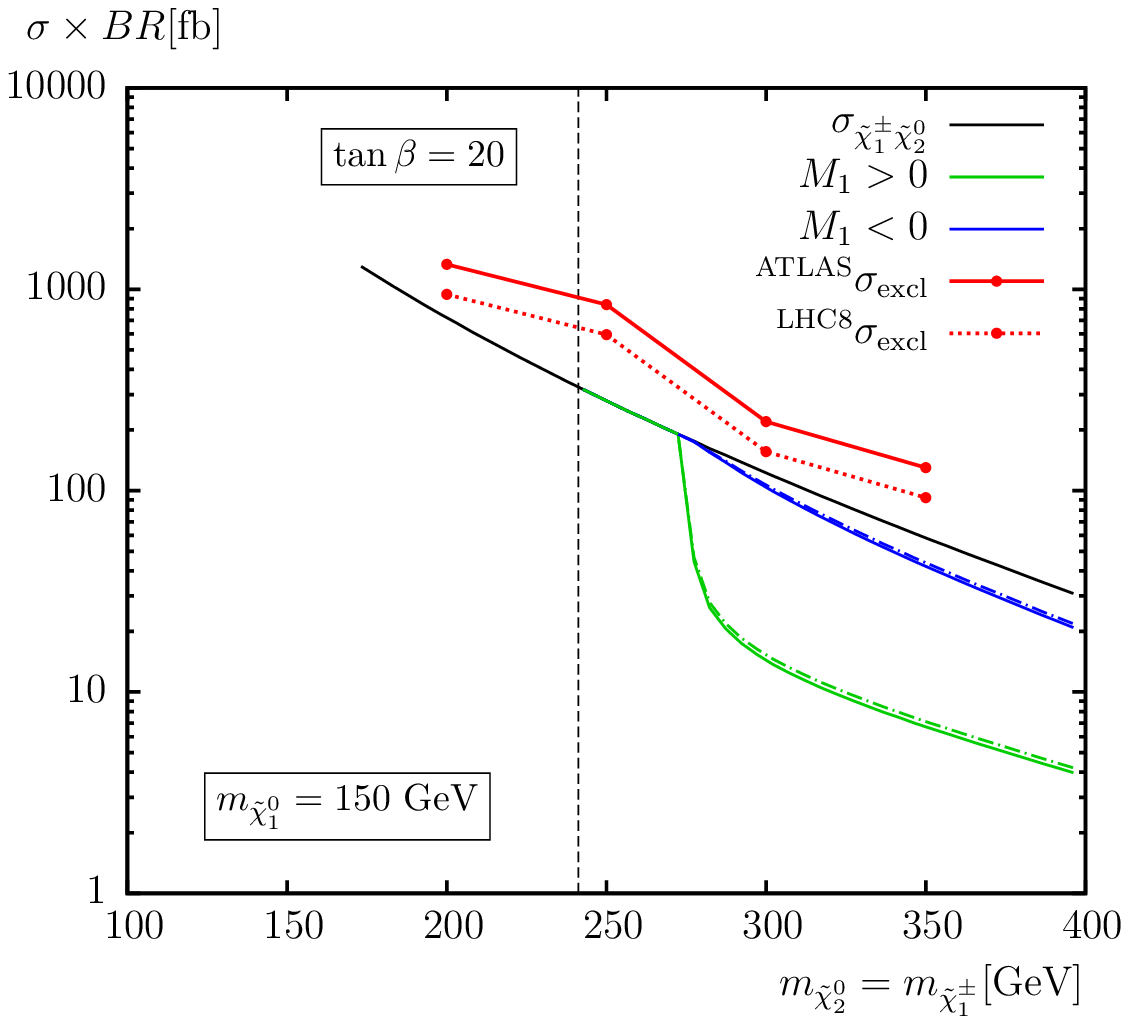} 
\end{tabular}
\caption{
As in \reffi{fig:xsbr.tb6.atlasexcl}, but for $\tb = 20$.
}
\label{fig:xsbr.tb20.atlasexcl}
\end{center}
\end{figure}

For an LSP of $50\gev$ the exclusion limits are found to be
very close to the threshold for Higgs decay for positive $\MOne$ 
and between $230 \gev$ and $250\gev$ for negative $\MOne$.
For LSP masses above $\sim 100\gev$ no chargino masses are excluded once
the decay of the second lightest neutralino to the Higgs boson is open. 
In this region a more meaningful quantity to consider is 
the ratio of the excluded cross section divided by the theoretical
production cross section times branching ratios, indicating the
required ``improvement'' necessary for an exclusion.
For instance, for $\mneu{1}=100\gev$, $\mneu{2}=250\gev$ and $\tb=6$
this ratio is, respectively,
 $17.7$ and $2.1$ for $\MOne$ positive and negative. 
For the same masses but for $\tb=20$, this ratio is smaller, 
$5.1$ and $1.1$, respectively.

\medskip
The results presented in \reffis{fig:xsbr.tb6.atlasexcl} and
\ref{fig:xsbr.tb20.atlasexcl} are summarized in \reffi{fig:contour-LHC8}, 
where we show the exclusion region in the $\mneu{2}$--$\mneu{1}$ plane
for $\tb=6$ (upper plot) and $\tb=20$ (lower plot). 
The solid lines (shaded areas) correspond to the currently analyzed
data. The dashed lines are the projection for the combination of ATLAS
and CMS LHC8 data, where the exclusion limit is calculated as for the
dotted red line in \reffi{fig:xsbr.tb6.atlasexcl}. 
The red lines show the ATLAS analysis, the green lines take
into account the decays $\neu{2} \to \neu{1} \He$ for $\MOne > 0$, and
the blue ones for $\MOne < 0$. 
The exclusion curves are not smooth, reflecting the fact that excluded
cross sections obtained from ATLAS are only available for a sparse grid of
points in the $\mneu{2}$--$\mneu{1}$ plane, and are given by the
points in \reffi{fig:xsbr.tb6.atlasexcl} and
\ref{fig:xsbr.tb20.atlasexcl} where the red lines cross the black, blue
and green lines, both for the ATLAS data and for LHC8 combined data. 
Technically, this is achieved by interpolating the
cross section as a function of $\mneu{1}$ for fixed values of $\mneu{1}$.
Note that above the light (dark) gray line the on-shell
decay $\neu{2} \to \neu{1} Z (\He)$ is kinematically forbidden.
Above the light gray line only off-shell decays of $\neu{2}$ are
allowed, which is discussed in \refse{sec:offshell}. 
\begin{figure}[t!]
\begin{center}
\begin{tabular}{c}
\includegraphics[width=0.70\textwidth]{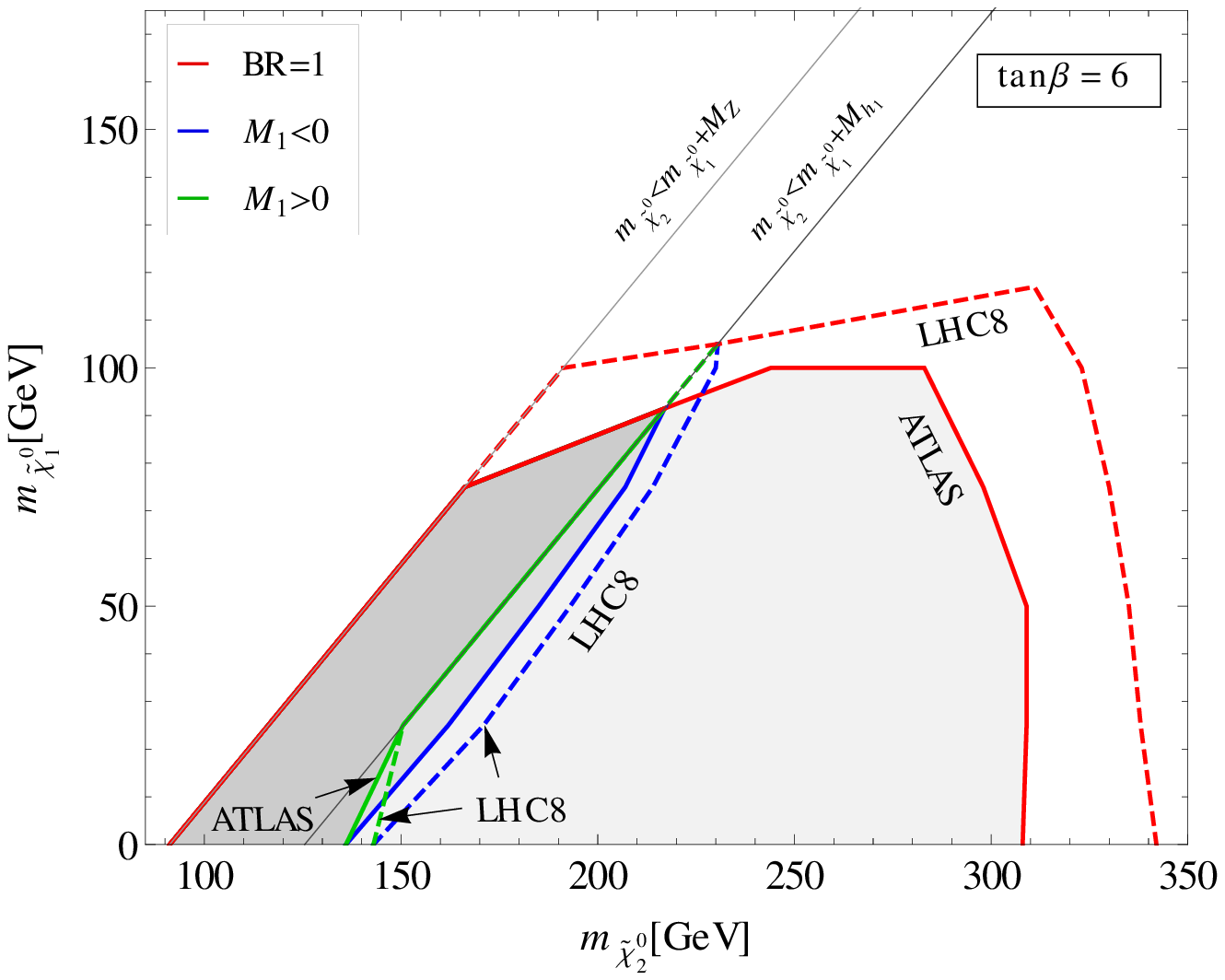}
\end{tabular}
\begin{tabular}{c}
\includegraphics[width=0.70\textwidth]{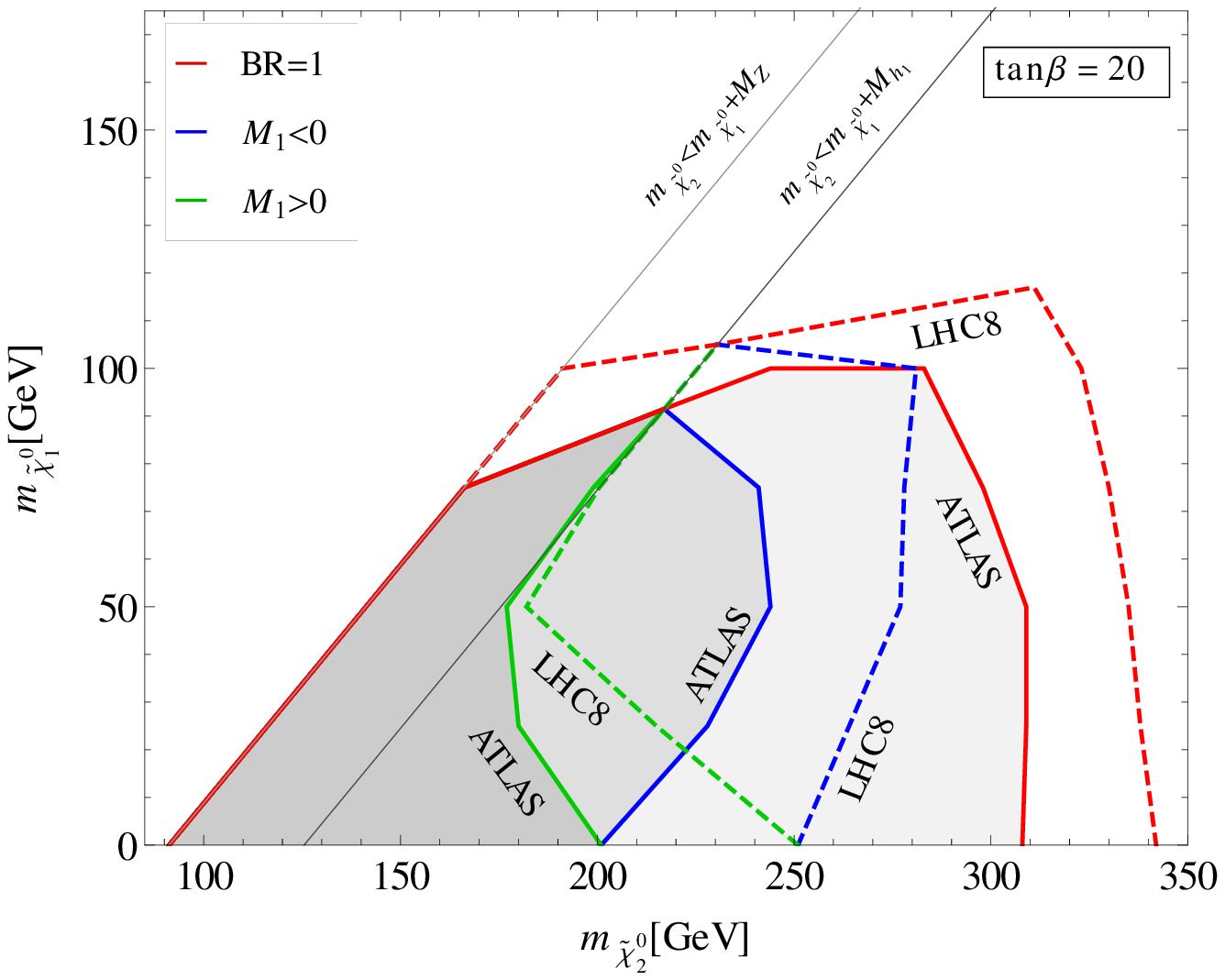}
\end{tabular}
\vspace{-1em}
\caption{
Contours showing the approximate excluded region from \Satlas\
in the $\mneu{2}$--$\mneu{1}$ plane
with $\tb = 6$ (upper plot) and  $\tb = 20$ (lower plot), 
The solid lines (shaded areas)
correspond to the exclusion for the currently analyzed ATLAS data, and
the dashed lines indicate the projection for the combined LHC8 data,
both for the case where it is assumed $\br(\neu{2}\to\neu{1}Z)=1$,  
and where the decays $\neu{2} \to \neu{1} \He$ are taken into account
for $\MOne > 0$ (green), and for $\MOne < 0$ (blue) as indicated,
calculating $\br(\neu{2}\to\neu{1}Z)$ at NLO.  
Above the light (dark) gray line the on-shell
decay $\neu{2} \to \neu{1} Z (\He)$ is kinematically forbidden.
}
\label{fig:contour-LHC8}
\end{center}
\vspace{-2em}
\end{figure}
The results for 
scenario \Satlas, i.e.\ with $\tb=6$, are shown in the upper figure. 
The dramatic reduction of the excluded area from the ATLAS result in
comparison when the decay $\neu{2} \to \neu{1} \He$ is taken into
account is clearly visible. Only the region where 
$\neu{2} \to \neu{1} \He$ is kinematically forbidden, extended by small
strips close to the kinematic limit can be excluded by the
current ATLAS analysis. The excluded area grows only marginally taking
into account the projection for the LHC8 full data set, i.e.\ the 
projected combination of ATLAS and CMS data.

The results for scenario \Stb, i.e.\ with $\tb=20$,
are displayed in the lower figure of \reffi{fig:contour-LHC8}.
While, by definition the curves with $\br(\neu{2} \to \neu{1} Z) = 1$
are identical for $\tb = 6$ and $\tb = 20$,
the regions excluded taking $\neu{2} \to \neu{1} \He$ into account are
somewhat larger for $\tb = 20$.
Still a substantial reduction of the excluded regions remains visible.
Again, the observations in \reffi{fig:contour-LHC8} can easily be understood in
terms of \refeq{eq:nnh.approx} and (\ref{eq:nnz.approx}), where we see that for smaller $\tb$ and 
large $\mu$ the decay to the Higgs dominates, and the branching ratio to the
$Z$ boson substantially smaller than one.

Altogether these results show, on the one hand, how important it is to
look at a realistic spectrum (i.e.\ where the decays to a Higgs boson
are not neglected), and on the other hand that dedicated searches for
the $Wh+E^{\rm miss}_T$ channel are beneficial~\cite{Baer:2012ts}.


\subsection{Complex couplings}
\label{sec:complex}

As shown in \refse{sec:amplitudes}, the
partial decay width to Higgs bosons decreases with $\phiMe$, 
due to the decrease in the neutralino-Higgs coupling, which from
\refeq{eq:nnh.approx} is seen to be most dependent on the phase for large
$\tb$. 
The $\phiMe$ dependence could have an interesting impact on the
exclusion bounds on $\MOne$ and $\MTwo$.
This is illustrated in \reffi{fig:contour-phiMe}, where we interpret the
ATLAS exclusion 
limits in the $\phiMe$--$\MOne$ plane, for $\tb = 6$ (left)
and $\tb = 20$ (right),
and for $\De := \MTwo - \MOne = \MHe,\,130,\,150,\,180 \gev$
(defining the value of $\MTwo$ in the plots).
The values of $\Delta$ here correspond approximately to diagonal lines in the
$\mneu{2}$--$\mneu{1}$ plane in \ref{fig:contour-LHC8}, starting from the
Higgs threshold at $\Delta=M_{h_1}$.
The solid (dotted) lines correspond to the NLO (tree-level) calculation.
We also indicate the limits in red, given by the requirement that the EDMs 
for thallium and mercury, $d_{\rm Tl}(\msusy)$ and $d_{\rm Hg}(\msusy)$, calculated using \texttt{CPsuperH~2.3}~\cite{Lee:2012wa,Lee:2007gn,Lee:2003nta}%
\footnote{
Similar results can be obtained with \texttt{FeynHiggs}~\cite{feynhiggs,mhiggslong,mhiggsAEC,mhcMSSMlong}.}%
~for a specific value of $\msusy$ within \Ssusy, is below the upper
limit, i.e.\  
$d^{\rm exp}_{\rm Tl} =9.0\,10^{-25}\, e\,\mathrm{cm}\,(90\%~\mathrm{CL})$ or 
$d^{\rm exp}_{\rm Hg} =3.1\,10^{-29}\, e\,\mathrm{cm}\,(95\%~\mathrm{CL})$~\cite{Regan:2002ta,Griffith:2009zz}.
We adopt a common mass scale $\msusy = \Msqez = \Msqd = \Msl$, although
the EDMs depend mainly on $\Msqez$ and $\Mslez$. Note that the
predictions rely on atomic or hadronic matrix elements which can have
large theoretical uncertainties, ranging from 
$\sim 10\%$ to $50\%$ (see e.g.~\citere{Demir:2003js}).
Of the numerous options available in \texttt{CPSuperH~2.3} for the parametrization of the Schiff moment contribution 
to $d_{\rm Hg}$, as described in \citere{Ellis:2011hp}, we choose the result of \citere{Pospelov:2005pr}.
We display the limit for the EDM that provides the
strongest bound, i.e.\ from $d_{\rm Tl}$ for $\tb=6$ and from
$d_{\rm Hg}$ for $\tb=20$. Although 
$\msusy=0.8\tev$ is disfavored at the LHC, the lines are
indicated for comparison, as there is no exclusion 
from the EDMs for higher values of $\msusy$ (i.e. in \Satlas) for the case
$\tb=6$. 

From \reffi{fig:contour-phiMe} one can clearly see the effect of
$\phiMe$ on the exclusion limit on $M_1$, 
being much higher for $\phiMe=\pi$ than for $\phiMe=0$, as also seen in
\reffi{fig:contour-LHC8} and discussed in \refse{sec:Scomp}.
Furthermore, as discussed above, it can be observed that the
effect of the phase is much more pronounced for lower values of $\tb$.
On changing $\phiMe$ from 0 to $\pi$, for $\De =130\gev$, the limit on $\MOne$ 
changes by $\sim 80\gev$ for $\tb=6$ as opposed to $50\gev$ for $\tb=20$.
Note that the exclusion disappears completely for $\De >135\gev$ for
$\tb=6$ and $\De >200\gev$ for $\tb=20$. The $\De = \MHe$ line is also
shown, illustrating that below the Higgs threshold, 
the dependence on $\phiMe$ vanishes. 
Further effects of $\phiMe$ will be discussed in \refse{sec:dm}.

The relevance of the one-loop corrections, i.e.\ the difference
between solid and dotted lines, is clearly visible for $\tb = 20$,
in the right plot of \reffi{fig:contour-phiMe}, via the shift in the 
excluded $\MOne$ value (for fixed $\phiMe$), which ranges
from $\sim 0$ for $\De = \MHe$ and $\sim 12 \gev$ for $\De = 180 \gev$.
A more detailed discussion of the impact of NLO corrections can be
found in \refse{sec:nlo}.

\begin{figure}[t!]
\begin{center}
\begin{tabular}{c}

\includegraphics[width=0.49\textwidth]{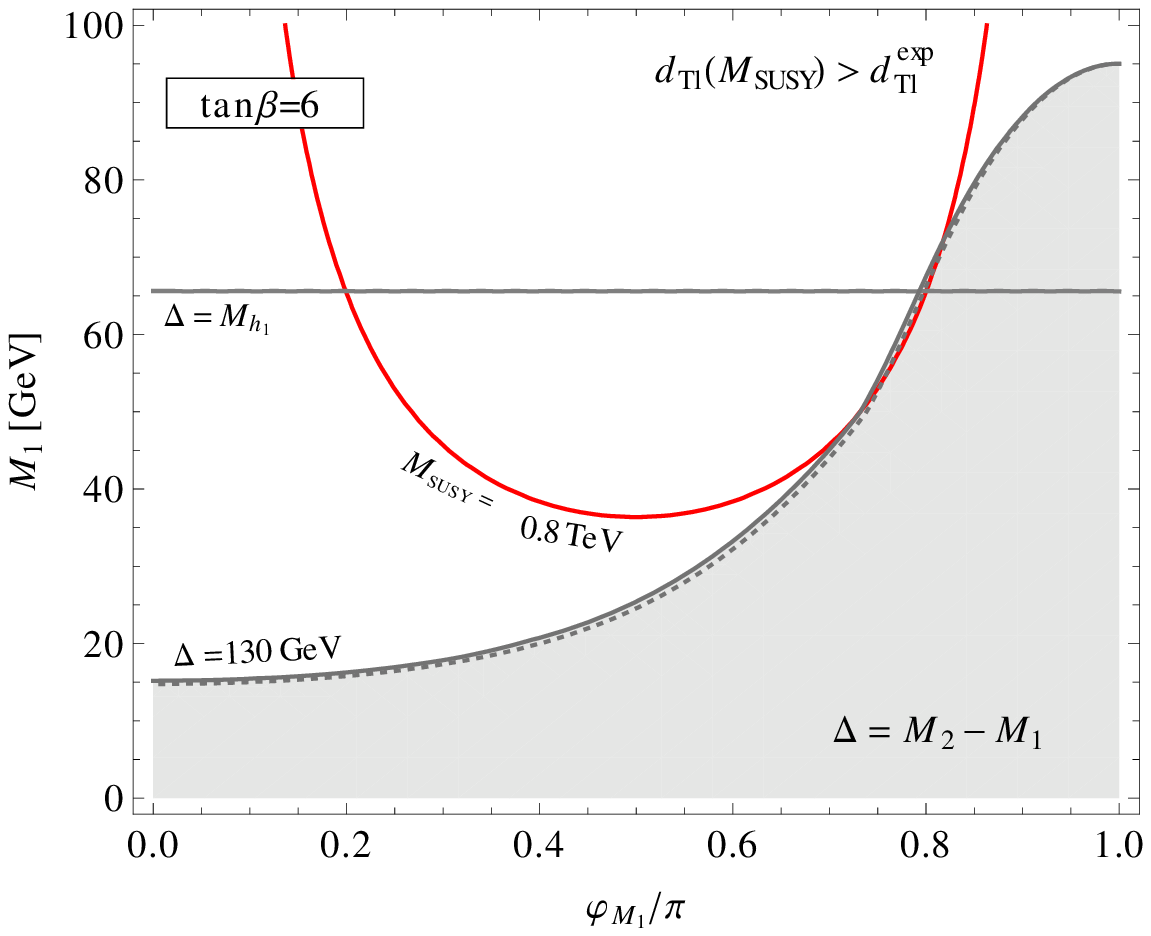}\hspace{.3cm}
\includegraphics[width=0.49\textwidth]{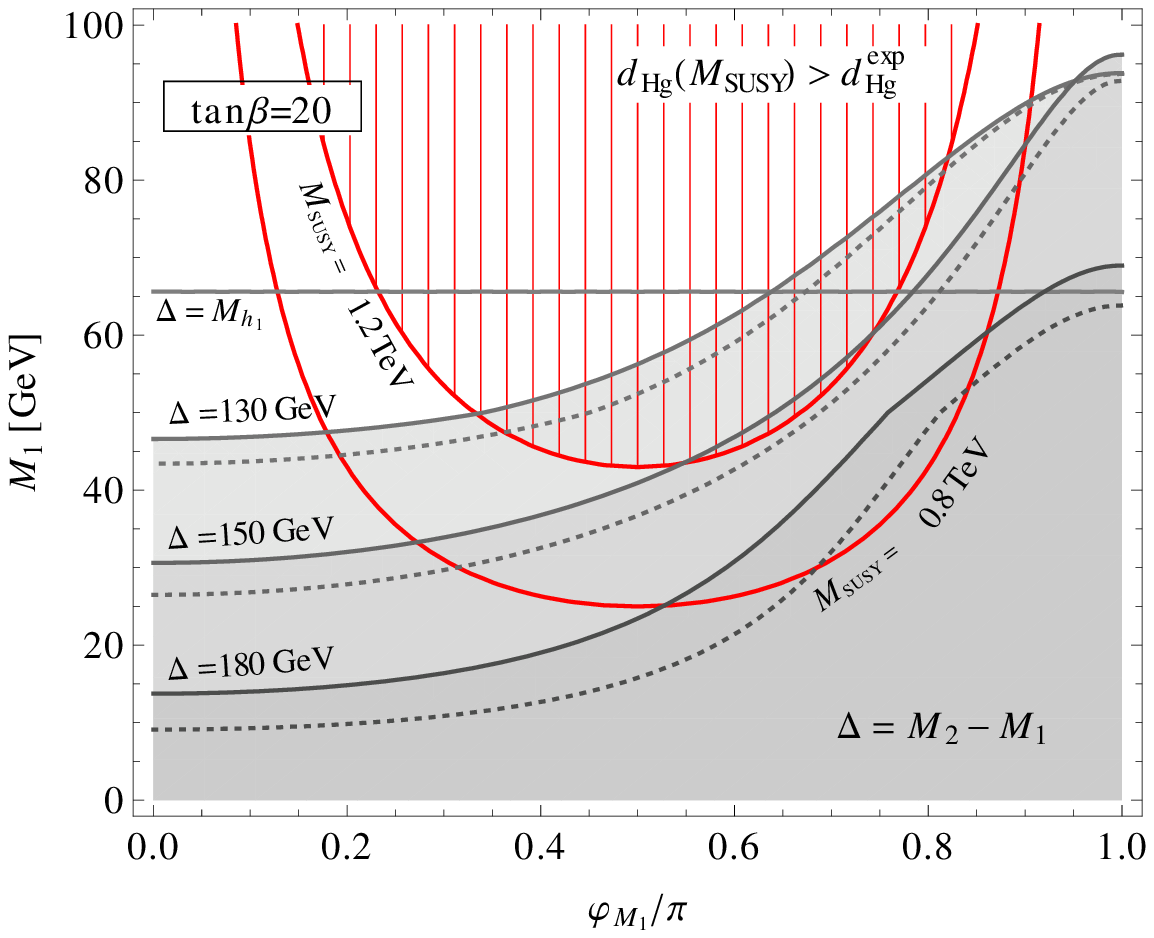}

\end{tabular}
\caption{
Contours showing the excluded region from currently analyzed ATLAS data for
\Satlas\ in the $\MOne$--$\phiMe$ plane, with $\tb = 6$ (left) and 
$\tb = 20$ (right). $\MTwo$ is fixed via $\De = \MTwo - \MOne$, which
corresponds approximately to  
diagonal lines in e.g.\ \reffi{fig:contour-LHC8}, parallel to the Higgs
threshold given by $\De = \MHe$. 
The solid (dotted) lines indicate that the exclusion
contours are calculated using NLO (tree-level) branching ratios for 
the $\neu2$ decays.
At $\De = 150 \gev$ for $\tb = 6$ and $\De = 210\gev$ for $\tb = 20$
there is no exclusion from ATLAS. 
The red lines define exclusion contours for \Ssusy, where $\msusy$ is
indicated (see text), from the EDMs of thallium ($d_{\rm Tl}$) and mercury
($d_{\rm Hg}$).}
\label{fig:contour-phiMe}
\end{center}
\end{figure}


\subsection{The DM scenario: effect of a low scalar tau mass}
\label{sec:dm}

In this section we briefly analyze the effect of making the 
low scalar tau nearly degenerate with the LSP, i.e.\ scenario \Sstau.
In this way the $\neu1$ provides the correct amount of relic Cold Dark
Matter~\cite{cdm,micromegas}. In this scenario 
the new $\neu2$ decay channel, $\neu2 \to \Staue \tau$ opens up.
The appearance of this decay channel not only reduces the 
$\br(\neu2 \to \neu1 Z)$, i.e.\ the channel taken into account by
the ATLAS analysis, but also results in a new channel to be analyzed, as in 
\citere{CMS:2012ewa,ATLAS:2012ku,CMS:2012ewb,ATLAS:2012-154,ATLAS:2013-028,ATLAS:2013-035}.
A combination of the experimental analyses for these two channels is clearly
beyond the scope of our paper. Consequently in this scenario, we analyze
the branching ratios, but 
do not attempt to re-evaluate the exclusion limits in the
$\mneu{2}$--$\mneu{1}$ plane.
The problem of combination of analyses due to the appearance of new
channels becomes even more important in this scenario, since the
channel $\cha1 \to \Staue \nu_\tau$ also opens up, which can have a
non-negligible branching ratio~\cite{LHCxC}. However, we will not
discuss this point further here.

The decays of gauginos to a lepton-slepton pair strongly depends on the
character of the sleptons: 
while the decay to left-handed sleptons is unsuppressed, 
that to right-handed sleptons is proportional to the Yukawa coupling,
which is strongly suppressed by the small mass of the leptons.
As here we are interested in the interplay of the $Z$, $\He$ and $\Staue$
channels, we focus on the region of parameter space where the suppressed
neutralino decays to $Z$ and Higgs bosons (see \refse{sec:amplitudes})
are competitive with the decay to $\Staue\tau$. 
Unsuppressed decays to a left-handed $\Stau$ and a $\tau$ would
strongly dominate all other decay channels and are thus of limited
interest here.
Decays to right-handed staus, on the other hand, are potentially
interesting from a phenomenological point of view, and we consider the
possibility that the right-handed soft SUSY-breaking parameter
$\MstauR$ is much smaller than the left-handed one, resulting in an
almost purely right-handed lightest stau, as given in the
  definition of \Sstau.
The decay to a pure $\Stau_R$ results in the minimum possible BR to 
$\Staue \tau$ of the gaugino-like neutralino. 
However, the stau mixing induced by the non-diagonal entry in the stau 
mass matrix, given by $\mtau \times \mu\,\tb$ (using $\Atau = 0$), 
adds a small $\Stau_L$ admixture, strongly enhancing the decay 
$\neu2 \to \Staue \tau$. We find 
a stau mixing angle of $\theta_{\Stau} \simeq 0.25 \times 10^{-3}$ 
for $\tb=6$ and $\theta_{\Stau}\simeq 0.7\times 10^{-3}$ for $\tb=20$.
Both the increased left-handed component, as well as the
$\tb$-dependence of the Yukawa couplings 
enhance decay $\neu2 \to \Staue \tau$ with growing $\tb$.

\begin{figure}[t!]
\begin{center}
\begin{tabular}{c}
\includegraphics[width=0.49\textwidth,height=7.5cm]{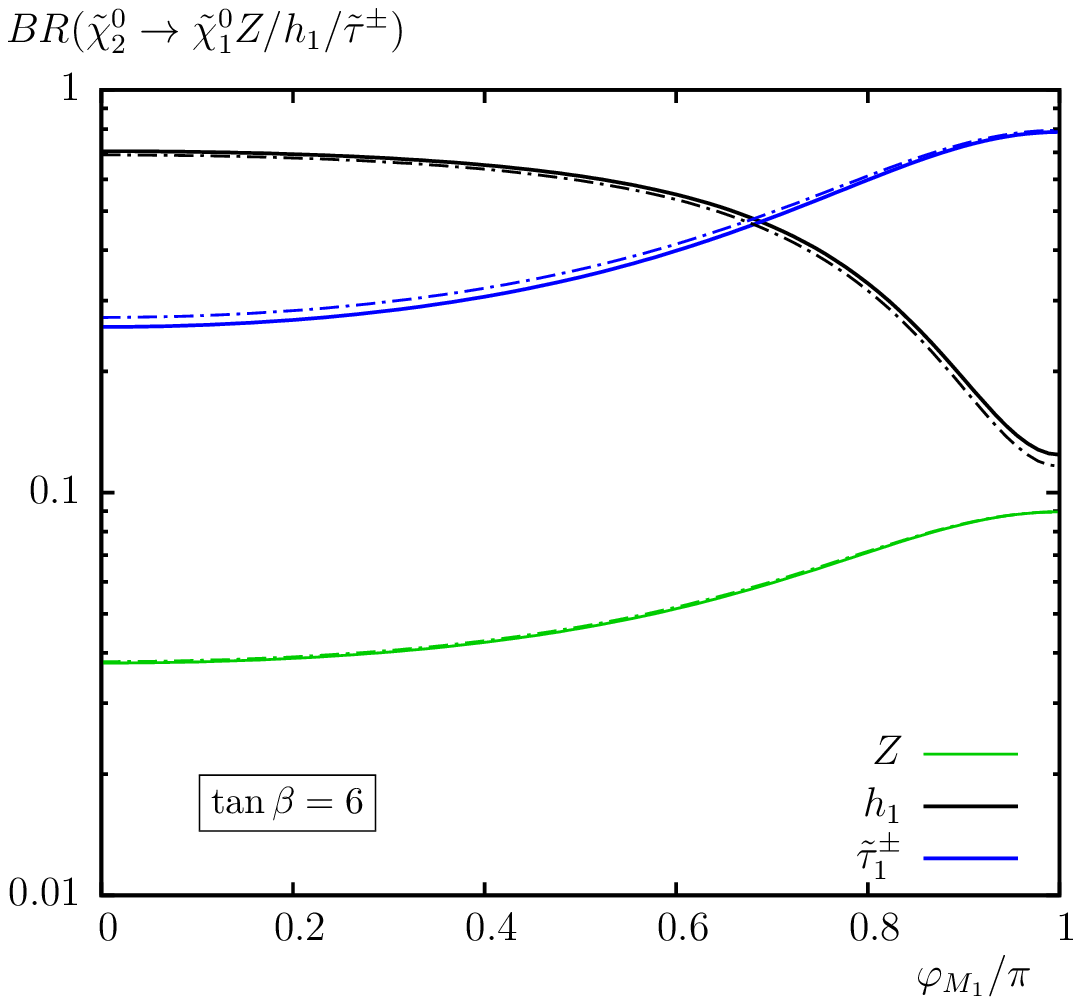} 
\hspace{4mm}
\includegraphics[width=0.49\textwidth,height=7.5cm]{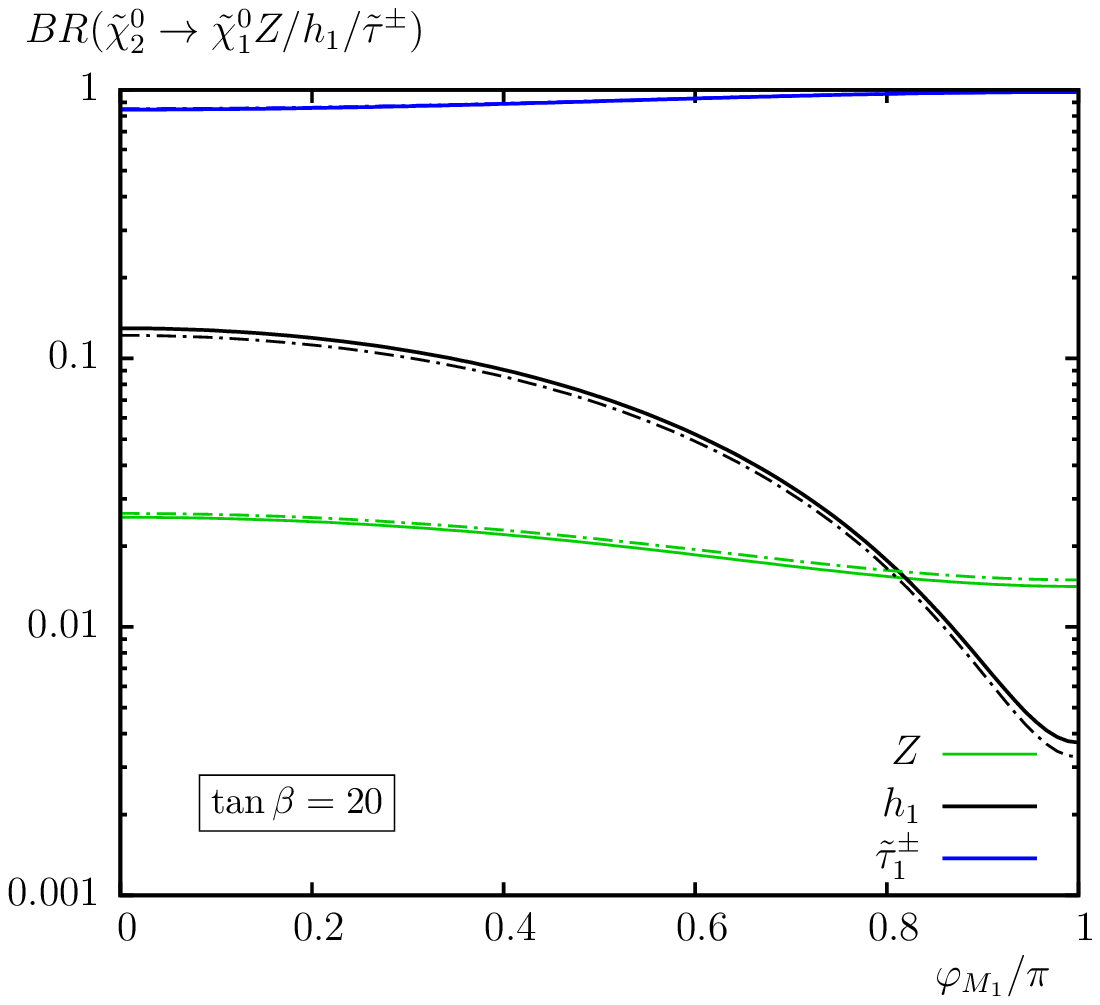}
\end{tabular}
\caption{
$\br(\neu2 \to \neu1 Z)$, 
$\br(\neu2 \to \neu1 \He)$ and $\br(\neu2 \to \Staue \tau)$ in the
\Sstau\ scenario as a function of $\phiMe$ for $\tb = 6$ (left) and~$20$
(right). Solid lines show the tree-level result, whereas
dash-dotted lines display the NLO branching ratios.
}
\label{fig:BRfphiM1neu2staur}
\end{center}
\end{figure}

The results are shown in \reffi{fig:BRfphiM1neu2staur}, where we
display $\br(\neu2 \to \neu1 Z)$, $\br(\neu2 \to \neu1 \He)$ and 
{$\br(\neu2 \to \Staue^\pm \tau^\mp)$} as a
function of $\phiMe$ for $\tb = 6$ and~$20$ in the left and right plot,
respectively. Solid lines show the tree-level result, whereas
dash-dotted lines display the NLO branching ratios. 
We see that the branching ratio for the decay to $\neu1 Z$ is reduced to
the percent level, in general lying between $\sim 1.5$ and $\sim 3\%$
and reaching at most $\sim 9\%$ for $\tb = 6$ and $\phiMe = \pi$. 
This strong reduction would make this decay challenging at the LHC13/LHC14.
Note that in our benchmark scenarios with a bino-like LSP, the decay of the 
second lightest neutralino to a lepton-slepton pair has a negligible
dependence on $\phiMe$, and the increase of $\br(\neu2 \to \Staue \tau)$
with $\phiMe$ is  due to the decrease in the partial decay widths to $Z$
and Higgs bosons, as discussed in \refse{sec:amplitudes}.
There we observed that these couplings are largest when $\phiMe=0$,
see \refeq{eq:nnh.approx}, 
and that the dependence on the phase is largest
for large $\tb$, in agreement with \reffi{fig:BRfphiM1neu2staur}.


\subsection{Gaugino vs.\ higgsino production: the low \boldmath{$\mu$} case}
\label{sec:higgsino}

The limits on searches for electroweak SUSY particles presented by ATLAS
and CMS assume that the relevant particles are gaugino-like,
corresponding to a relatively large value of~$\mu$.
In this region of the SUSY parameter space the production cross sections
for $\cha{1}\neu{2}$ are largest due to the unsuppressed coupling to the
$W$-boson. On the other hand, when
$\MOne < \mu < \MTwo$ the second and third neutralino, as well as the
lightest chargino, are higgsino-like and roughly degenerate in mass,
changing the phenomenology of the particles under investigation. 
In this section we briefly analyze the
phenomenology in the scenario \Sgauge\ with $\MOne < \mu < \MTwo$ with 
$\mu = 100 \ldots 400 \gev$ and $\MTwo = 500 \gev$.

In \reffi{fig:xsbr.higbinomn.atlasexcl} we show the results obtained in
\Sgauge, in analogy to \reffi{fig:xsbr.tb6.atlasexcl}, for $\tb = 6$ and
for $\mneu{1} = 0, 50, 100 \gev$ in the top, middle and lower row,
respectively. Contrary to \reffi{fig:xsbr.tb6.atlasexcl}, besides the
results for $\cha1 \neu2$ production shown in the left column as
function of $\mneu2$, we present in the right column the analysis for
$\cha1 \neu3$ production as a function of $\mneu3$.  
As discussed above, in this scenario we have 
$ \mcha1 \approx \mneu2 \approx \mneu3 $. 
The mass differences range from a few to a few tenths of GeV,
not allowing a clean separation of these channels at the LHC.
This spectrum also forbids two-body decays between these states.
A full experimental analysis will combine the two production
channels. However, such a combination of different channels goes beyond
the scope of our paper, and we discuss the mass limits obtained for the
two production channels individually. 

\begin{figure}[ht!]
\begin{center}
\begin{tabular}{c}
\includegraphics[width=0.49\textwidth,height=6.7cm]{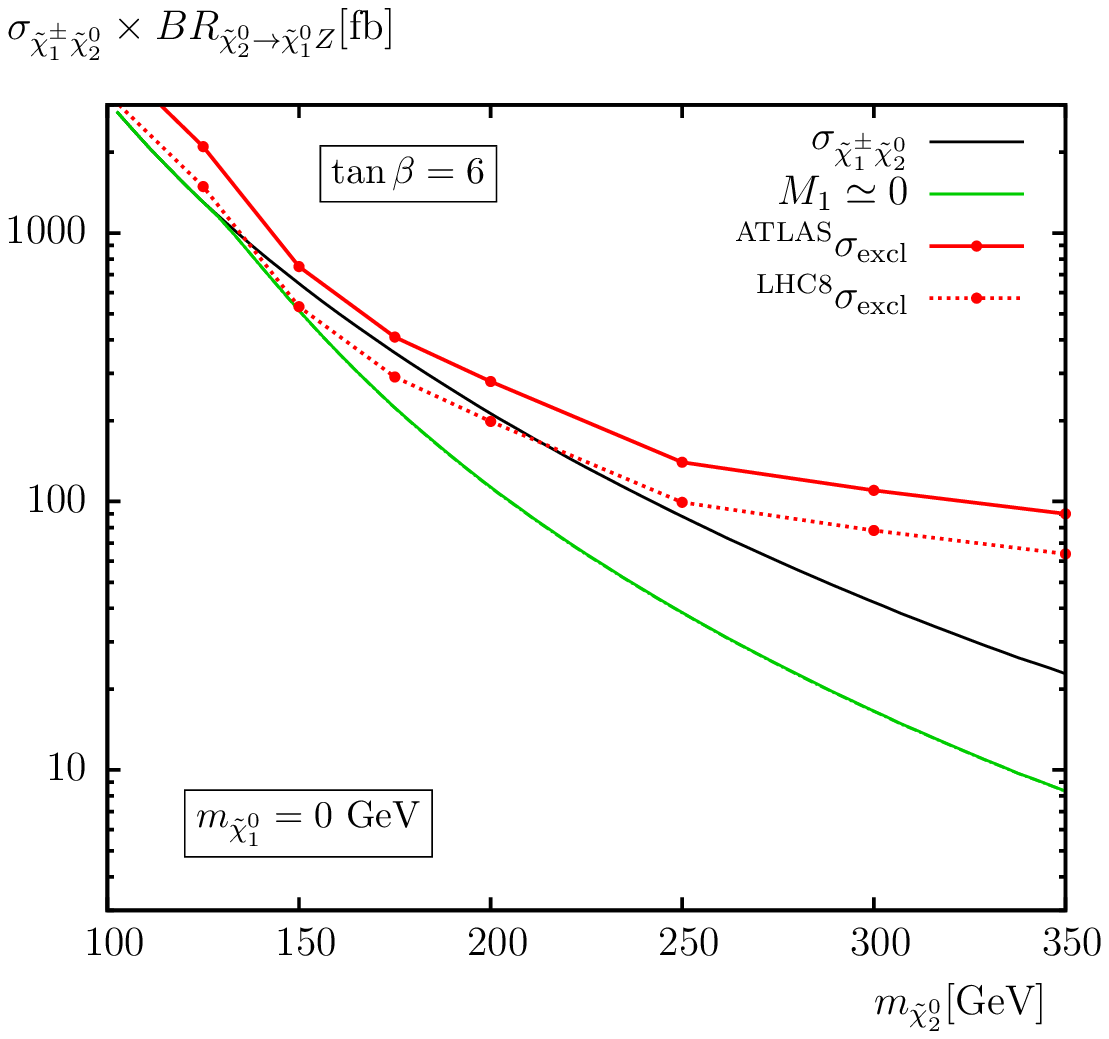}
\hspace{-4mm}
\includegraphics[width=0.49\textwidth,height=6.5cm]{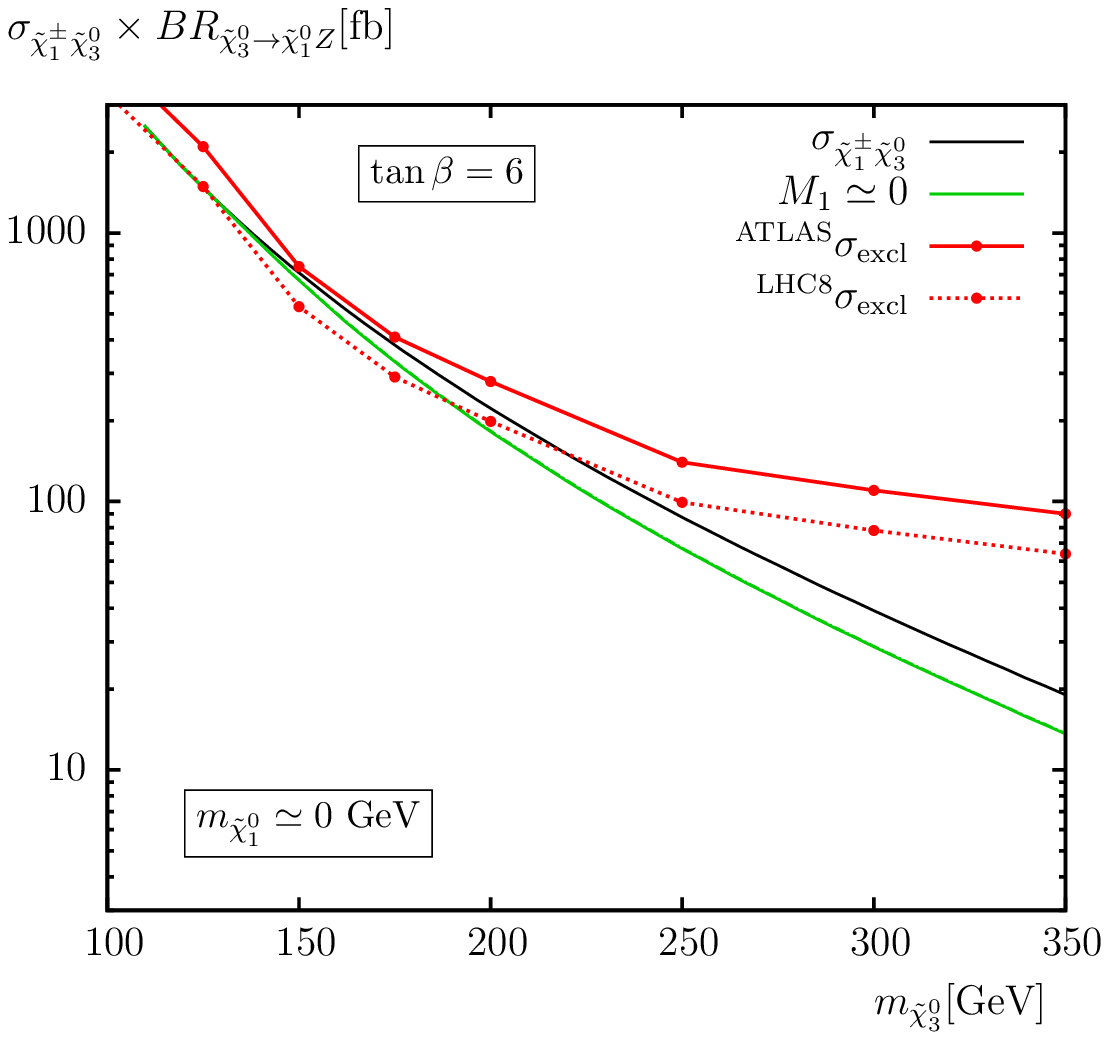}
\end{tabular}
\begin{tabular}{c}
\includegraphics[width=0.49\textwidth,height=6.5cm]{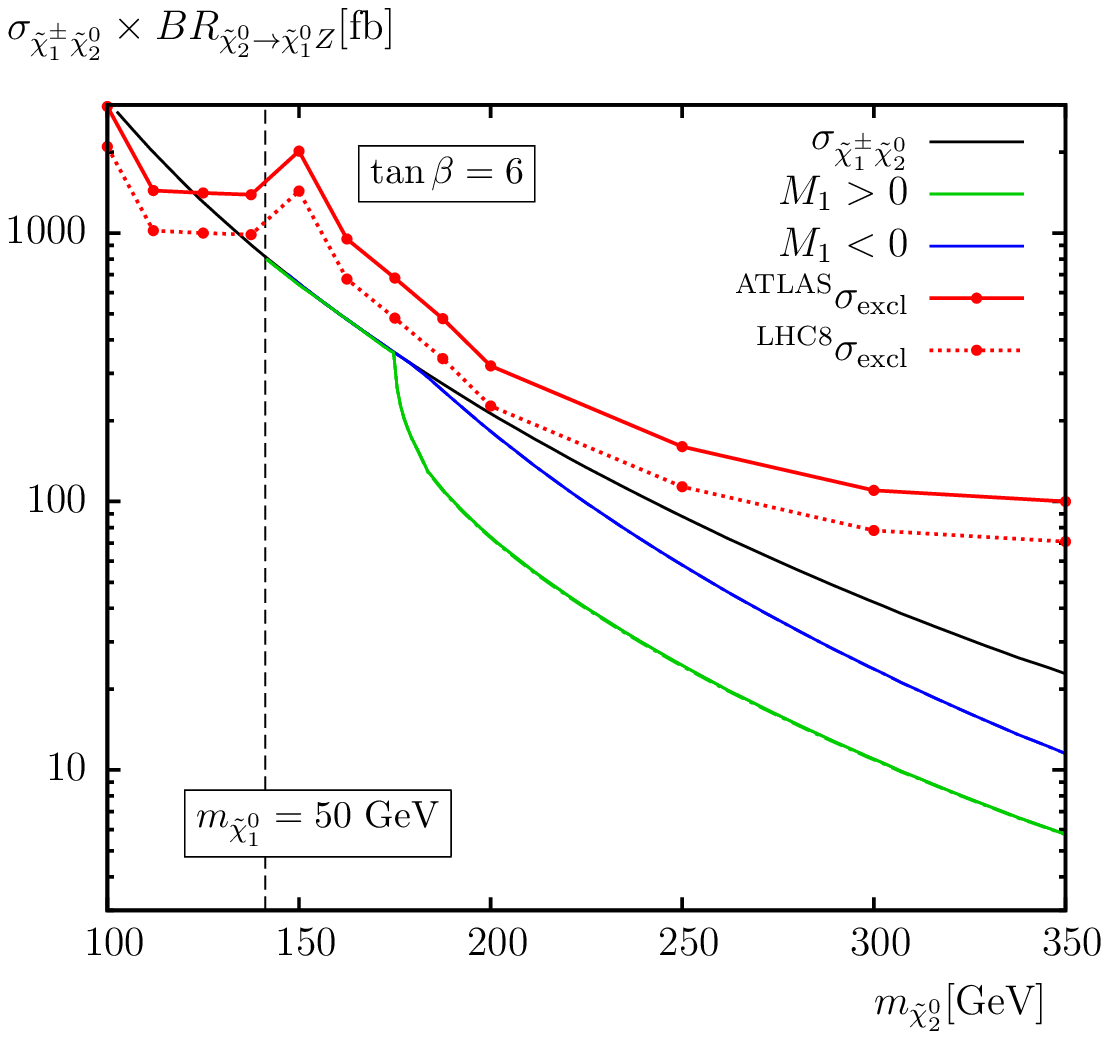}
\hspace{-4mm}
\includegraphics[width=0.49\textwidth,height=6.5cm]{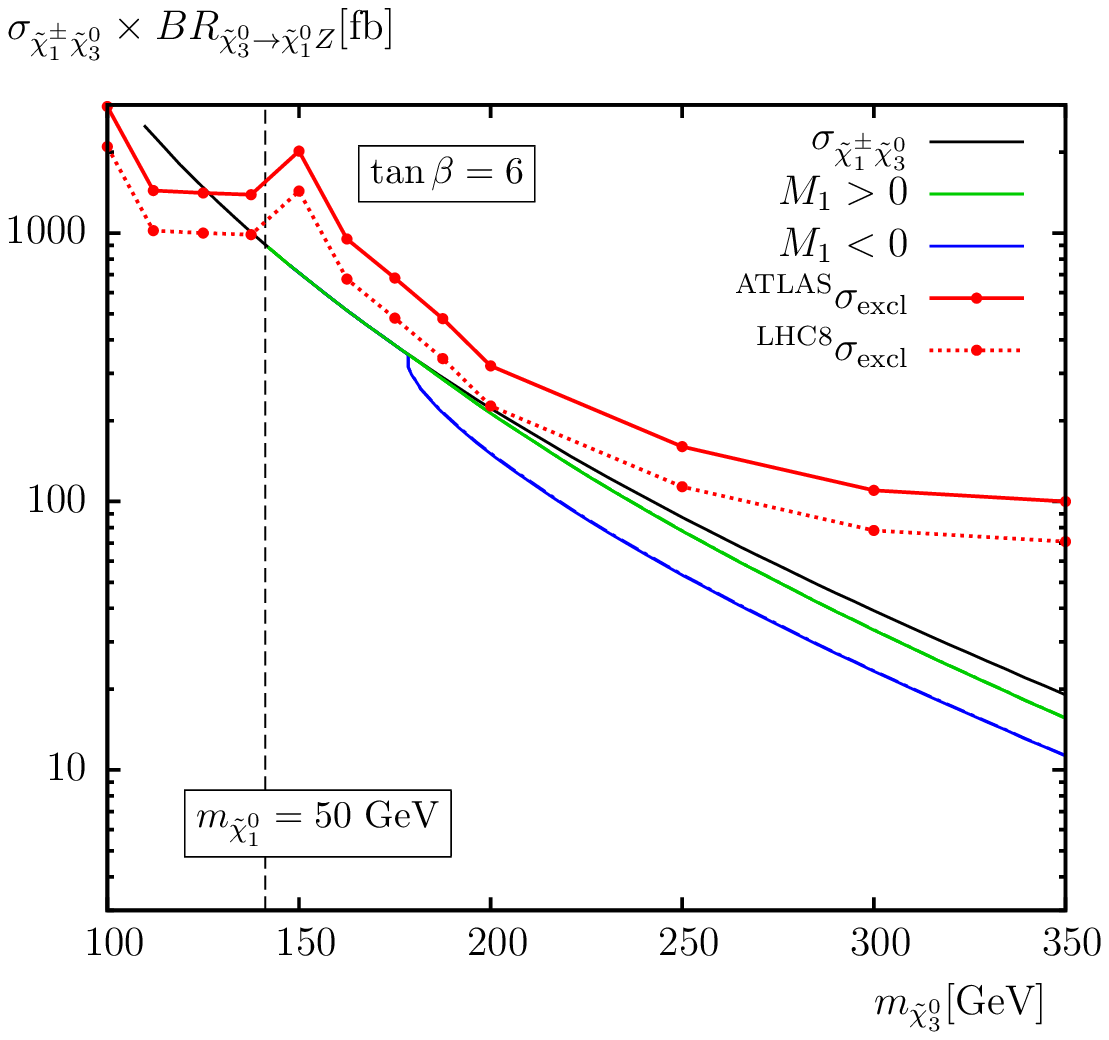}
\end{tabular}
\begin{tabular}{c}
\includegraphics[width=0.49\textwidth,height=6.5cm]{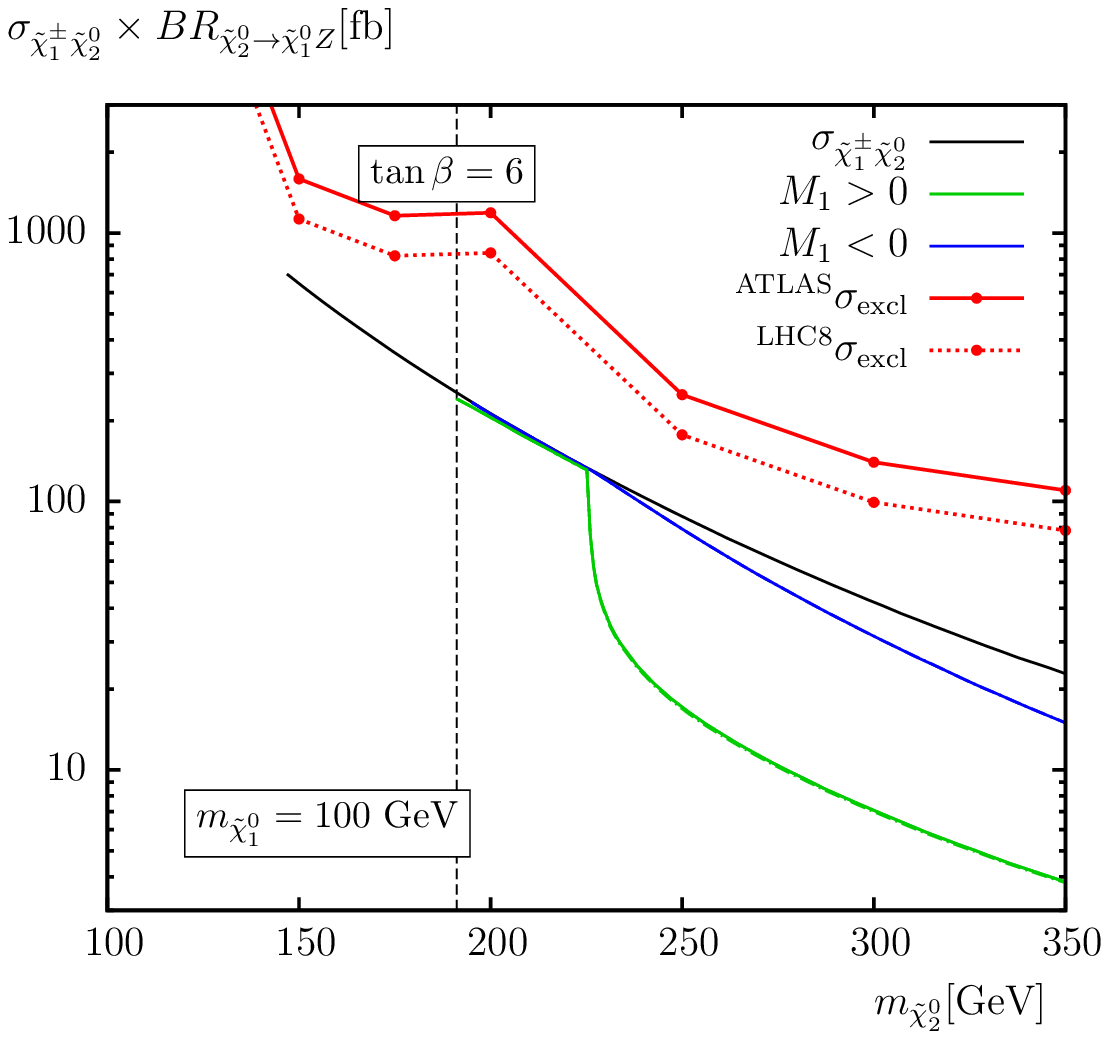}
\hspace{-4mm}
\includegraphics[width=0.49\textwidth,height=6.5cm]{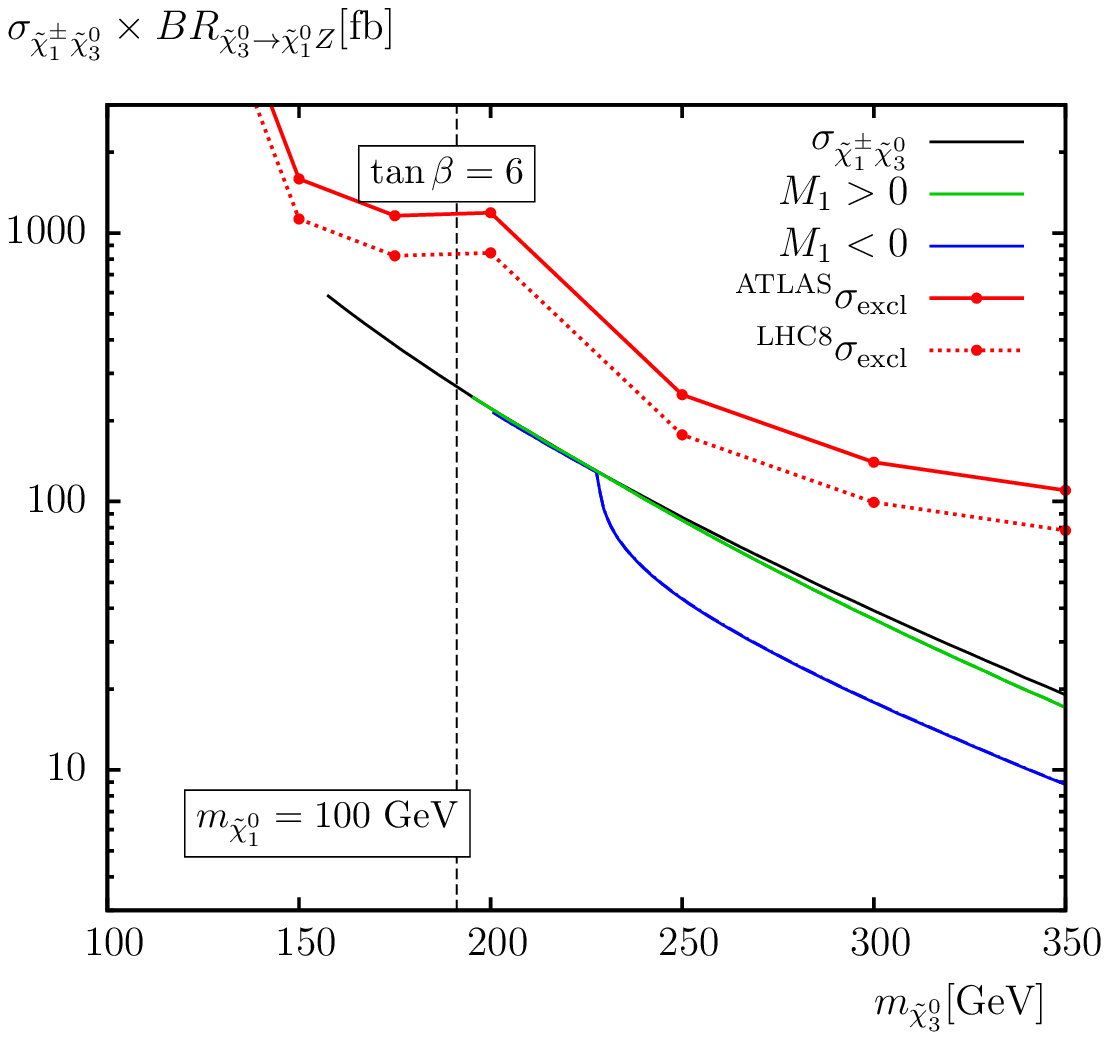}
\end{tabular}
\caption{
As in \reffi{fig:xsbr.tb6.atlasexcl}, but for 
$\cha{1}\neu{2}$ (left)
and $\cha{1}\neu{3}$ (right)
in scenario \Sgauge , with
$M_1$ is chosen such that $\mneu{1} = 0, 50, 100\gev$, 
in the top, middle and bottom plot, respectively. 
}
\label{fig:xsbr.higbinomn.atlasexcl}
\end{center}
\end{figure}

As a first general observation it should be 
noted that the smaller couplings to gauge bosons suppress the 
production cross section by almost a factor of~$4$ for low chargino
masses to $\sim 2$ for high masses (where a substantial mixing with
gauginos increases the coupling to the $W$-boson). 
Notice that this suppression could be partially overcome 
by a combination of the two production cross sections.
In order to interpret the ATLAS limits~\cite{ATLAS:2013-035}
in this scenario we must take into account that the chargino is not exactly
degenerate with the two neutralinos. 
We interpret the limits as a function of the corresponding neutralino mass,
which is larger than that of the chargino.
In our simple interpretation of the ATLAS limits, as shown in
\reffi{fig:xsbr.higbinomn.atlasexcl}, no mass value for 
$\mneu2$ or $\mneu3$ can be excluded. Only the combination of ATLAS and
CMS, shown as red-dotted line could yield an exclusion for $\mneu1 = 0$
for $\mneu3 \lsim 200 \gev$. 

The most interesting observation in this scenario is the complementarity
between  
$\cha1 \neu2$ and $\cha1 \neu3$ production. For $\MOne$ positive, shown
as the green lines, only $\br(\neu2 \to \neu1 Z)$ shows a strong
suppression, whereas $\br(\neu3 \to \neu1 Z)$ is close to one,
as noticed in \citere{Arbey:2012fa}. 
This complementary behavior 
is due largely to the Higgs-higgsino-gaugino couplings discussed in
\refse{sec:amplitudes} and the fact that $\neu2$ and $\neu3$ are opposite in
their $\cp$ behavior (being the SUSY partners of the $\cp$-even and -odd
Higgses).
This is also reflected in the difference between $\MOne$ positive or
negative. While for $\cha1 \neu2$ a negative $\MOne$ results in a larger
$\br(\neu2 \to \neu1 Z)$, it is exactly opposite for $\cha1 \neu3$,
where {\em only} for $\MOne$ negative a strong reduction of 
$\br(\neu3 \to \neu2 Z)$ is found. Consequently, for $\MOne < 0$ both
channels, $\cha1 \neu2$ and $\cha1 \neu3$ are suppressed. 

In summary: the case of a low $\mu$ parameter appears to be more challenging
for ATLAS and CMS than the case of a large TeV--scale $\mu$. The
production cross sections are suppressed for small $\mu$, and in
particular for negative $\MOne$ the $\br(\neu2/\neu3 \to \neu1 Z)$ lead
to a further suppression and hardly any limit can be derived from the 2012
data. Only a combination of the two search channels
might partly overcome this reduction.


\subsection{Impact of radiative corrections}
\label{sec:nlo}

It is of interest to calculate the predicted cross sections and the
resulting exclusion bounds at NLO (the full one-loop level), in
order to assess if there are regions of parameter space where these 
are large or of importance and must be taken into account.
A detailed study of the NLO corrections to neutralino decays was carried 
out for the complex MSSM in \citere{LHCxN}. There it was found that the
size of the corrections for $\neu{i} \to \neu1 Z$ and $\neu{i} \to \neu1 \He$ 
were greater in general, when the decaying neutralino was higgsino-like
rather than wino-like (assuming that $\neu1$ is predominantly bino-like).
However, in the case of the decaying neutralino being wino-like, 
the radiative corrections were strongly dependent on $\phiMe$.
This arises when the tree-level decay is suppressed, due to the coupling
or due to a p-wave suppression as described in \refse{sec:complex},
whereas the radiative corrections may not be suppressed by these mechanisms,
particularly in mixed scenarios.
In such cases at different values of $\phiMe$, the tree-level decay is
no longer suppressed, and therefore the relative size of the NLO
contribution is smaller.
For \Satlas\ the situation is different, as $\mu$ is at the TeV scale,
and these  effects are not so pronounced.
In \refta{tab:loopcorr}, we have summarized the percentage contribution of
the loop corrections for the decay $\neu2 \to \neu1 Z$ for
representative parameter points of each of the scenarios defined in
\refta{tab:para}. Here we use 
\begin{align}
\De\Ga^{\rm loop} := \frac{\Ga^{\rm NLO} - \Ga^{\rm tree}}{\Ga^{\rm tree}}\,,
\quad
\De\br^{\rm loop} := \frac{\br^{\rm NLO} - \br^{\rm tree}}{\br^{\rm tree}}~.
\end{align}
One can observe the following: 

\begin{itemize}

\item Our baseline scenario,
motivated by the ATLAS analysis, has a high $\mu$ and $\msusy$ as
well as small $\tb$, resulting in a change  
in the branching ratio to $Z$ bosons of the order of 8\%. As seen from
the relative correction to $\Ga(\neu2 \to \neu1 Z)$, small due to a cancellation
between the chargino/neutralino and sfermion loops, 
this 8\% arises due to the corrections to the decay width 
$\Ga(\neu2 \to \neu1 \He)$.

\item As seen in \refeq{eq:nnh.approx}, 
$\Ga(\neu2 \to \neu1 \He)$ is maximum
at $\phiMe=0$, where the loop corrections amount to $-11\%$, and
reaches a minimum when the coupling is smallest 
at $\phiMe=\pi$, where the corrections are maximal, i.e.~$-15\%$.

\item On decreasing $\msusy$, the correction to the branching ratio
increases by $50\%$, as the correction 
to $\Ga(\neu2 \to \neu1 Z)$ increases, due to the sfermion loop contribution,
to $-4\%$ at $\phiMe=0$ and $-2\%$ at $\phiMe=\pi$. There is also 
a large increase in the magnitude of the loop corrections to  
$\Ga(\neu2 \to \neu1 \He)$ from $\sim -18\%$ at $\phiMe=0$ to $-25\%$ at 
$\phiMe=\pi$. Overall corrections to $\br(\neu2 \to \neu1 Z)$
larger than 10\% can be observed.

\item Finally, the effect of $\mu$ was investigated in both scenarios
\Satlasmu\ and \Sgauge, and found to play an important role. 
As $\mu$ increases, as in \Satlasmu, the chargino/neutralino
contribution to the corrections decouples. 
Therefore the cancellation seen for \Satlas\ no longer exists, 
and the correction to the decay width (mostly due to sfermion loops) 
becomes $-5\%$.  
This means that the corrections to $\Ga(\neu2 \to \neu1 Z)$ and 
$\Ga(\neu2 \to \neu1 \He)$ partially compensate each other. 
For low $\mu$ (\Sgauge), the chargino/neutralino loop contribution
increases, resulting in the effect on the BR canceling out nearly
completely and is found to be at the level of $-1\%$.
\end{itemize}

Note that the impact of loop corrections at large $\tb$ is 
nicely illustrated in \reffi{fig:contour-phiMe}, where the difference
between the tree-level and NLO exclusion line reaches up to
$\sim 12 \gev$, resulting in a 30\% change in the excluded $\MOne$.
Therefore the NLO corrections can have a substantial effect and should
eventually be included 
when interpreting exclusion limits (or discovery results) for 
MSSM parameters.

\begin{table}[ht!]
\renewcommand{\arraystretch}{1.5}
\BC
\begin{tabular}{|c||c|c|c|c|c|c|c|r|r|
}
\hline
Scenario & $|M_1|$ & $M_2$&$\phiMe$ & $\mu$& $\tb$ &
$\msusy$&$M_{\tilde{\tau}_R}$& $\De \br^{\mathrm{loop}}$ & $\De \Ga^{\mathrm{loop}}$\\ \hline\hline
\Satlas &$100$&$250 $&$0$& $ 1000  $ & $ 6$ & $ 2000 $ & $\msusy$& 8\%\quad\quad&$<1\% $ \quad\\ 
\Satlas &$100$&$250 $&$\pi$& $ 1000  $ & $ 6$ & $ 2000 $ & $\msusy$& 4\%\quad\quad& $1\% $ \quad\\ \hline
\Scompl &$100$&$250 $& $\pi/2$ & $ 1000   $ & $ 6 $ & $ 2000 $ & $ \msusy$\quad& 8\%\quad\quad& $<1\% $ \quad\\ \hline
\Stb  &$100$&$250$&$   0$ & $ 1000$ & 20 & $ 2000 $ & $\msusy$& 8\%\quad\quad& $<1\%$ \quad\\ 
\Stb  &$100$&$250$&$  \pi$ & $ 1000$ & 20 & $ 2000 $ & $\msusy$& 4\%\quad\quad& $1\%$ \quad\\ \hline
\Satlasmu &$100$&$250$& $ 0$ & $2000$ & $ 6$ & $ 2000 $ & $\msusy$& 7\%\quad\quad& $-5$\% \quad\\ \hline
\Ssusy &$100$&$250 $&$  0 $ & $ 1000 $ & $ 6 $ & $ 1200$ & $\msusy$& 12\%\quad\quad&$-4$\% \quad\\
 \Ssusy &$100$&$250 $& $\pi$ & $ 1000 $ & $ 6 $ & $ 1200$ & $\msusy$& 11\%\quad\quad&$-2$\% \quad\\\hline\hline
\Sstau  &$100$&$250$&$  0$ & $ 1000$ & $ 6$ & $ 2000 $ & $|M_1|$ &5\%\quad\quad& $-1$\% \quad\\
\Sstau  &$100$&$250$&$  \pi$ & $ 1000$ & $ 6$ & $ 2000 $ & $|M_1|$ &5\%\quad\quad& $-1$\% \quad\\\hline
\Sgauge  &$100$&$500$& $ 0$ & $250$ & $ 6$ & $ 2000 $ & $ \msusy$ & $-1$\%\quad\quad& 2\% \quad\\  
\Sgauge  &$100$&$500$& $ 0$ & $350$ & $ 6$ & $ 2000 $ & $ \msusy$ & $-1$\%\quad\quad& 4\% \quad\\ \hline   
\end{tabular}
\caption{Percentage contribution of the one-loop corrections to the
  branching ratio and decay width for $\neu{2}\to\neu{1}Z$, for each of
  the scenarios defined in \refta{tab:para}. For those scenario defined in terms of a range in a particular parameter, we specify the value of this 
parameter used given, with the exception of $\tb$, in $\gev$.}
\label{tab:loopcorr}
\EC
\renewcommand{\arraystretch}{1.0}
\end{table}

\newpage

\subsection{LHC13 expected sensitivity}
\label{sec:lhc13}

In this section we make a simple projection of the LHC8 results to
the future upgrade to the LHC13 assuming a luminosity $\mathcal{L}_{\rm LHC13}=100\,\ifb$.
In order to estimate the LHC reach without making a full dedicated analysis
we need to make an extrapolation of the relevant background.
The main irreducible background for the chargino/neutralino direct production 
with subsequent decays to gauge bosons
is diboson production, see e.g.~\cite{ATLAS:2012-154, ATLAS:2013-035}.
We therefore rescale the number of background events with 
\begin{align}
\label{eq:scaling13bkgr}
R_{\rm bkg} = \frac{\sigma_{WZ}(13\tev)}{\sigma_{WZ}(8\tev)}
\times
\frac{\mathcal{L}_{\rm LHC13}}{\mathcal{L}_{\rm LHC8}}~.
\end{align}
where $\sigma_{WZ}(x\tev)$ denotes the inclusive $W^\pm Z$ production cross section at LHCx.
This projection  neglects any effects from the different environment at the higher energy LHC run, 
for instance the larger pile up. 
Notice that similar assumptions have been made for projections 
for scalar top searches at CMS~\cite{OlsenTalk}, 
where the main background is related to the $t\bar{t}$ production cross section.
Diboson inclusive cross sections has been evaluated at
NLO~\cite{Campbell:2011bn}. 
The $W^\pm Z$ cross section increases
from 
$23$~pb at LHC8 to $47$~pb at the LHC13, 
i.e., roughly by a factor of two.%
\footnote{The extrapolation of production cross sections to higher energies
 of both the background as well as the signal
will depend on the specific choice of kinematical cuts. For instance, the ratio $R_{\rm bkg} $ increases by up to
$~20\%$ if one applies large transverse momentum cuts~\cite{Campbell:2011bn}. 
However, also the signal is expected to grow by a similar factor. Therefore we neglect these effects here.
}

The naive estimate of the sensitivity reach of the LHC13 in this channel is
obtained by scaling the 
expected exclusion sensitivity at $8\tev$ c.m.e., $^{\rm LHC8}\sigma_{\rm excl}$,
by a factor 
\begin{align}
\label{eq:scaling13}
R_{13/8}={\sqrt{R_{\rm bkg}}}
\times
\frac{\mathcal{L}_{\rm LHC8}}{\mathcal{L}_{\rm LHC13}}~,
\end{align}
where the first term on the right hand side takes into account the increase of
 the background, which decreases the sensitivity, while the second term the increase 
of the signal with the luminosity.
In our specific case this results in a factor of
$\sqrt{2}\times \sqrt{21/100}\approx 0.65$, i.e.\ an improvement of $35\%$ w.r.t. the current sensitivity.

\medskip

\begin{figure}[t!]
\begin{center}
\begin{tabular}{c}
\includegraphics[width=0.49\textwidth,height=7.5cm]{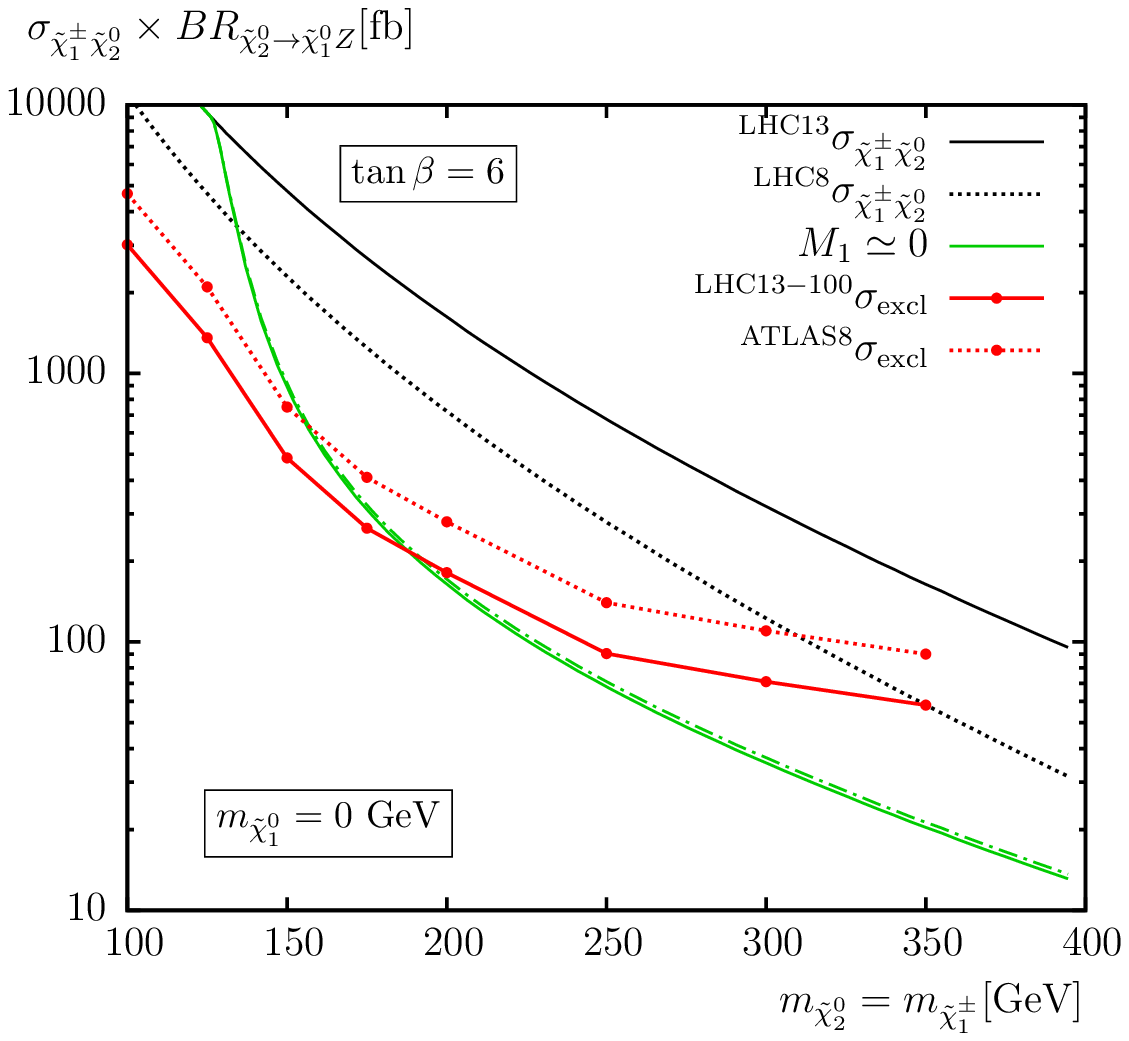}
\hspace{-4mm}
\includegraphics[width=0.49\textwidth,height=7.5cm]{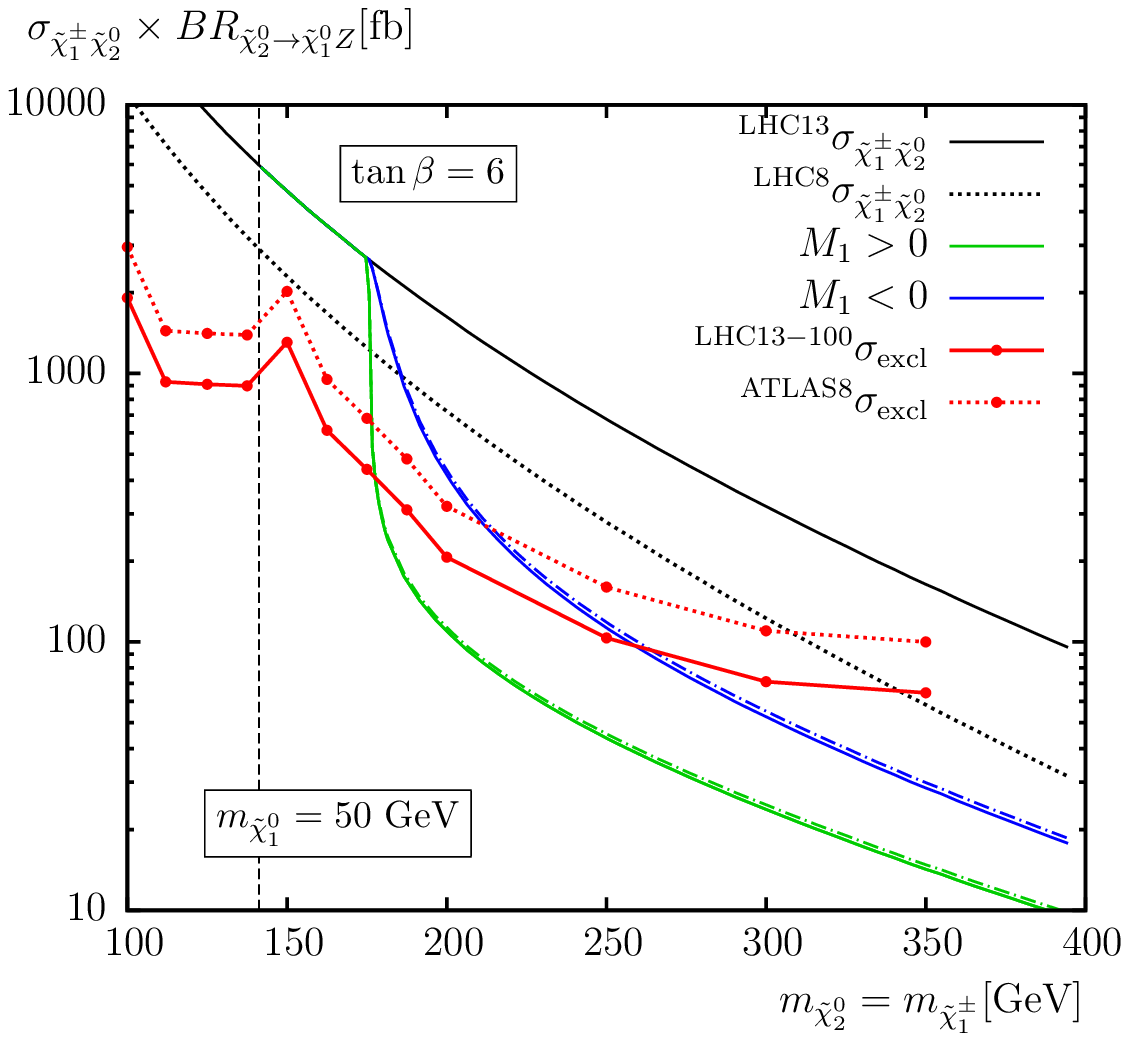} 
\end{tabular}
\begin{tabular}{c}
\includegraphics[width=0.49\textwidth,height=7.5cm]{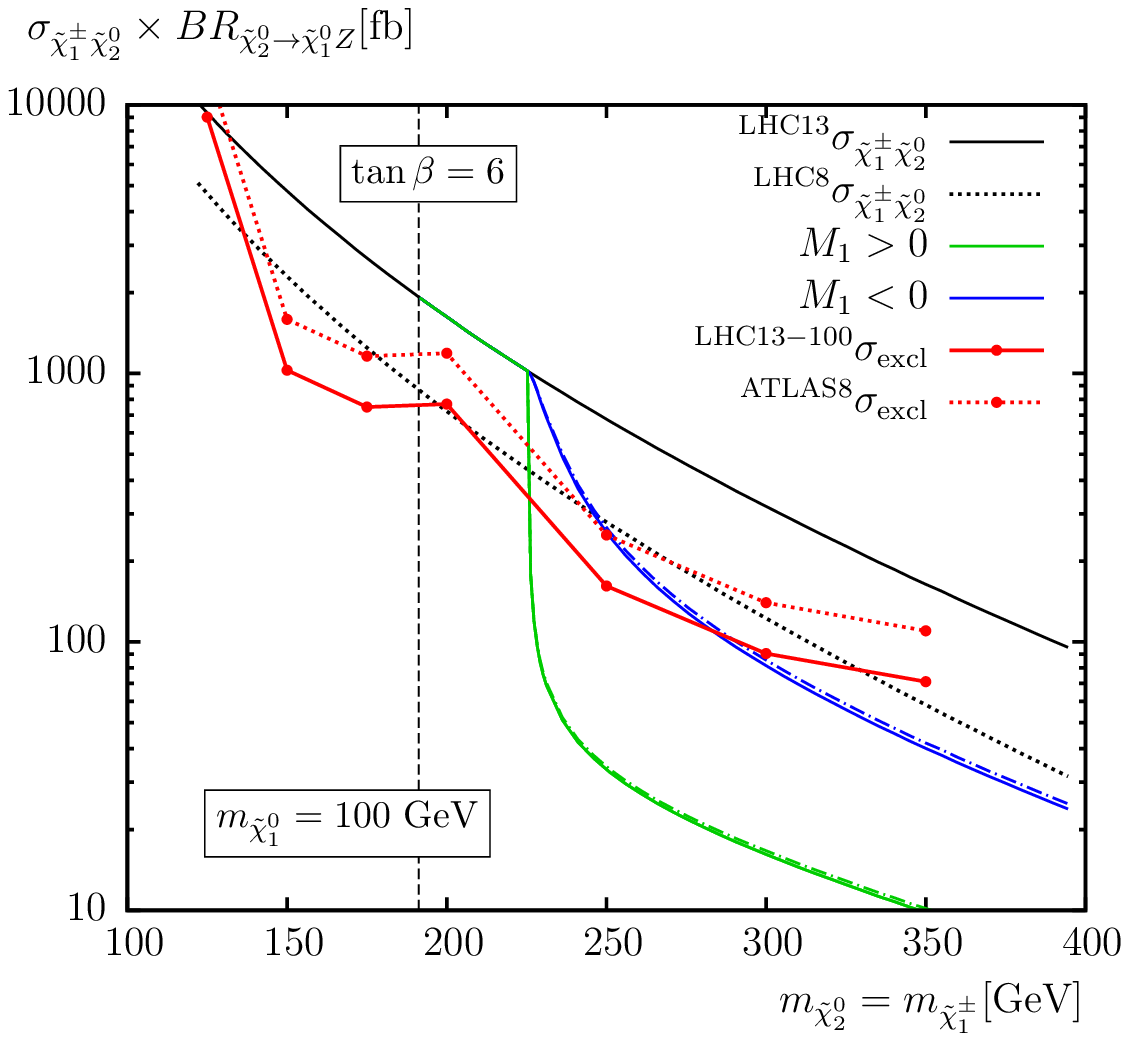}
\hspace{-4mm}
\includegraphics[width=0.49\textwidth,height=7.5cm]{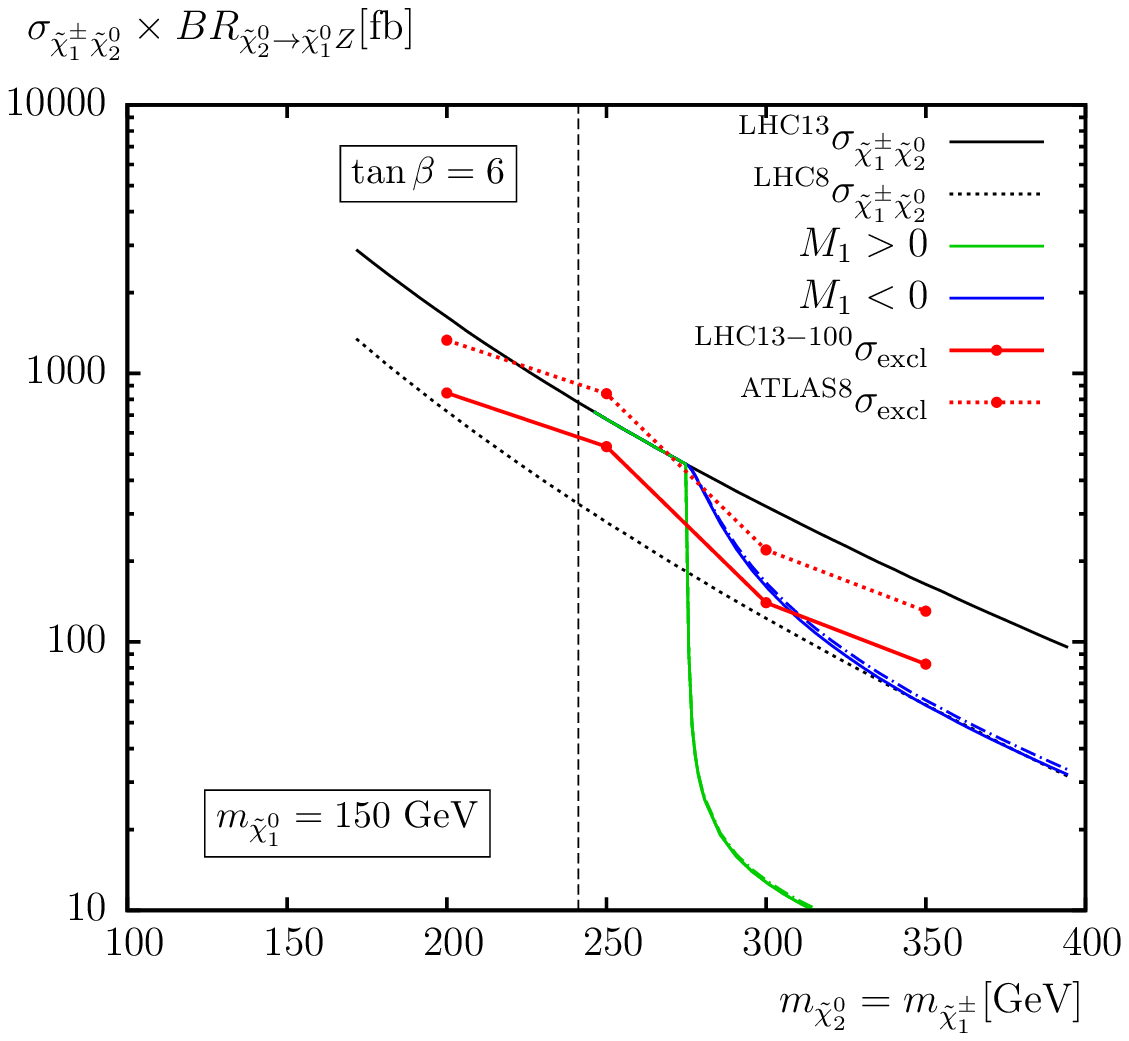} 
\end{tabular}
\caption{
$\cha{1},\neu{2}$ production cross section at the LHC13 evaluated with
 {\tt Prospino~2.1}\cite{prospinoNN} (black) times the BR$(\neu{2}\to Z\neu{1})$ 
at tree (solid) and one-loop level (dashed-dotted)
for $M_1>0$ (green) and $M_1<0$ (blue). 
Also shown is the LHC8 production cross section (thin black).
$M_1$ is chosen such that
 $\mneu{1}=0,50,100,150\gev$, top left and right, bottom left and right, respectively.
The $95\%$ CL exclusion cross sections from  ATLAS at $8\tev$ and $21~\ifb$ (solid red),
  \cite{ATLAS:2013-035} and the projection for $100$ and $300~\ifb$ at LHC13 (dashed and dot-dashed red).
}
\label{fig:xsbr.atlasexcl.lhc13tb6}
\end{center}
\end{figure}
\begin{figure}[t!]
\begin{center}
\begin{tabular}{c}
\includegraphics[width=0.49\textwidth,height=7.5cm]{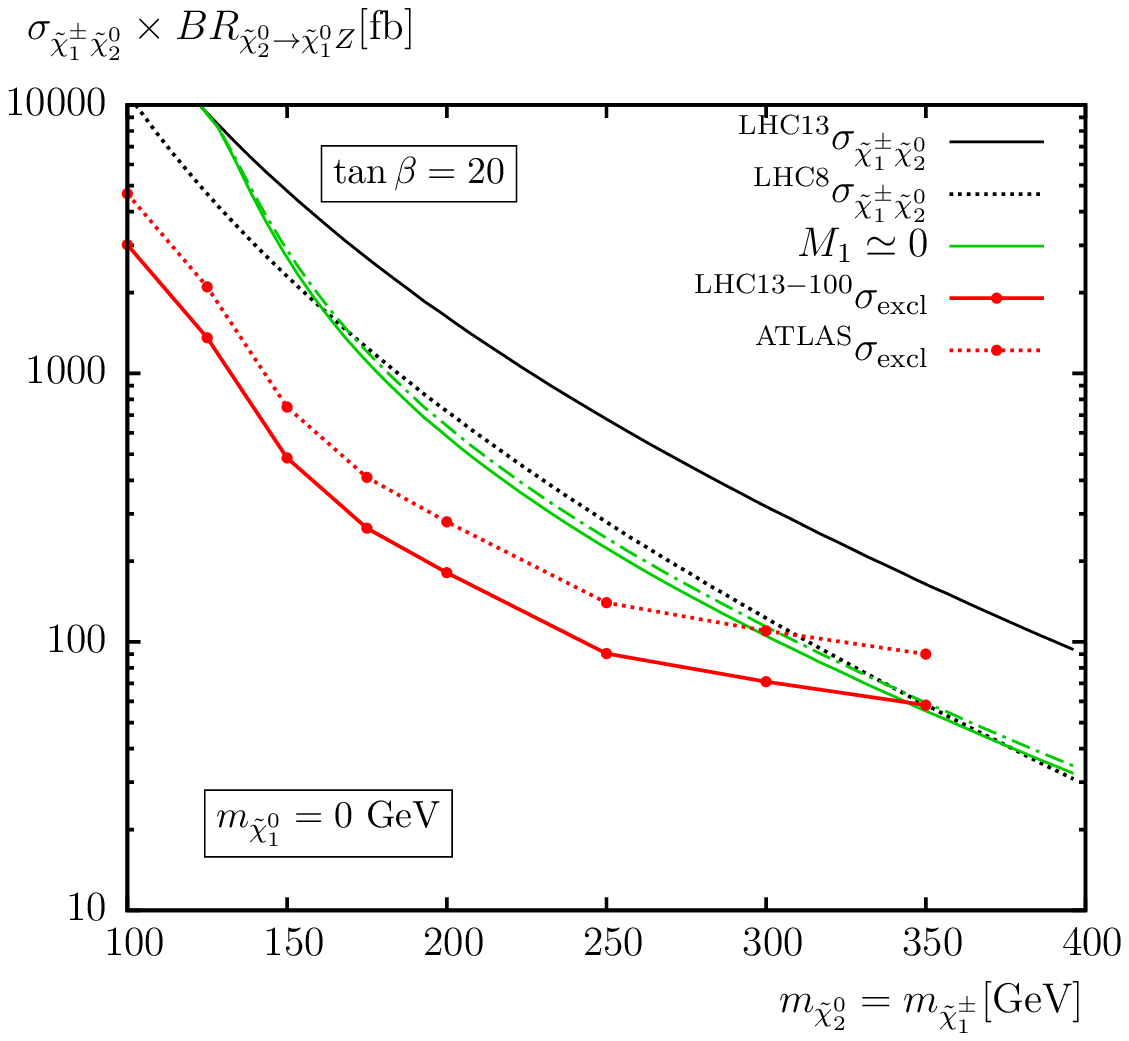}
\hspace{-4mm}
\includegraphics[width=0.49\textwidth,height=7.5cm]{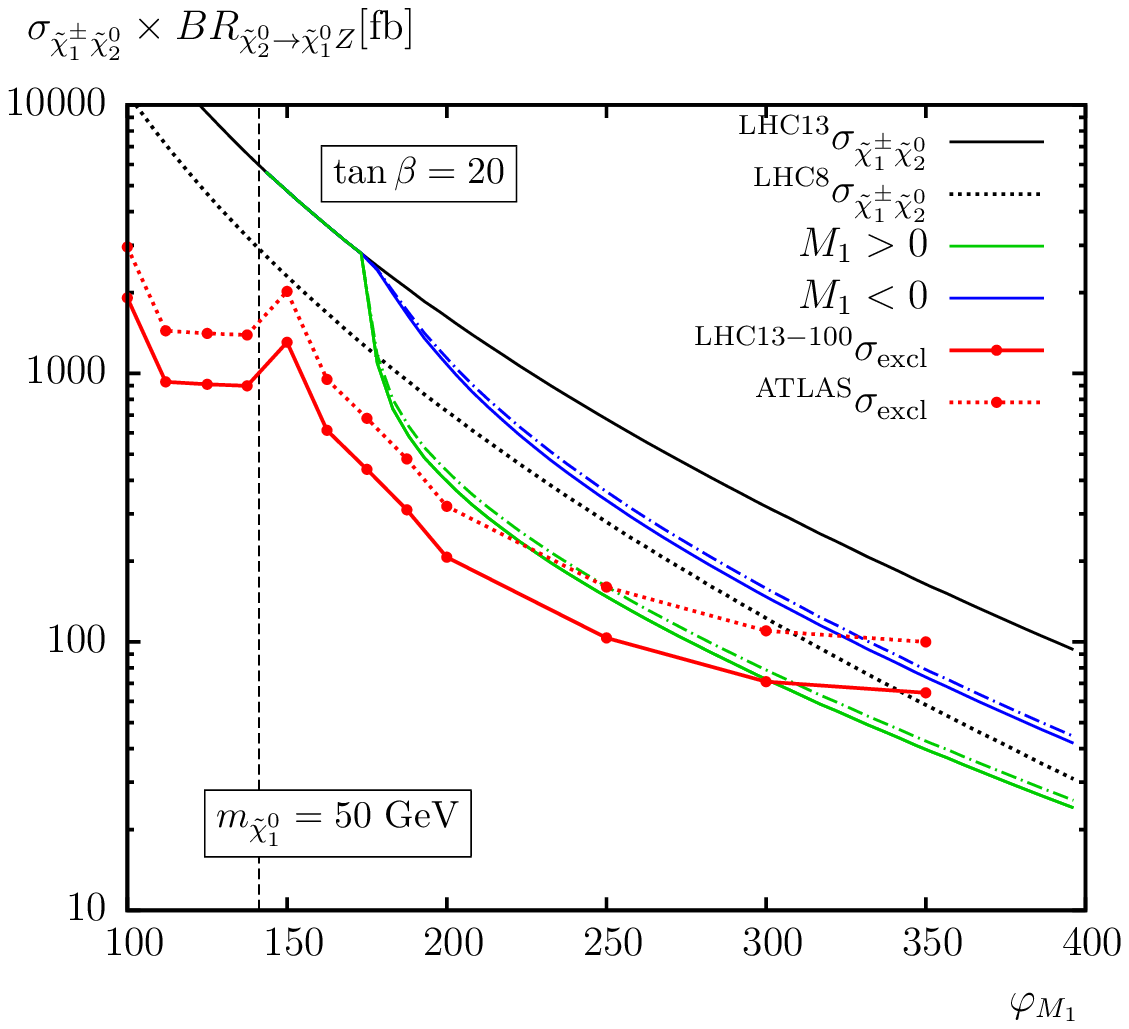} 
\end{tabular}
\begin{tabular}{c}
\includegraphics[width=0.49\textwidth,height=7.5cm]{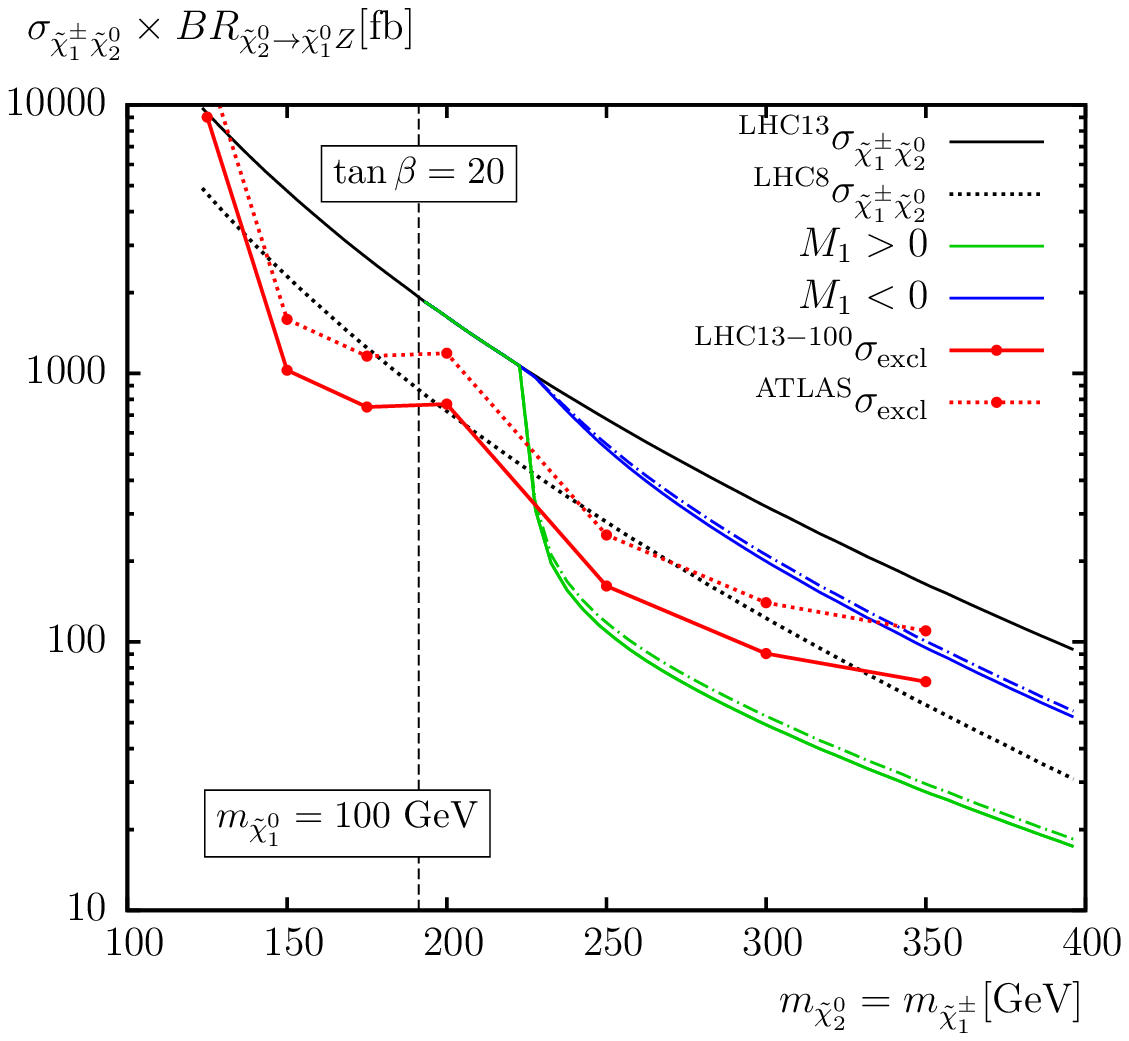}
\hspace{-4mm}
\includegraphics[width=0.49\textwidth,height=7.5cm]{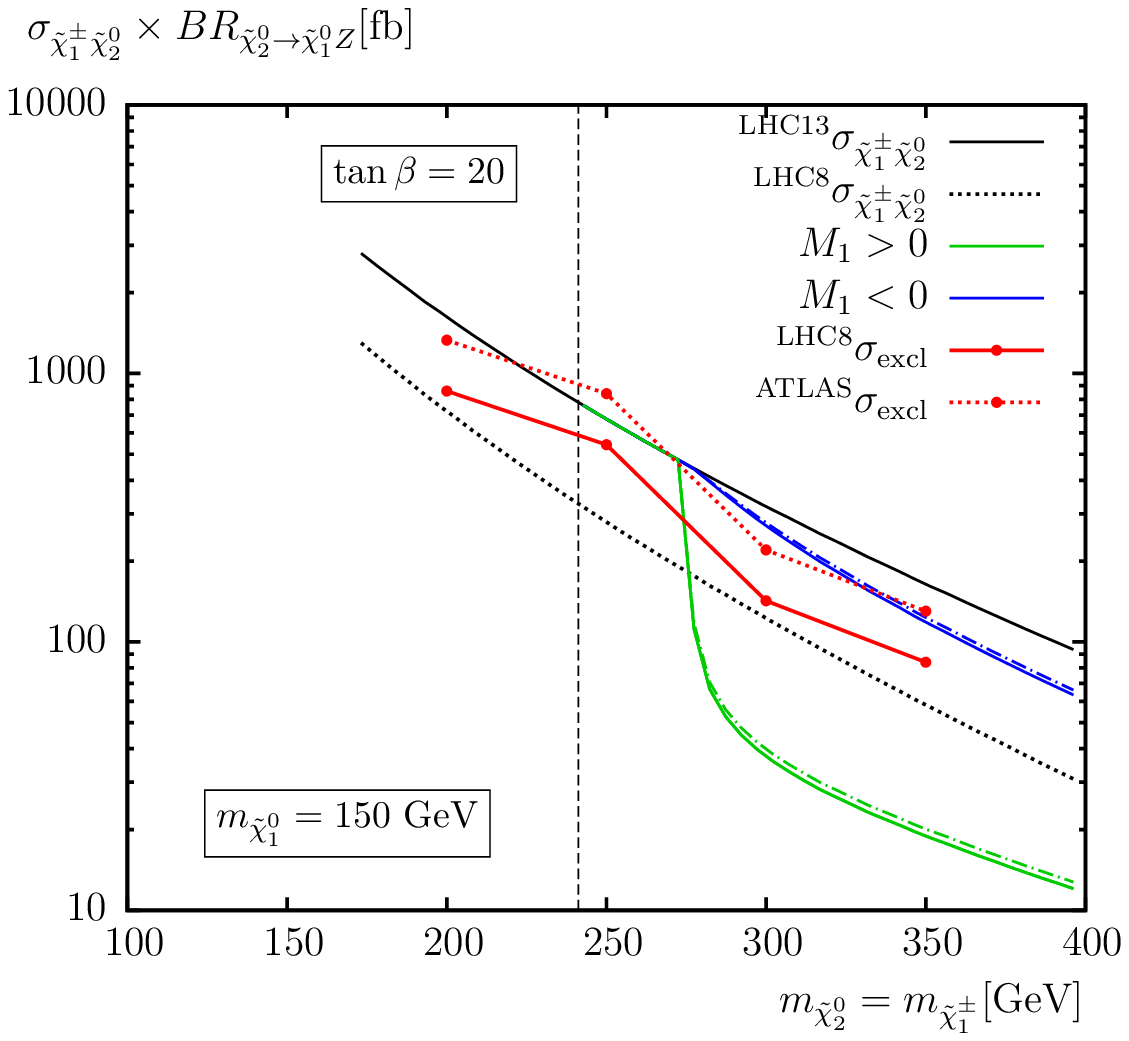} 
\end{tabular}
\caption{
As in \reffi{fig:xsbr.atlasexcl.lhc13tb6}, but for $\tb = 20$.
}
\label{fig:xsbr.atlasexcl.lhc13tb20}
\end{center}
\end{figure}

In \reffi{fig:xsbr.atlasexcl.lhc13tb6} we show the LHC13 expectations
corresponding to the results shown in
\reffi{fig:xsbr.tb6.atlasexcl}. To guide the eye the dashed black line
shows the LHC8 production cross section, and the dotted red line
corresponds to the current exclusion, (i.e.\ the solid red in
\reffi{fig:xsbr.tb6.atlasexcl}), whereas the solid red line in the four
plots of \reffi{fig:xsbr.atlasexcl.lhc13tb6} show the LHC13 expected
exclusion obtained as described above. 
One can observe that the analysis neglecting $\neu2 \to \neu1 \He$ would
exclude all $\mneu2$ values in the parameter space analyzed. Taking the
decay to $\neu1 \He$ into account, however, the bound lies roughly between 
$\mneu2 \sim 200 \gev$ and $\sim 300 \gev$, depending on $\mneu1$ and
the sign of $\MOne$.

The LHC13 expectations are similar for $\tb = 20$ as shown in
\reffi{fig:xsbr.atlasexcl.lhc13tb20}. In agreement with
\reffi{fig:xsbr.tb20.atlasexcl} the excluded regions are larger than for
$\tb = 6$, and for $\mneu1 \ge 50 \gev$ and $\MOne$ negative no bound on
$\mneu2$ can be read of (without extrapolation). 

As for the LHC8 analysis we summarize the results of
\reffis{fig:xsbr.atlasexcl.lhc13tb6}, \ref{fig:xsbr.atlasexcl.lhc13tb20}
in contour plots in the $\mneu2$--$\mneu1$ plane shown in
\reffi{fig:contour-LHC13} for $\tb = 6$ (upper plot) and $\tb = 20$
(lower plot). Here we compare the most recent ATLAS results with the 
LHC13 projections. The color coding is as in
\reffi{fig:contour-LHC8}. In the low $\tb$ case, even at the LHC13 only
relatively small strips going beyond the region where $\neu2 \to \neu1 \He$ 
is kinematically forbidden can be excluded. In the high $\tb$ case the
excluded values of $\mneu2$ for $\mneu1 \sim 0$ for LHC13, taking 
$\neu2 \to \neu1 \He$ into account, are similar to
the limits currently published in \citere{ATLAS:2013-035}, i.e.\ only
with the LHC13 the current simplified models exclusion can be reached. The same
holds for positive $\MOne$ for larger values of $\mneu1$.

\begin{figure}[t!]
\begin{center}
\begin{tabular}{c}
\includegraphics[width=0.73\textwidth]{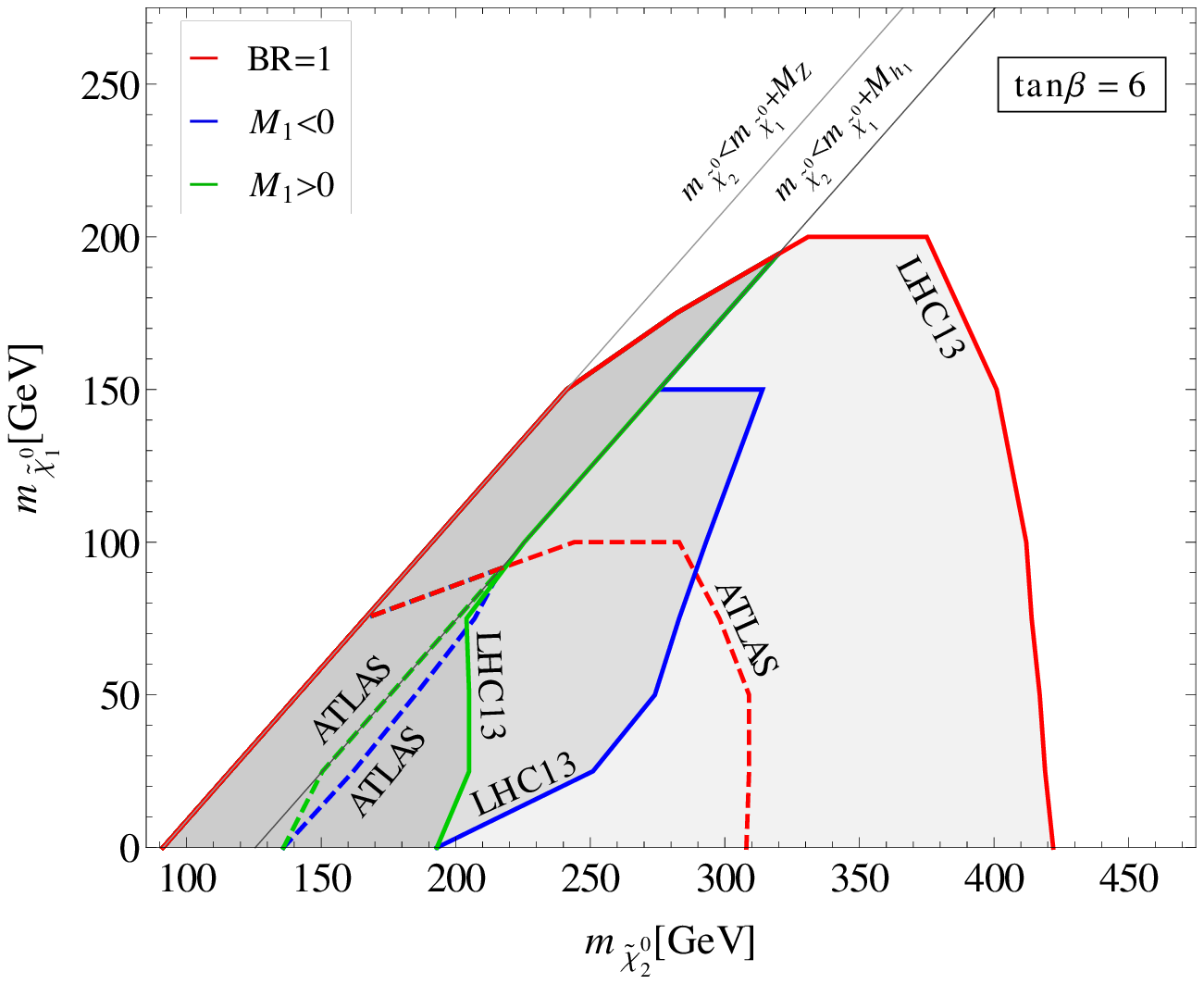}
\end{tabular}
\begin{tabular}{c}
\includegraphics[width=0.73\textwidth]{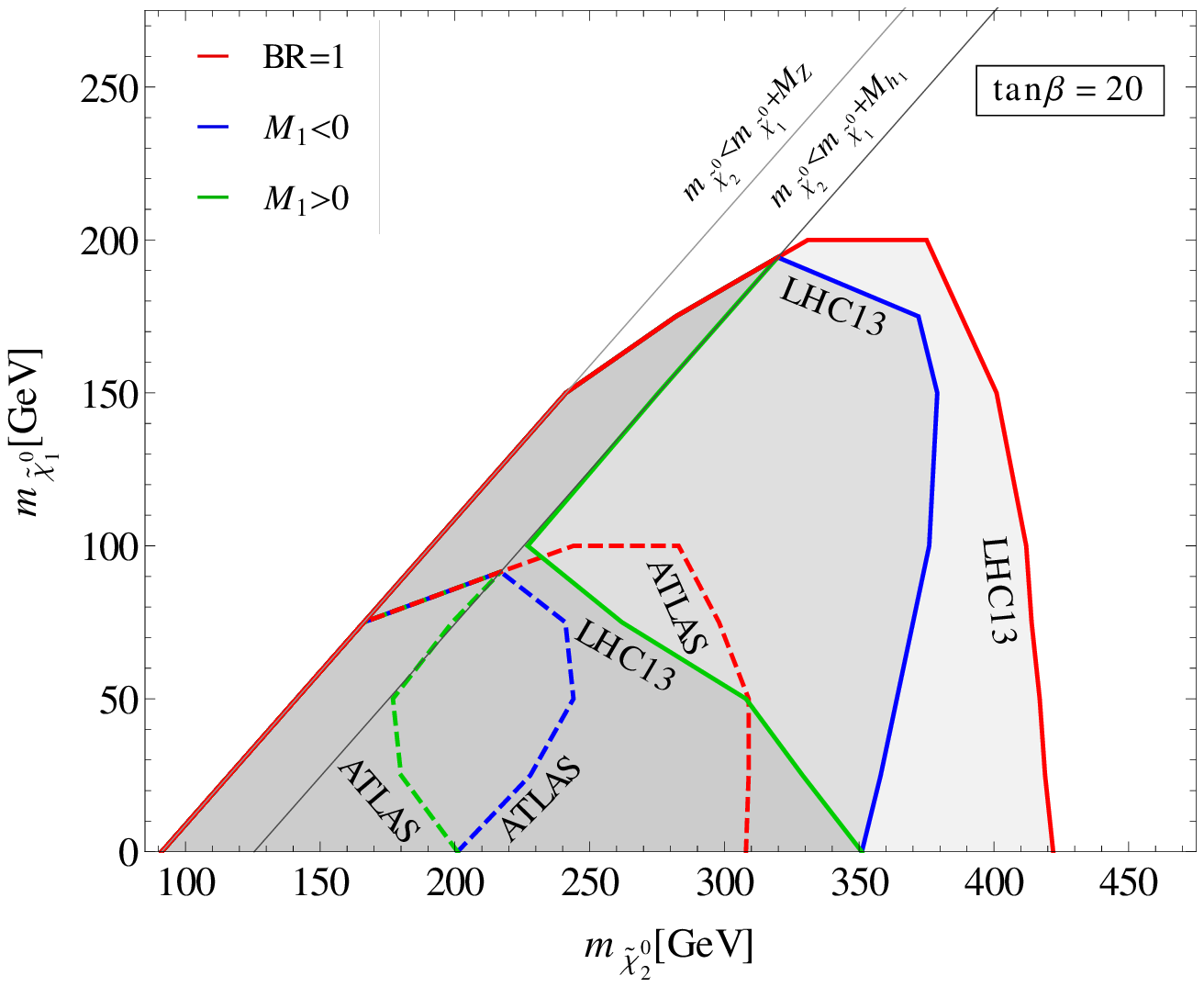}
\end{tabular}
\caption{
Contour plots showing approximate excluded region from \Satlas\
in the $\mneu{2}$--$\mneu{1}$ plane, 
with $\tb = 6$ (upper plot) and  $\tb = 20$ (lower plot), 
showing 
the projected exclusions for $100~\ifb$ at LHC13. The color
coding is as in \reffi{fig:contour-LHC8}.
}
\label{fig:contour-LHC13}
\end{center}
\end{figure}
%

\medskip
The sensitivity of chargino/neutralino searches at the LHC13 has
been discussed in the existing literature and can be related to our
estimates. Searches for the decay $\neu{2}\to\neu{1}\He$ 
will start to become sensitive~\cite{Baer:2012ts,Byakti:2012qk,Ghosh:2012mc,Howe:2012xe,Arbey:2012fa}.
This will provide an interesting complementary channel, which may well
dominate completely over $\neu{2}\to\neu{1}Z$. It has been shown in
\citeres{Baer:2012ts,Ghosh:2012mc} that with 
$100\,\ifb$, a $5\,\si$ discovery should be possible for 
$\mneu{2}\sim 400 \ldots 500 \gev$ (where the GUT relation between $\MOne$ and
$\MTwo$ has been assumed). This might overcome the reduction in sensitivity that
we found in our analysis.
It should finally be noticed that the branching ratio 
$\br(\neu2 \to \neu1 \He)$
may also be strongly dependent on $\cp$ phases and may in this way
provide unique information on the neutralino and Higgs sectors.

\medskip

\clearpage
\newpage

\subsection{Three-body decays of \boldmath{$\neu2$}}
\label{sec:offshell}

In the region where the third lepton (the lepton not part of
the SFOS lepton pair) is softest, ATLAS is supposedly sensitive
to the decays via off-shell gauge bosons, and limits have been set,
allowing the region close to the diagonal in the plane of the masses of
the lightest neutralino and chargino to be probed. 
\footnote{This area was not marked in our contour plots in
\reffi{fig:contour-LHC8}, \ref{fig:contour-LHC13}.}
The relevant decay channels are
\begin{align}
\neu2 \to \neu1 Z^* \to \neu1 l^+l^- \mbox{~and~}
\neu2 \to \Sl^{\pm *} l^\mp \to \neu1 l^+ l^-~,
\label{def:RZ}
\end{align}
where it is crucial that the experimental searches are optimized for the
$\neu2 \to \neu1 Z^*$ channel.

In this section we discuss some problems in the
re-interpretation of the ATLAS limits from~\cite{ATLAS:2013-035}.
We would mainly like to
point out that in the region below the $Z$~threshold, while it is a
reasonable approximation to neglect the 
contribution via off-shell decays to Higgs bosons, there is substantial
destructive interference between the off-shell $Z$ and slepton
channels for slepton masses below approximately $5 |\mu|$ (here $5\tev$). 
This was previously discussed in \citere{Baer:1992dc} and references
therein.
Within the ATLAS analysis the slepton channel was neglected,
effectively by pushing the slepton masses to the multi-$10 \tev$
scale. However, within most realizations of low-energy SUSY such high
slepton mass scales are considered unrealistic, and masses
not far above the LEP limits of \order{100 \gev} are experimentally
permitted -- i.e.\
the interpretation of these bounds in terms of concrete models is difficult. 

To illustrate this point, we have calculated the
ratio of the partial decay width of $\neu{2}\to \neu{1} l^+l^-$ either
excluding those diagrams involving sleptons or excluding diagrams
involving $Z$/$\He$ bosons, to the partial width including all possible
diagrams, 
\begin{align}
 R_{Z}=\frac{\Ga^{{\rm no}\,\,\tilde{l}}(\neu{2}\to \neu{1} l^+l^-)}
            {\Ga^{\rm total}(\neu{2}\to \neu{1} l^+l^-)}
 \quad\mbox{and} \qquad R_{\tilde{l}}=
  \frac{\Ga^{{\rm no}\,\,Z\He}(\neu{2}\to \neu{1} l^+l^-)}
       {\Ga^{\rm total}(\neu{2}\to \neu{1} l^+l^-)}
\end{align}
for $l=e$ and $\mu$. 
The results are shown for the ATLAS baseline scenarios in
\reffi{fig:1to3} for $\tb = 6$ ({\Satlas,} left) and $\tb = 20$ (\Stb, right).
$R_Z$ ($R_{\Sl}$) is shown as solid (dashed) lines for $\MOne > 0$ (blue) and
$\MOne < 0$ (green). Note that this interference is
sensitive to $\mu$, and the position of the peak of this quantity is
close to the value of $\mu$, which in \Satlas\ is $1 \tev$. 
Note furthermore that the coupling
$\neu{2}\to \neu{1} Z$ is inversely proportional to $\mu^2$, see
\refeq{eq:nnz.approx}.
On the other hand the slepton propagator renders a factor
proportional to $1/\Msl^2$ in the amplitude for the
decay via sleptons. Consequently, in the region where $\mu\sim \Msl$,
the interference is maximal, as can be observed in \reffi{fig:1to3}. 
Further, as
the slepton mass increases, the contribution of the sleptons decouples,
and $R_Z$ approaches~1, whereas $R_{\Sl}$ approaches zero, as
expected. 
A similar interference effect might also occur below the threshold
for the chargino decay $\cha{1}\to\neu1 W^-$, however a detailed study is 
beyond the scope of this paper.
In summary, \reffi{fig:1to3} shows that the slepton mass scale
has a dramatic effect on the composition of the decay channel 
$\neu2 \to \neu1 l^+l^-$, which is crucial for the experimental
analysis. Consequently, the excluded regions should be viewed with care
and not taken at face value. The published limits are only valid for
(normally deemed unrealistic) relatively high slepton mass scales.

\begin{figure}[t!]
\begin{center}
\begin{tabular}{c}
\includegraphics[width=0.5\textwidth]{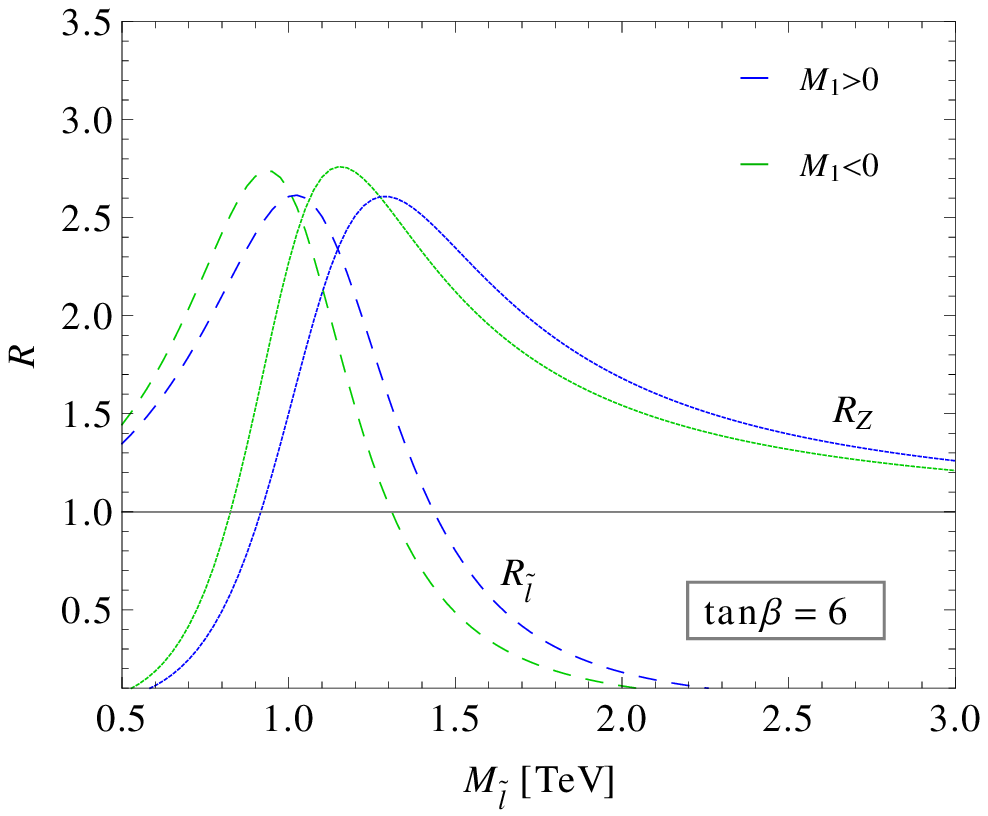}
\includegraphics[width=0.5\textwidth]{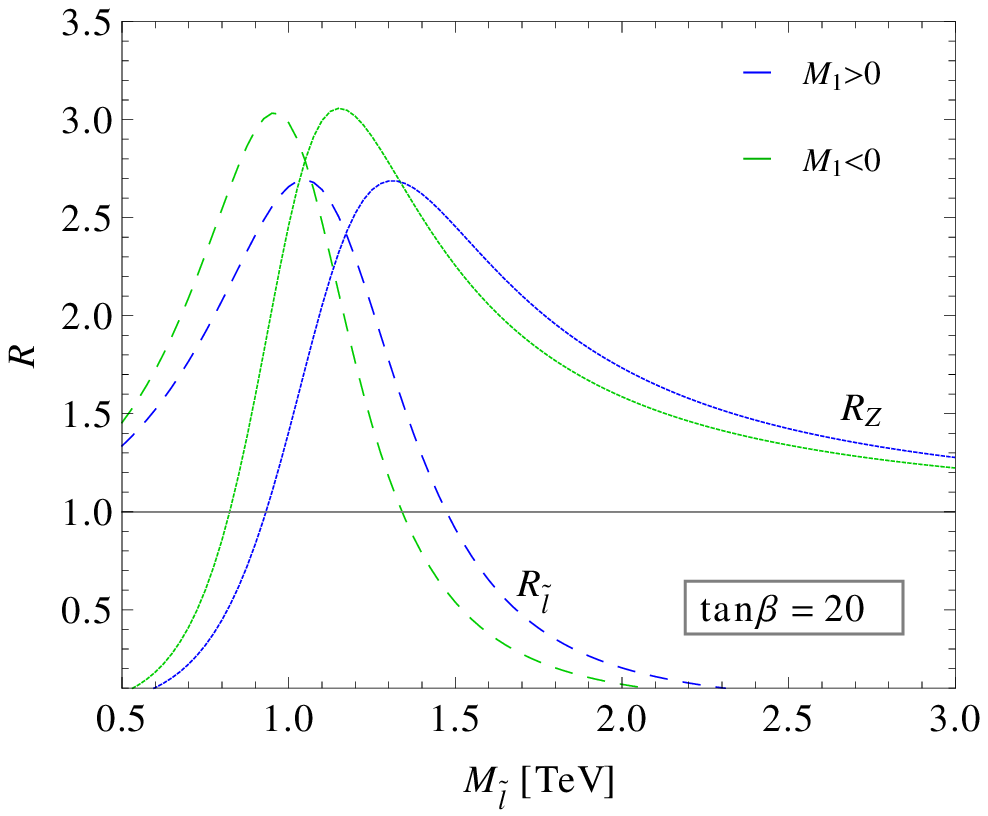}\\
\end{tabular}
\caption{
We show the ratios $R_{Z}$ (solid) and $R_{\tilde{l}}$ (dashed), as
defined in \refeq{def:RZ} as a function of $\Msl$ for $\tb = 6$ (left)
and $\tb = 20$ (right). Results for 
$\MOne > 0$ are shown in blue, for $\MOne < 0$ in green.
All other parameters are as for our ATLAS baseline scenario.
}
\label{fig:1to3}
\end{center}
\end{figure}


\section{Conclusions}
\label{sec:conc}

We re-assessed the exclusion limits on the parameters describing the
supersymmetric (SUSY) electroweak sector of the MSSM obtained from
the search for direct $\cha1 \neu2$ production via $WZ+E_T^{\rm miss}$
at the LHC.
Our starting point is the baseline scenario used by ATLAS in their most 
recent public note
using $21~\ifb$~\cite{ATLAS:2013-035}. This analysis is carried out
for a simplified model scenario where it is assumed that the 
$\cha1$ and $\neu2$ decay 100\% via $\cha1 \to \neu1 W^\pm$ and 
$\neu2 \to \neu1 Z$, and 
limits on $\mneu2$ of up to $\sim 300 \gev$ are derived, mostly
displayed in the $\mneu2$--$\mneu1$ plane.
In our analysis we investigated how these limits change on using NLO results both for the SUSY production cross
sections as well as for the branching ratio calculations. 

The first step in our analysis was the inclusion of the decay 
$\neu2 \to \neu1 \He$, which can substantially lower 
$\br(\neu2 \to \neu1 Z)$, thereby strongly reducing the excluded parameter space. 
Besides the region where
$\neu2 \to \neu1 \He$ is kinematically forbidden, only a very small
strip in the $\mneu2$--$\mneu1$ plane can be excluded. 
As a second step we allowed the gaugino mass parameter $\MOne$
to take negative values (corresponding to $\phiMe = \pi$). In this case
slightly larger regions in the $\mneu2$--$\mneu1$ plane can be
excluded. Going from the baseline value $\tb = 6$ to $\tb = 20$ again
leads to somewhat larger excluded regions, but the decay
$\neu2 \to \neu1 \He$ is still clearly seen to have a substantial effect on
the limits.  

We additionally assessed how much a combination of ATLAS and CMS data could
change 
the results (effectively assuming a doubling of the analyzed ATLAS
data). Only in the most ``favorable'' case, $\tb = 20$ and 
$\phiMe = \pi$ the combined ATLAS/CMS analysis nearly reaches the
exclusion region reported by ATLAS alone for the simplified model case.

As the next step we investigated the dependence of the excluded mass
regions on the phase of $\MOne$. By projecting the results onto the
$\phiMe$--$\MOne$ plane a strong dependence on $\phiMe$ becomes
evident. 
In the future, limits on $WZ+E_T^{\rm miss}$ and $Wh+E_T^{\rm miss}$ 
could also be exploited as a method to constrain $\phiMe$,
complementary to the EDMs.
Furthermore, the relevance of the loop corrections in the
branching ratio calculations was shown to reach more than $10 \gev$ in
the limit on $\MOne$, amounting up to a change of $30\%$.
The overall size of the NLO corrections in the branching ratio
calculations was found to reach the level of up to $12\%$. 

Another interesting deviation from the ATLAS baseline scenario is given by
the inclusion of a light scalar tau,  
with $\mstaul - \mneu1 \lsim 10\gev$, such that
the $\neu1$ provides the correct amount of relic Cold Dark
Matter~\cite{cdm,micromegas}. This 
opens up the decay modes $\neu2 \to \Staue \tau$ and 
$\cha1 \to \Staue^\pm \nu_\tau$, further reducing the desired branching
ratios for $\neu2 \to \neu1 Z$. We have shown that the new decay modes
can strongly influence the parameter dependences of 
$\br(\neu2 \to \neu1 Z)$ and thus require a new analysis in this scenario
where the searches for different relevant decays modes are combined.

The final scenario analyzed has $\MOne \le \mu \le \MTwo$, leading to 
higgsino-like neutralinos $\neu2$, $\neu3$, and a higgsino-like
chargino $\cha1$ with $\mneu2 \approx \mneu3 \approx \mcha1$. 
In this scenario the lower value of $\mu$ results in strongly reduced
production cross sections, and no mass value can be excluded by
re-analyzing the published bounds. In this scenario a combination not
only of ATLAS and CMS data, but also of the production modes $\cha1 \neu2$ 
and $\cha1 \neu3$ will have to be performed to reach sensitivity for
higgsino masses below $\sim 200 \gev$. 

As a last step we presented the exclusion regions expected for $100~\ifb$
analyzed by ATLAS during the next LHC run at $\sqrt{s} = 13 \tev$,
assuming the absence of any signal. We showed that with the strong
increase in the integrated luminosity as well as with the increase in
the production cross sections, the current simplified model exclusion
regions can roughly be reached, where details depend on the $\phiMe$ and $\tb$. 
This would result in an important advance into MSSM parameter space,
and any hints of low charginos--neutralinos seen could further be
investigated at the linear collider~\cite{Bharucha:2012ya}.

Finally, we briefly investigated the regions where ATLAS claims
sensitivity to $\cha1 \neu2$ production via the off-shell decay 
$\neu2 \to \neu1 l^+l^-$, i.e.\ below the kinematic threshold for
$\neu2 \to \neu1 Z$. We showed that more realistic values for the
slepton mass scale can strongly enhance the interference between
$\neu2 \to \neu1 Z^* \to \neu1 l^+l^-$ and 
$\neu2 \to \Sl^\pm l^\pm \to \neu1 l^+l^-$. This in turn will have a
strong impact on the lepton distributions and thus on the experimental
analysis in this kinematic region.

\medskip
In summary we have re-analyzed the latest ATLAS limits on direct
electroweak SUSY production. We have found that translating the public
limits obtained in simplified scenarios to realistic scenarios, where in
particular the decay $\neu2 \to \neu1 \He$ is accounted for, strongly
reduced the excluded parameter regions.  
Conversely, we encourage the LHC
experiments to include the channel $\neu2 \to \neu1 \He$ into their analysis,
as this would provide access to SUSY-Higgs couplings and strengthen the
electroweak SUSY searches.


\subsection*{Acknowledgments}

We thank  
A.~Calder\'on,
J.~Dietrich, 
M.~Elsing,
A.~H\"ocker,
N.~Kauer,
F.~Moortgat,
G.~Moortgat-Pick,
T.~Potter,
W.~Waltenberger
and
G.~Weiglein
~for helpful discussions. 
A.B.\ gratefully acknowledges support of the DFG through the grant SFB 676,
``Particles, Strings, and the Early Universe''. 
The work of S.H.\ was partially supported by CICYT (grant FPA
2010--22163-C02-01). 
F.v.d.P.\ was supported by 
the Spanish MICINN's Consolider-Ingenio 2010 Programme under grant
MultiDark CSD2009-00064. 
We thank the GRID computing network at IFCA for 
technical help with the OpenStack cloud infrastructure.



\begin{thebibliography}{99}
\bibitem{feynarts-mf} The couplings can be found in the files
                      {\tt MSSM.ps.gz}, {\tt MSSMQCD.ps.gz} and 
                      {\tt HMix.ps.gz} as part of the
                      \fa~package~\cite{feynarts}.
\end{thebibliography}
\end{document}